\documentclass[fleqn,usenatbib]{mnras}
\usepackage{graphicx}
\usepackage{natbib}
\usepackage{epstopdf}
\usepackage{amssymb}
\usepackage{amsmath}
\usepackage{mathptmx}
\usepackage{multirow}
\usepackage{array,booktabs}
\usepackage{hyperref}
\usepackage{float}
\usepackage{subfig}

\newcommand{\msun}{\,{\rm M_{\odot}}}

\newcommand{\cm}{\,{\rm cm}}

\newcommand{\tdeff}{\,{t_{d,\rm eff}}}
\newcommand{\Ljiso}{\,{L_{j,\rm iso}}}
\newcommand{\Ejiso}{\,{E_{j,\rm iso}}}
\newcommand{\qj}{\,{\theta_{j,0}}}
\newcommand{\al}{\,{\alpha}}

\newcommand{\Eej}{\,{E_{ej}}}
\newcommand{\Eejtot}{\,{E_{ej,{\rm tot}}}}
\newcommand{\Ejtot}{\,{E_{j,{\rm tot}}}}
\newcommand{\vej}{\,{v_{ej,h}}}
\newcommand{\vejh}{\,{v_{ej,h}}}
\newcommand{\mmin}{\,{m_{\rm min}}}
\newcommand{\vmin}{\,{v_{\rm min}}}
\newcommand{\rmin}{\,{r_{\rm min}}}
\newcommand{\vmax}{\,{v_{\rm max}}}
\newcommand{\rmax}{\,{r_{\rm max}}}
\newcommand{\rhcol}{\,{r_{h,{\rm col}}}}
\newcommand{\vhcol}{\,{v_{h,{\rm col}}}}
\newcommand{\tti}{\,{\tilde{t}}}
\newcommand{\E}{\,{\tilde{E}}}
\newcommand{\Etot}{\,{\tilde{E}_{\rm tot}}}
\newcommand{\NEtot}{\,N_{\tilde{E}_{\rm tot}}}
\newcommand{\tc}{\,{\tti_{\rm col}}}
\newcommand{\Ncrit}{\,{N_{\rm crit}}}
\newcommand{\Ncol}{\,{N_{\rm col}}}
\newcommand{\ta}{\,{\tti_{\rm a}}}
\newcommand{\etauc}{\,{\eta_{\rm uncol}}}
\newcommand{\etac}{\,{\eta_{\rm col}}}
\newcommand{\etacav}{\,{\eta_{\rm cav}}}
\newcommand{\etacrit}{\,{\eta_{\rm crit}}}
\newcommand{\Lt}{\,{\tilde{L}}}
\newcommand{\Ca}{\,{\it \alpha_2-190}}
\newcommand{\Nc}{\,{\it \alpha_2n-240}}
\newcommand{\Wc}{\,{\it \alpha_2w-390}}
\newcommand{\Cb}{\,{\it \alpha_2-1.7}}
\newcommand{\Cc}{\,{\it \alpha_2-1}}
\newcommand{\Cd}{\,{\it \alpha_2-0.5}}
\newcommand{\Ce}{\,{\it \alpha_2-{0.15}}}
\newcommand{\Cf}{\,{\it \alpha_2-{0.05}}}
\newcommand{\Oa}{\,{\it \alpha_{1}-370}}
\newcommand{\Ob}{\,{\it \alpha_{1}-1}}
\newcommand{\Ta}{\,{\it \alpha_{3}-39}}
\newcommand{\Tb}{\,{\it \alpha_{3}-1}}
\newcommand{\Tc}{\,{\it \alpha_{3}-0.05}}
\newcommand{\Fa}{\,{\it \alpha_{4}-50}}
\newcommand{\Fb}{\,{\it \alpha_{4}-1}}
\newcommand{\Gh}{\,{\Gamma_h}}
\newcommand{\bmin}{\,{\beta_{\rm min}}}
\newcommand{\bmax}{\,{\beta_{\rm max}}}

\title[Jet propagation in expanding media]{The propagation of relativistic jets in expanding media}
\author[Gottlieb \& Nakar]{
    Ore Gottlieb$^{1,2}$\thanks{ore@northwestern.edu},
	Ehud Nakar$^{1}$
	\\
	$^{1}${School of Physics and Astronomy, Tel Aviv University, Tel Aviv 69978, Israel}\\
	$^{2}${Center for Interdisciplinary Exploration \& Research in Astrophysics (CIERA), Physics \& Astronomy, Northwestern University, Evanston, IL 60201, USA}
}
\pubyear{2021}
\begin{document}
	\label{firstpage}
	\pagerange{\pageref{firstpage}--\pageref{lastpage}}
	\maketitle	
\begin{abstract}
We present a comprehensive analytic model of relativistic jet propagation in expanding homologous media (ejecta). This model covers the entire jet evolution as well as a range of configurations that are relevant to binary neutron star mergers. These include low and high luminosity jets, unmagnetized and mildly magnetized jets, time-dependent luminosity jets, and Newtonian and relativistic head velocities. We also extend the existing solution of jets in a static medium to power-law density media with index $\alpha<5$. Our model provides simple analytic formulae (calibrated by 3D simulations) for the jet head propagation and breakout times.  We find that the system evolution has two main regimes: strong and weak jets.  Strong jets start their propagation immediately within the ejecta. Weak jets are unable to penetrate the ejecta at first, and breach it only after the ejecta expands significantly, thus their evolution is independent of the delay between the onset of the ejecta and the jet launching. After enough time, both strong and weak jets approach a common asymptotic phase. We find that a necessary, but insufficient, criterion for the breakout of unmagnetized [weakly magnetized] jets is $E_{j,{\rm iso,tot}} \gtrsim 3[0.4]\Eejtot\left({\qj}/{0.1{\rm~rad}}\right)^2$, where $E_{j,{\rm iso,tot}}$ is the jet total isotropic equivalent energy, $\qj$ is its opening angle, and $\Eejtot$ is the ejecta energy. Applying our model to short GRBs, we find that there is most likely a large diversity of ejecta mass, where mass $ \lesssim 10^{-3}~{\rm M}_{\odot} $ (at least along the poles) is common.
\end{abstract}

	\begin{keywords}
		{ gamma-ray bursts | neutron star mergers | methods: analytical | methods: numerical }
	\end{keywords}

\section{Introduction}

Relativistic jets appear to be common in a wide range of astrophysical systems. The jet forms following accretion of infalling mass onto a compact object through an accretion disk or via fast rotation of a highly magnetized neutron star. In a variety of astrophysical environments the jet encounters dense media during its propagation. These can be the infalling mass, outflows from the compact object or the disk, or some other mass that surrounds the compact object. The interaction between the jet and the media is an essential part of the jet evolution, affecting the jet velocity, collimation and stability. This encounter ultimately dictates whether the jet breaks out from the media and it strongly affects its structure in case that it does.

The first works that studied the jet-medium interaction considered the simplest setup of hydrodynamic (unmagnetized) jets propagating in a static medium, addressing it both analytically \citep[e.g.,][]{Blandford1974,Begelman1989,Meszaros2001,Matzner2003,Lazzati2005,Bromberg2011}, and numerically \citep[e.g.,][]{Marti1995,Marti1997,Aloy2000,MacFadyen2001,Zhang2004,Mizuta2006,Morsony2007,Wang2008,Rossi2008,Lazzati2009,Mizuta2009,Morsony2010,Nagakura2011,Lopez-camara2013,Lopez-camara2016,Matsumoto2013,Matsumoto2019,Mizuta2013,Ito2015,Toma2017,Harrison2018,Gourgouliatos2018,Gottlieb2019,Gottlieb2020a,Gottlieb2021a}. These studies found that the interaction of the jet with the medium leads to the formation of a hot cocoon that envelops the jet, and has a fundamental role in the collimation and stability of the jet.

In recent years additional numerical works also included magnetic fields in order to study how they alter the hydrodynamic picture \citep[e.g.,][]{Meliani2009,Mizuno2009,Mizuno2012,Tchekhovskoy2016,Bromberg2016,Bromberg2019,Kathirgamaraju2018,Kathirgamaraju2019,Gottlieb2020b,Gottlieb2021b,Gottlieb2022b,Gottlieb2022,Matsumoto2021,Nathanail2021}. In general, the effect of the magnetization depends on the magnetic field configuration (e.g., poloidal, toroidal etc.), and its strength compared to the internal energy, $ \sigma \equiv \frac{B'^2}{4\pi h\rho c^2} $, where $ B' $ is the proper magnetic field, $ h $ is the specific enthalpy and $ \rho $ is the proper mass density. It was shown that while low $ \sigma $ jets ($10^{-2} \lesssim \sigma \lesssim 10^{-1} $) are still hydrodynamically dominated, their magnetization might be sufficient to stabilize the jet against hydrodynamic instabilities, thereby supporting faster jets and less energetic cocoons \citep{Gottlieb2020b,Gottlieb2021a,Matsumoto2021}. In high $ \sigma $ jets ($ \sigma \gtrsim 1 $), current-driven instabilities such as kink emerge, rendering the jet structure globally unstable and considerably altering the hydrodynamical picture \citep[e.g.,][]{Bromberg2016,Gottlieb2022b,Gottlieb2022}.

The first detection of a neutron star merger, GW170817, by gravitational waves (GW) was followed by detections across the entire electromagnetic spectrum \citep[see reviews in][and references therein]{Nakar2019,Margutti2020}. The kilonova optical signal, powered by radioactive decay of $r$-process elements, revealed the ejection of subrelativistic massive ($ \sim 5\times 10^{-2}~\msun $) component from the merger, slightly more massive than the predictions of numerical relativity simulations which find mass ejection of $ 10^{-3}-10^{-2} ~\msun $. Months later, the multi-band afterglow light curve and VLBI observations revealed  a relativistic jet that is similar to the one expected in short gamma-ray burst (sGRB). According to any reasonable chain of events, at least a non-negligible fraction of the subrelativistic outflow was ejected before the launching of the jet. Thus, the observations of GW170817 have shown that the jet has propagated through the expanding subrelativistic ejecta that generated the kilonova emission, and it stands to reason that a similar propagation also takes place in sGRBs that are detected via their $ \gamma $-rays. Additional astrophysical system where there may be a jet that propagates in expanding medium is hydrogen-poor superluminous supernovae \citep[e.g.,][]{Margalit2018}.

The propagation of relativistic jets in expanding medium started to draw special attention in anticipation of the first GW signal from binary neutron star (BNS) mergers \citep{Murguia-Berthier2014,Murguia-Berthier2017,Lazzati2017,Gottlieb2018a}, and it became the focus of many studies following GW170817 \citep{Kasliwal2017,Gottlieb2018b,Gottlieb2021a,Duffell2018,Kathirgamaraju2018,Lazzati2019,Geng2019,Gottlieb2020c,Gottlieb2021c,Gottlieb2022c,Ito2021,Klion2021,Murguia-Berthier2021,Pavan2021,Urrutia2021,Lamb2022,Nativi2022}. Most of these studies used numerical simulations showing that the jet-ejecta interplay shapes the emerging jet-cocoon structure, and thus has important observational implications on  the prompt emission and the afterglow light curve, and possibly also on the kilonova emission.

First attempts of deriving an analytic model for the propagation of relativistic jets in expanding media have started recently with the works of
\citet{Duffell2018,Matsumoto2018,Lyutikov2020,Margalit2018,Nakar2019,Hamidani2020,Hamidani2021}. These studies have provided a major step-forward in understanding the effect of expanding medium on the jet propagation.
However, the analytic description in these studies is incomplete and in some of the studies it is also not fully accurate. Moreover, these studies focused only on part of the evolutionary regimes and considered only a limited range of the jet-medium properties phase space, leaving  many aspects of the jet propagation unresolved, some of which are directly relevant for BNS mergers and sGRBs. In this paper we derive an analytic description of jet propagation within expanding homologous media that covers all the phases of the evolution and a wide range of initial conditions. These include: (i) strong jets that breach the ejecta immediately after launch as well as weak jets that are  initially stalled by the ejecta, and can only penetrate it following a considerable expansion; (ii) the entire evolution of the jet, starting on timescales that are much shorter than the delay time between the subrelativistic mass ejection and the jet launching; (iii) the initial collimation phase of the jet; (iv) relativistic jet head; (v) various jet magnetizations; (vi) time dependent luminosity. We calibrate  our analytic formula using 3D relativistic magnetohydrodynamic simulations. Previous analytic models either did not use numerical simulations or made use of axisymmetric simulations. It was shown in the past that in significant part of the parameter phase space 3D models  produce different results than those of 2D ones \citep{Gottlieb2021a}, as we also find for our models here. Thus, even in regimes that were explored by previous studies we obtain different calibrations.

The paper starts with an overview of the jet dynamics (\S\ref{sec:overview}) followed by an analytic model of the jet propagation in the Newtonian regime through all the various phases and the entire range of initial conditions (\S\ref{sec:analytic}). We derive a very simple breakout criterion and a formula for the breakout time in the Newtonian regime in \S\ref{sec:E_tilde_and_BO}. We generalize our solution to a time-dependent luminosity in \S\ref{sec:varying_L} and to the relativistic head regime in \S\ref{sec:relativistic}. We test our analytic model using numerical simulations and find calibration coefficients in \S\ref{sec:numerical}. In \S\ref{sec:cavity} we examine the effect of the density in the cavity, which may form in the ejecta before the jet is launched, on the jet propagation.
In \S\ref{sec:magnetization} we test numerically the effect of non-negligible sub-dominant magnetization (i.e., $\sigma \sim 0.01-0.1 $) on the jet propagation velocity showing that the presence of magnetic fields might be equivalent to an increase of an order of magnitude in the jet luminosity. In \S\ref{sec:previous} we review the results of previous studies and compare them with ours. In \S\ref{sec:BNS} we discuss the implications of our results on the understanding of the sGRB population in general and on GW170817 in particular. We summarize and conclude in \S\ref{sec:summary}.

\section{An overview of the jet dynamics}\label{sec:overview}

\subsection{Initial conditions}
Consider a bi-polar relativistic jet. The jet is launched with a half opening angle $\theta_{j,0}$ and the jet power is assumed to be constant over the entire engine work-time (in \S\ref{sec:varying_L} we relax this assumption) with a total (two-sided\footnote{Note that in many studies of jet propagation (including \citealt{Bromberg2011}) the jet luminosity is defined to include only the energy launched in one direction.}) luminosity $L_j$. The jet launching starts with a delay $t_d$ with respect to the merger time, which is also assumed to be the time at which the subrelativistic ejecta is launched. The subrelativistic ejecta is assumed to be homologous (i.e., $r=vt)$ with a power-law density profile, 
\begin{equation}\label{eq:rho_ej}
    \rho_{ej}=K r^{-\alpha} t^{\alpha-3} = K t^{-3} v^{-\alpha}~~~;~~~\vmin<v<\vmax
\end{equation} 
where $K$, $\vmin$ and $\vmax$ are constants and $t$ is the time since the merger.  The power-law index at $\vmin<v<\vmax$ is constant $\alpha < 5$, so that most of the energy is carried by the fast part of the ejecta. We assume that if there is ejecta with $v>\vmax$ and/or $v<\vmin$, then its density is significantly lower than the power-law extrapolation of the density at $\vmin<v<\vmax$. We denote the mass in the velocity range given by Eq. \ref{eq:rho_ej} as the ``bulk" of the ejecta and consider the jet as propagating inside the ejecta only when the ejecta velocity at the jet head location is within this velocity range. The values of $\vmin$, $\vmax$ and $\alpha$ that represent the ejecta in astrophysical explosions may vary significantly between different systems. Observations of the BNS merger GW170817, as well as many numerical simulations, indicate that in BNS mergers $\vmin \approx 0.05$c and $\vmax \approx 0.3$c \citep[][and references therein]{Nakar2019}. The value of $\alpha$ in BNS mergers is less certain, although there are some observational and numerical results that suggest that at least in some mergers $\alpha \approx 3$ gives a decent description of the ejecta density profile (see \S\ref{sec:BNS}). 

Note that we approximate the ejecta as being spherical. However, also If the ejecta is not spherically symmetric, then this approximation is good as long as the angular profile of the ejecta at the poles does not vary strongly over an angle that is significantly larger than $\qj$. The adujstment needed for the application of our model to a non-spherical ejecta are discussed in  \S\ref{sec:nonisotropic}. 

\subsection{General properties of the jet evolution}
Before deriving an analytic approximated solution for the propagation of a jet in such a setup, we describe the properties of the solution in the various regimes based on general considerations. Here and the following several sections we restrict our discussion to Newtonian head velocities\footnote{Note that while the jet is launched at relativistic velocities the head itself under a wide range of conditions propagates at Newtonian velocities.}. The propagation of a jet with a relativistic head is discussed in \S\ref{sec:relativistic}. A key parameter of the problem is the ratio between the velocity of the jet head as seen in the local ejecta frame and the ejecta velocity at the location of the head:
\begin{equation}\label{eq:eta_definition}
    \eta \equiv \frac{v_h-\vej}{\vej}~,
\end{equation}
where $v_h$ is the head velocity and $\vej$ is the ejecta velocity at the location of the head, both measured in the lab frame.
Below we show that the evolution of $\eta$ with time can be approximated using very general considerations, and once $\eta(t)$ is known, one can derive the head location $r_h(t)$ by integrating the following equation:
\begin{equation}\label{eq:dlogr_dlogt}
    \frac{d\log(r_h)}{d\log(\tti)}=\frac{dr_h}{d\tti}\frac{\tti}{r_h}=\frac{v_h}{\vej}=\eta+1~,
\end{equation}
where here and elsewhere we use normalized time
\begin{equation}\label{eq:tti}
    \tti \equiv \frac{t}{t_d}.
\end{equation}
Then, for a given initial condition of $ r_h $ at some time $\tti_0$, the head location is:
\begin{equation}\label{eq:r_h}
    r_h(\tti)=r_h(\tti_0)\exp \left[\int_{\tti_0}^{\tti}  \frac{\eta(\tti')+1}{\tti'}d\tti' \right]~.
\end{equation}

To understand the evolution of $\eta$ with time, it is instructive to look at the two extreme regimes of the parameter $\eta$.
When $\eta \gg 1$ the expansion of the ejecta can be neglected and the propagation is similar to the one obtained for a jet in a static medium with a power-law density gradient $\rho \propto r^{-\alpha}$ \citep{Bromberg2011}. In this solution the evolution of the shock in the Newtonian regime satisfies $v_h \propto (t-t_d)^\frac{\alpha-2}{5-\alpha}$ and $r_h \propto (t-t_d)^{\frac{3}{5-\alpha}}$, so $v_h \propto r_h^\frac{\alpha-2}{3}$. At the same time, homologous expansion dictates $\vej \propto r_h$. Thus, as long as $\eta \gg 1$, for $\alpha <5$ the value of $\eta$ decreases as the head propagates, $\eta \approx \frac{v_h}{\vej} \propto r_h^{-\frac{5-\alpha}{3}}$. On the other hand, when $\eta \ll 1$, the head can be approximated as being at rest in the local frame of the ejecta. Thus, the density seen by the head of the jet evolves as $\rho_{ej}(r_h) \propto r_h^{-3}$. The velocity of the head is determined by finding the frame in which the magnitudes of the momentum fluxes from the jet and from the ejecta are equal. This condition was derived by \cite{Matzner2003} for a static medium and modified for an expanding one by moving to the forward shock local upstream frame, namely, the frame of the ejecta at the head location \citep[e.g.,][]{Ioka2018,Hamidani2020}:
\begin{equation}\label{eq:v_Lt}
    \frac{v_h-\vej}{c} \approx \frac{1}{1+\Lt^{-1/2}} \approx \Lt^{1/2}~~~(v_h, \vej \ll c),
\end{equation}
where 
\begin{equation}\label{eq:Lt}
    \Lt \simeq \frac{L_j}{\Sigma_j\rho_{ej}(r_h)c^3}~,
\end{equation}
and $\Sigma_j \approx 2\pi r_h^2 \theta_h^2$ is the jet cross-section at the location of the head (the factor of 2 accounts for both sides of the jet to match the two-sided luminosity $L_j$) and $\theta_h$ is the jet head opening angle. Thus, for a constant head opening angle, $\Sigma_j \propto  r_h^{2}$ and if $\eta \ll 1$ then $\Lt \propto r_h$ so $v_h-\vej$, and thus $\eta$, must grow with the radius. Even if the head opening angle is not constant, it is straight forward to show\footnote{ $\theta_h$ depends on the cocoon pressure. If the pressure is high enough the jet is collimated and $\theta_h<\qj$. Otherwise, the jet is roughly conical up to the head and $\theta_h \approx \qj$. Thus, in principle, $\Lt$ may grow faster or slower than $r_h$ if the jet collimation by the cocoon varies with time. If the collimation increases, $\Lt$ grows faster and so does $\eta$. If, instead, the collimation decreases, then there must be a time where the jet become uncollimated and $\theta_h \approx \theta_j$ cannot grow anymore. After this time  $\Lt$ and $\eta$ are guaranteed to increase with time until $\eta \sim 1$.} that as long as $\eta \ll 1$ it must start growing at some point until $\eta \sim 1$.

From the discussion above we learn that if $\eta \ll 1$ it must grow with time, whereas for $\eta \gg 1$ it is decreasing with time. Thus, we expect that regardless of the initial conditions, after enough time $\eta$ should converge to a constant asymptotic value, $\eta_a$. As we show next, the value of $\eta_a$ can be found by simple dimensional considerations.

\subsection{The temporal evolution during the asymptotic phase}\label{sec:eta_a}

During the asymptotic phase $t_d \ll t$ and therefore the initial timescale, $t_d$, is forgotten and the system has only a single timescale, $t$. Additionally, during this phase the jet has propagated a significant way in the ejecta, so $\vmin$ is forgotten and there is no velocity scale in the system. Thus, the only dimensional parameters of the system that are relevant during the asymptotic phase are $L_j$ and the ejecta density normalization constant $K$. Using dimensional analysis we see that the only combination that gives a length scale dictates
\begin{equation}\label{eq:rh_asymptotic}
    r_h \propto \left(\frac{L_j}{K}\right)^\frac{1}{5-\alpha} t^\frac{6-\alpha}{5-\alpha} ~.
\end{equation}
When $\eta=\eta_a$,
Eq. \ref{eq:r_h} reads $r_h \propto t^{1+\eta_a}$. We therefore conclude 
\begin{equation}\label{eq:eta_a}
    \eta_a=\frac{1}{5-\alpha}
\end{equation}

\subsection{The evolution before the asymptotic regime}
Had we been interested only in the asymptotic regime, all we should have done to provide a solution is finding the normalization factor for Eq. \ref {eq:rh_asymptotic} (as we do in \S\ref{sec:asymptotic}), which depends only on the dimensionless parameters $\qj$ and $\alpha$. However, there may be systems where the interesting part of the evolution takes place before the asymptotic phase.
For example, there are observational indications suggesting that in GW170817 the delay time, $t_d$, was of order of a second \citep[e.g.,][]{Metzger2018,Nakar2019}. In this event, the delay between the GW signal and the gamma-ray emission was 1.7s. Now, the propagation time of the jet in the merger ejecta until it breaks out, $t_{bo}$, must satisfy $t_{bo}<1.7s-t_d$. Therefore, if indeed $t_d$ was of order of a second then in GW 170817 $t_{bo} \lesssim t_d$. These indications are far from being conclusive but, if correct, then in GW170817 the jet breakout from the ejecta took place before its propagation reached the asymptotic phase.
Also, it is possible that in some BNS mergers the jets are choked before ever reaching the asymptotic phase. Therefore, we devote a significant part of this paper to study the evolution at early times, before it reaches the asymptotic phase.

The evolution before the asymptotic phase is also divided into two regimes - the collimation phase, during which the jet is being collimated by the cocoon pressure, and the evolution after the jet is fully collimated by the cocoon. As we show here, the collimation phase is, by itself, a transient and its properties (which may vary from one system to another) do not have a strong effect on the jet evolution after a full collimation is achieved. On the other hand, we find that the collimation phase can have an effect on the propagation of a jet in BNS merger ejecta over a significant part of the jet propagation, and therefore we need to solve for the evolution during this phase as well. Unfortunately, this phase may depend on the density in regions close to the jet launching site where the ejecta may not be homologous and the density may not follow Eq.~\ref{eq:rho_ej}. Yet, we find a description that constrains the evolution during this phase for a wide range of configurations (including a non-homologous ejecta at small radii). Luckily, as we show in \S\ref{sec:BNS}, for typical parameters the breakout from the ejecta in BNS mergers takes place near the end or after the completion of the collimation phase.  Thus, the criterion for a successful breakout that we derive, which is applicable only after the collimation phase is completed, is largely independent of the unknown conditions near the jet launching site. Next we describe our strategy to constrain the evolution during the collimation phase and the conditions under which it is valid.

\subsubsection{The effect of the density at $r<\rmin$ and non-homologous ejecta}\label{sec:cavity_intro}

If there is a significant delay between the time at which the bulk of the ejecta is launched and the time that the jet is launched, then the jet may encounter at first a low-density region. In BNS mergers such a delay can be a result of a delayed collapse of the central object to a black hole, as observations suggest for GW170817 \citep[][ and references therein]{Nakar2019} . We denote this low-density region as the ``cavity" and within our framework it extends at the time that the jet launch starts, $t_d$, between the jet launching site and $\rmin \equiv \vmin t_d$.

The definition of $\vmin$ is the minimal velocity at which a significant amount of mass is ejected, and therefore the density at $r<\rmin$ is often negligible. However, there are realistic scenarios where it is not. For example, it is possible that there is ejecta with $v<\vmin$ although its density is low compared to extrapolation of $\rho \propto v^{-\alpha}$ to low velocities. Another possibility is that there is a continuous mass ejection so a non-homologous outflow that was ejected at $t \sim t_d$ fills the cavity. Here we assume that if the latter is the case then the mass ejection rate at $t \sim t_d$ is such that 
the density in the cavity does not exceed an extrapolation of the power-law of the bulk of the ejecta to small radii, namely $\rho_{ej}(r<\rmin,t_d) \leq \rho_{ej}(r=\rmin,t_d)(r/\rmin)^{-\al}$.

The exact solution at early times, while the jet propagates at $r<\vmin t$ and also for a limited time after it starts its propagation in the bulk of the ejecta,  depends on the exact density profile at $r<\vmin t$. This may change from one system to another and therefore we cannot provide a general solution for that phase. However, we can provide solutions in two extreme cases, such that all other solutions must lie between them. The first is an empty cavity where we assume that $\rho(r<\rmin,t=t_d)=0$. This solution is a very good approximation for all scenarios where the mass ejection rate drops significantly at $t \ll t_d$ and there is a negligible amount of ejecta at $v<\vmin$. It is also the solution where the jet propagates at the highest possible velocity to $\rmin$ and starts its interaction with the bulk of the ejecta without being collimated at all. The analytic solution that we provide in the next section assumes an empty cavity and it follows the jet evolution from the first interaction of the jet with the base of the ejecta at $\rmin$ through the collimation phase and up to the asymptotic phase.
The second extreme possibility is the propagation of a jet in a ``full cavity", where the same power-law density of the ejecta is extended at $t_d$ to $r<\rmin$, namely $\rho(r<\rmin,t=t_d)=\rho(r=\rmin,t=t_d) (r/\rmin)^{-\al}$. This solution can be obtained simply by the empty cavity  solution in the limit of $\vmin\rightarrow 0$. Any solution of a partially filled cavity should lie between these two extreme solutions.

Another assumption of our model that may not be satisfied at $r<\rmin$ is that of a homologous ejecta. This may happen if ejecta continues to be launched at $t \sim t_d$ or later. We expect the effect of a non-homologous outflow close to the jet and ejecta launching sites to be minor, as long as the density in the cavity does not exceed our assumed upper limit (an extension of the power-law to $r<\rmin$) and the velocity of freshly launched ejecta does not increase over time (so there are no internal shocks that can generate high pressure within the ejecta). The reason is that if the jet is strong  ($\eta \gg 1$), then the motion of the ejecta does not play a significant role and it is unimportant whether it is homologous or not. If however the jet is weak compared to the ejecta, then it is stalled by the ejecta until it expands significantly, at which point the ejecta expansion becomes homologous and our solution is valid. 

Finally, we stress that the ejecta density at $r<\rmin$ affects only the initial phase of the jet evolution. After the jet propagates a significant way in the bulk of the ejecta, the initial density in the cavity is forgotten. This happens roughly when $r_h \gtrsim \theta_{j,0}^{-2/3} \vmin t$, as we show in \S\ref{sec:cavity}, where we discuss the effect of a non-empty cavity on the jet evolution based on analytic considerations as well as numerical simulations.
	
\section{Analytic propagation model of a jet with a Newtonian head}\label{sec:analytic}

We derive an analytic model of a Newtonian jet head propagation in expanding media and then (\S\ref{sec:E_tilde_and_BO}) deduce its breakout time and a breakout criterion. For convenience we concentrate in Table \ref{tab:equations} the  main propagation equations for $\al=2$ and $\al=4$, including the appropriate numerical coefficients (which are presented later in \S\ref{sec:numerical}). In \S\ref{sec:BNS} we provide the propagation equations for $\al=3$, normalized for canonical parameters of BNS mergers. For convenience we also assemble the main notations used throughout the paper in Table \ref{tab:keywords}.

\begin{table*}
	\setlength{\tabcolsep}{5pt}
	\centering
	\renewcommand{\arraystretch}{2}
	\begin{tabular}{ | l | c | c | c | c | c | }

        \hline
		Quantity & Regime & \multicolumn{2}{c|}{$\alpha=2$} & \multicolumn{2}{c|}{$\alpha=4$}\\
		\hline
		
		\multirow{6}{*}{$ r_h $} & \multirow{3}{*}{$ \etacrit > 1$} & $ \tti < \tc $: & $ \tti > \tc $: &
		$ \tti < \tc $: & $ \tti > \tc $: \\
		& &
        $ \vmin t_d \left[1+\frac{\theta_{j,0}^{-2/3}-1}{\tc-1}\left(\tti-1\right)\right]$
		&
		$\theta_{j,0}^{-2/3}\left(\frac{\tti-1}{\tc-1}\right)\left(\frac{\tti}{\tc}\right)^{1/4}$
		&
        $ \vmin t_d \left[1+\frac{\theta_{j,0}^{-2/3}-1}{\tc-1}\left(\tti-1\right)\right]$
		&
		$\theta_{j,0}^{-2/3}\left(\frac{\tti-1}{\tc-1}\right)^3\left(\frac{\tti}{\tc}\right)^{-3/2}$
		\\\cline{3-6}
		& & \multicolumn{2}{c|}{$ \tc = 2.5[1.3]\etacrit^{-2/3} $} & \multicolumn{2}{c|}{$ \tc = 9.4[4.7]\etacrit^{-2/3}\qj^{-4/9} $}
		\\
		\cline{2-6}
		& \multirow{3}{*}{$ \etacrit < 1$}
        & $ \tti < \ta $: & $ \tti > \ta $: &
		$ \tti < \ta $: & $ \tti > \ta $: \\
		& &
		$ \vmin t $
		&
		$ \vmin t^{4/3}t_a^{-1/3} $ 
		&
		$ \vmin t $
		&
		$ \vmin t^2/t_a $
		\\\cline{3-6}
        & & \multicolumn{2}{c|}{$ \ta = 16[2](\qj/\etacrit)^2 $} & \multicolumn{2}{c|}{$ \ta =  844[105](\qj/\etacrit)^2 $}
		\\
		\hline
		Asymptotic $ r_h $ & & 
		\multicolumn{2}{c|}{$0.25[0.5]{\cal L}^\frac{1}{3}\vmax t^{4/3}$} &
		\multicolumn{2}{c|}{$10^{-3}[7\times 10^{-3}]{\cal L}\vmax t^2$}
		\\
		\hline
		$ \tilde{E}_d $ & &
		\multicolumn{2}{c|}{$ 0.014[0.11]{\cal L}t_d$}
		&
		\multicolumn{2}{c|}{$ 0.006[0.05]{\cal L}t_d$}\\
		\hline
		\multirow{2}{*}{$ t_{bo} $} & $ \tilde{E}_d \gtrsim 1 $ &
		\multicolumn{2}{c|}{$ 5[2.5] t_d^{2/3}{\cal L}^{-1/3}$}
		&
		\multicolumn{2}{c|}{$ 5[3] t_d^{2/3}{\cal L}^{-1/3}$}\\
		\cline{2-6}
		& $ \tilde{E}_d\lesssim 1 $ &
		\multicolumn{2}{c|}{$ 287[36] {\cal L}^{-1}$}
		&
		\multicolumn{2}{c|}{$ 662[87]{\cal L}^{-1}$}\\
		\hline
	    Breakout criterion & & 
		\multicolumn{4}{c|}{$E_{j,{\rm iso,tot}} > 150[20] \left((t_d/t_e)^2+2\right)\Eejtot \theta_{j,0}^{2}$}
\\

		\hline
		
	\end{tabular}
	\hfill\break
	
	\caption{The main equations that describe the evolution of a subrelativistic head in the cases of $ \alpha = 2, 4 $ (including the appropriate numerical coefficients for each $\al$ value), using $ \etacrit = \frac{1}{t_d}\bigg(\frac{\Ljiso}{8\rho_{ej}(t_d,\rmin)\vmin^5}\bigg)^{1/2}$ and $ {\cal L} = \frac{\Ljiso}{\Eejtot\theta_{j,0}^2} $.
	If coefficients differ between unmagnetized and mildly magnetized jets, the numbers for magnetized jets are shown in square brackets. The same equations for $\al=3$, normalized for canonical parameters of BNS mergers, are given in \S\ref{sec:BNS}.
    For application of these formula to ejecta that is not spherically symmetric see \S\ref{sec:nonisotropic}.
    }
\label{tab:equations}
\end{table*}

\begin{table}
	\setlength{\tabcolsep}{0.01pt}
	\centering
	\begin{tabular}{ l  c c  }
		
		Notation & Definition & Equations \\ \hline
		& Jet notations & \\\hline
        $ \theta_{j,0} $ & jet launching half-opening angle\\
        $ \Ljiso $ & jet isotropic equivalent luminosity\\
        $ L_j $ &  jet total (two-sided) luminosity $(\Ljiso\theta_{j,0}^2/2)$\\
        $ \Ejiso(t) $ & $ \int_{t_d}^t \Ljiso~ dt $\\
        $ E_j(t) $ & $ \int_{t_d}^t L_j~ dt $\\
        $ \Ejtot $ & $ \int_{t_d}^{t_d+t_e} L_j~ dt $\\
        $ t_e $ & working time of the jet engine\\
        $ r_h $ & jet head location & \ref{eq:r_h},\ref{eq:rh_strong},\ref{eq:rh_asymptotic_full},\ref{eq:rh_weak},\ref{eq:rh_full_cavity}\\
        $ v_h $ & jet head velocity & \ref{eq:v_Lt},\ref{eq:velocity_difference}\\
        \hline
        
		& Ejecta notations & \\\hline
        $ \vej $ & ejecta velocity at the location of the jet head\\
        $ \vmin[\bmin] $ & minimal ejecta velocity [in units of c]\\
        $ \vmax[\bmax] $ & maximal ejecta velocity [in units of c]\\
        $ \rmin $ & $ \vmin t_d $\\
        $ \rmax $ & $ \vmax t_d $\\
        $ w $ & $ \vmax/\vmin $\\
        $ \rho_{ej} $ & ejecta density &
        \ref{eq:rho_ej}\\
        $K$ & normalization of the ejecta density profile  & \ref{eq:rho_ej}\\
        $ A_\rho $ & normalization of the ejecta density profile at $t_d$ & \ref{eq:Arho}\\
        $\mmin $ & $ \rho_{ej}(t_d,\rmin) \rmin^3$ &\\
        $ \alpha $ & power-law of the ejecta density profile\\
        $ \Eej $ & energy carried by ejecta with $ v < \vej $ & \ref{eq:eta_tmp}\\
        $ \Eejtot$ & total ejecta energy \\
        \hline
        
 		& Jet-ejecta relation notations & \\\hline
 		$ \eta $ & $ (v_h-\vej)/\vej $ & \ref{eq:eta_definition},\ref{eq:eta_delta},\ref{eq:eta_weak},\ref{eq:eta_tmp},\ref{eq:eta_E}\\
 		$ \eta_0 $ & $\eta$ at $\rmin$ if the jet is fully collimated & \ref{eq:eta0}\\
 		$ \etauc $ & $ \eta $ at $\rmin$ if the jet is uncollimated& \ref{eq:eta_uncol} \\
 		$ \etacav $ &  $ \eta $ when the jet collimation is cavity dominated & \ref{eq:eta_cav} \\
 		$ \etacrit $ & critical $ \eta $ to determine if the jet is strong or weak & \ref{eq:eta_crit} \\
        $ \eta_a $ & $ \eta $ during the asymptotic regime& \ref{eq:eta_a}\\
 		$ \rhcol $ & jet head location when the jet is collimated & \ref{eq:rh_col} \\
 		$ \vhcol $ & jet head velocity when the jet is fully collimated & \ref{eq:vh_col} \\
 		$ \etac $ & $ \eta $ when the jet becomes fully collimated & \ref{eq:eta_col} \\
        $ \tilde{E} $ & dimensionless parameter proportional to $\frac{E_j}{\Eej} \theta_{j,0}^{-4}$ & \ref{eq:tilde_E},\ref{eq:tilde_E(t)}\\
        $ \E_a $ &  $ \tilde{E} $ during the asymptotic regime& \ref{eq:E_eta_a}\\
        $ \Etot $ & $\E$ with the total jet and ejecta energies & \ref{eq:Etot},\ref{eq:Etot_crit},\ref{eq:Etot_crit_simple}\\
        $ \tilde{E}_d $ & $ \Etot $ where $ \Ejtot$ is replaced by $L_jt_d$ & \ref{eq:Ed} \\
        $ {\cal L}$ & $\Ljiso\qj^{-2}/\Eejtot$\\
        
        \hline
		& Time notations & \\\hline
		$ t $ & time from the merger\\
        $ t_d $ & delay time between the merger and the jet launch\\
        $ \tti $ &  $ t/t_d $ & \ref{eq:tti}\\
        $ \delta $ & $ t/(t-t_d) = \tti/(\tti-1) $ & \ref{eq:delta_def}\\
        $ \tc $ & time for jet to reach full collimation & \ref{eq:t_col} \\
        $ \ta $ & time of the transition to the asymptotic regime  $ \eta_a $ & \ref{eq:ta_strong},\ref{eq:ta_weak}\\
        $ \tdeff $ & effective $ t_d $ of weak jets & \ref{eq:tdeff}\\
		$ t_{bo} $ & breakout time since jet launching & \ref{eq:t_bo1},\ref{eq:t_bo2}\\
        $ t_{e,bo} $ & minimal engine work-time needed for breakout & \\
    \hline
		
	\end{tabular}
	\hfill\break
	
	\caption{Notations throughout the paper.}
	\label{tab:keywords}
\end{table}

\subsection{Critical parameters of the jet evolution}
 Altogether, under the assumptions described above, the evolution of the system depends on six parameters: $L_j$, $\qj$, $\vmin$, $t_d$, $\al$, and $K$. However, we find that most of the jet evolution depends only on three parameters, $\qj$, $\al$ and another dimensionless parameter that we denote $\eta_0$, which is related to the initial value of $\eta$. At very early times there may also be a minor dependence on $\vmin$. Below, we first define $\eta_0$ using the fact that for $\eta \gg 1$ the motion of the ejecta can be neglected and the solution is approaching that of a static medium. Later, we show that the various phases of the jet evolution depend on various combinations of $\eta_0$, $\qj$ and $\al$. 

Consider a jet that propagates in a hypothetical static medium with a density profile that is a snapshot of the ejecta density at $t_d$, with a single modification: the density profile of the hypothetical static medium is a single power-law, $\rho_{stat} \propto r^{-\alpha}$ not only at $r>\rmin$ but also at $r<\rmin$. We can therefore write the density profile of the static medium as\footnote{For some values of $\alpha$ the density profile must have a cut-off at small radius. We assume that this radius is much smaller than $\rmin$ and then its value has a negligible effect on the hypothetical jet dynamics that we consider in order to define $\eta_0$ when its head reaches $\rmin$.}
$ \rho_{stat}=A_\rho r^{-\alpha}$,
where this equation is applicable starting at $r \ll \rmin$ and $A_\rho$ is related to the expanding medium density (Eq. \ref{eq:rho_ej}) via
\begin{equation}\label{eq:Arho}
    A_\rho=K t_d^{\alpha-3}=\rho_{ej}(t_d,\rmin) \rmin^\alpha.
\end{equation}
 Under the assumption that $\rho_{stat} \propto r^{-\alpha}$ is starting at $r \ll r_{min}$, the jet that propagates in the hypothetical static medium is fully collimated by the cocoon when the head reaches $\rmin$, and its evolution is described well by \cite{Bromberg2011} and \cite{Harrison2018}.
We define the velocity $v_{h,0}$ as the velocity of the jet head that propagates in the static medium when it reaches $\rmin$.
$v_{h,0}$ can be found using equations A2 and A3 of  \cite{Harrison2018}:
\begin{equation}\label{eq:vh0}
    v_{h,0} \approx \frac{0.4}{(5-\alpha)^{1/3}} \left(\frac{L_j}{A_\rho \theta_{j,0}^4}\right)^{1/3} \rmin^{\frac{\alpha-2}{3}}~.
\end{equation}

Now, consider a moving ejecta, with a profile that is similar to that of the original system, except for the modification that the density profile $\rho_{ej}=K t_d^{-3}v^{-\alpha}$ is extended also to  $v \ll \vmin$, such that $\rho_{ej}(r,t_d)=\rho_{stat}(r)$. If $v_{h,0}\gg \vmin$, then the jet arrives to $\rmin$ at a time which is much shorter than the ejecta dynamical time, $t_d$, namely at $t-t_d \ll t$. Thus, as far as the propagation of the jet head is concerned, the expanding ejecta can be approximated as static up to the time that the jet head arrives to $\rmin$ with a velocity that is approximately $v_{h,0}$. The ejecta velocity at this location is approximately $\vmin$ so the corresponding value of $\eta$ at this location, in case that $\eta \gg 1$, can be approximated by Eq. \ref{eq:vh0}. We use this result to define
\begin{equation}\label{eq:eta0}
\eta_0 \equiv  \left(\frac{L_j t_d^{\alpha-2}}{A_\rho \theta_{j,0}^4\vmin^{5-\alpha}}\right)^{1/3}=\left(\frac{L_j t_d}{\mmin v_{\rm min}^2 }\right)^{1/3} \theta_{j,0}^{-4/3}.
\end{equation}
which satisfies, up to a factor of order unity,  $\eta_0 \approx v_{h,0}/\vmin$. In the second expression we rearranged the parameters to highlight that the dimensionless factor $\eta_0$ can be expressed using the ratio between the injected jet energy over $t_d$,  $L_j t_d$, and the kinetic energy carried by ejecta with velocity $\sim \vmin$, 
$\mmin v_{\rm min}^2$, where we define $\mmin \equiv \rho_{ej}(t_d,\rmin) \rmin^3$.

In the real system the value of $\eta$ at $\rmin$ is not necessarily $\eta_0$. First, in the derivation of $\eta_0$ we neglected the collimation phase, which depends on the density in the region $r<\rmin$ at $t_d$. Second, if $\eta \lesssim 1$, then the static approximation used to estimate $\eta_0$ is not valid and $\eta(\rmin) \neq \eta_0$. Thus, next we explore the evolution of the jet during the collimation phase for all values of $\eta_0$

\subsubsection{The collimation phase}\label{sec:uncollimated}

	\begin{figure*}
		\centering
		\includegraphics[scale=0.33]{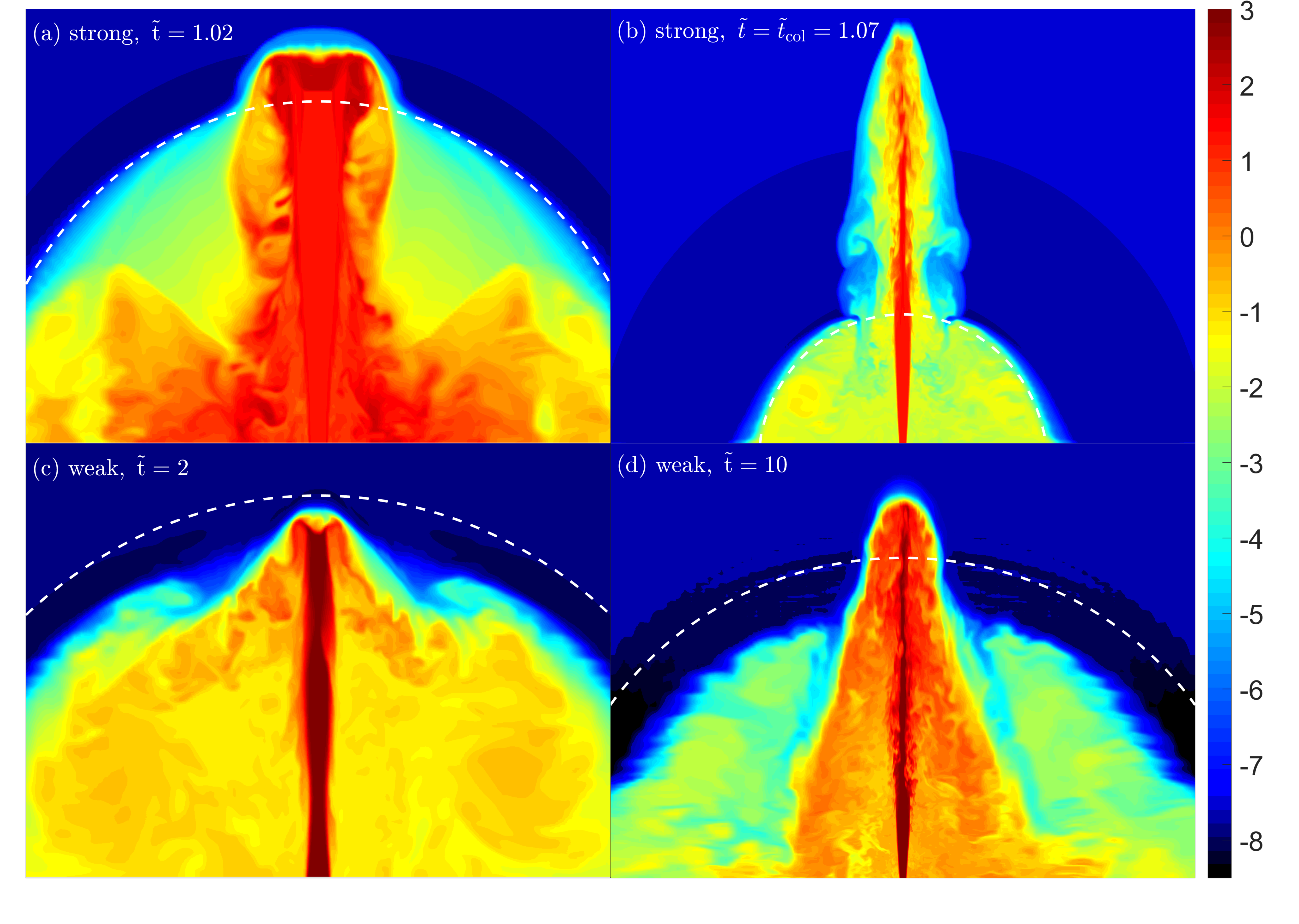}
		\caption[]{
		Illustration in arbitrary scales of the different phases in strong (a,b) and weak (c,d) jets, using the logarithmic specific entropy meridional maps from simulations (models $ \Ca $ and $ \Ce $, see \S\ref{sec:numerical} for details). Left: the jet is partially collimated and its head is inside the cavity moving at $ \vmin $. Right: the jet is collimated and its head is breaching the ejecta from the cavity. Black dashed lines mark the radius of the cavity, $ \vmin t$.
		}
		\label{fig:illustration}
	\end{figure*}

The jet evolution depends strongly on the collimation by the cocoon pressure. However, it takes time for this pressure to build up after the jet launching starts. Thus, the early jet evolution goes through a phase of collimation, during which its propagation is affected by the collimation process. The collimation process depends on the density of the medium that lies between the launching site and $\rmin$. Here we consider a setup where there is a negligible ejecta mass with $v<\vmin$ and that the ejection of subrelativitic material drops significantly at $t \ll t_d$, so that at $t_d$ there is  effectively an empty cavity at $r<\rmin$ and we can approximate $\rho_{ej}(t_d,r<\rmin)=0$. In \S\ref{sec:cavity} we discuss the jet propagation when the cavity is not empty and the density at $r<\rmin$ affects its evolution.

In an empty cavity, the jet propagates to $\rmin$ at a velocity close to the speed of light. Since there is no pressure in the cavity, the jet is conical (i.e., uncollimated) up to $\rmin$, and its cross-section when it starts interacting with the medium is $\Sigma_j = 2 \pi r_{\rm min}^2\theta_{j,0}^2$ (the factor of 2 accounts for the two sides of the jet). Plugging this cross-section into Eq.~\ref{eq:Lt}, while taking $\rho(\rmin,t_d)$ from Eq.~\ref{eq:rho_ej}, we obtain the value of $\Lt$. Then we use Eq.~\ref{eq:v_Lt} to find the value of $\etauc(t_d)$ while the head starts propagating through the dense medium at $\rmin$ (we neglect the head propagation time in the empty cavity). A comparison to $\eta_0$ shows that:
\begin{equation}\label{eq:eta_uncol}
    \etauc(\tti=1) \approx \beta_{\rm min}^{1/2}\eta_0^{3/2}\qj ~,
\end{equation}
where $\bmin=\vmin/c$.
If $ \etauc(\tti=1) \gg 1$, the jet promptly starts propagating in the medium. An example of a jet in that phase is shown in Fig. \ref{fig:illustration}a. However, if $\etauc(\tti=1) \ll 1$, the jet is pounding at first on the slow end of the ejecta without penetrating it almost at all. We focus first on the latter case, in which the location of the jet head at any given time is roughly $\vmin t$. Before the jet collimation, the jet is conical and the cross-section of the uncollimated jet head increases as  $\Sigma_j \propto r_h^2 \propto \tti^2$ while the density of the ejecta at the head location drops as $\rho_{ej}(r_h)) \propto \tti^{-3}$. Therefore, we obtain that as long as the jet is uncollimated $\etauc(\tti) \propto (\Sigma_j \rho_{ej}(r_h))^{-1/2} \propto \tti^{1/2}$.

As the jet tries unsuccessfully to propagate into the ejecta, the gas that goes into the jet head (mostly through the reverse shock) spills sideways and fills the spherical cavity at $r<\vmin t$. The pressure in the cavity increases as the energy deposited by the jet accumulates:
\begin{equation}\label{eq:P_cavity}
    P \approx \frac{L_j (t-t_d)}{4\pi \vmin^3 t^3}.
\end{equation}
When the pressure is high enough, it starts collimating the jet so the head cross-section is \citep{Bromberg2011}:
\begin{equation}\label{eq:Sig_cavity}
    \Sigma=\frac{L_j \qj^2}{4cP}.
\end{equation}
Plugging Eqs. \ref{eq:Sig_cavity} and \ref{eq:P_cavity} into Eqs. \ref{eq:Lt} and \ref{eq:v_Lt}, and using the fact that $\vej \approx \vmin$ at this stage, one obtains that once the jet collimation starts, $\eta$ satisfies,
\begin{equation}\label{eq:eta_cav}
    \etacav(\tti) \approx \eta_0^{3/2}\qj (\tti-1)^{1/2}.
\end{equation}
The jet propagates at first at $\eta=\etauc$ and once $\etacav=\etauc$, the collimation starts being significant and the jet propagates at $\eta=\etacav$. Comparing Eqs. \ref{eq:eta_uncol} and \ref{eq:eta_cav}, while taking into consideration the relations $\etauc(\tti)  \propto \tti^{1/2}$ and $\bmin \ll 1$, one finds that by $\tti=2$ all jets with $\etauc(\tti=1) \ll 1$  maintain $\eta=\etacav$. This implies that within one dynamical time of the ejecta since the jet launching starts, the pressure in the cavity is always high enough to start collimating the jet. The discussion above implies that we can define a critical value for the system parameters that dictates its evolution:
\begin{equation}\label{eq:eta_crit}
    \etacrit \equiv \etacav(\tti=2)=\Ncrit\eta_0^{3/2}\qj=\Ncrit \left(\frac{L_j\theta_{j,0}^{-2} t_d}{\mmin v_{\rm min}^2 }\right)^{1/2}~,
\end{equation}
where $ \Ncrit $ is a numerical coefficient of order unity that we extract from the simulations described in \S\ref{sec:numerical}.
If $ \etacrit \ll 1$, the jet is unable to propagate into the ejecta for many dynamical times. In that case $\eta$ grows slowly until it reaches $\eta_a$. We refer to this regime as the ``weak jet" regime. If, however, $\etacrit \gg 1$, then $\eta$ becomes much larger than unity on a time that is much shorter than the ejecta dynamical time ($t_d$) and the jet starts its propagation through the ejecta almost immediately after it is launched. In such a case, at early time $\eta \gg 1$ and it starts decreasing a short time after the jet launch starts, until it reaches $\eta_a$. We refer to this case as the ``strong jet" regime. We note that although we use the terms 'strong' and 'weak' to describe the initial propagation regime of a jet, the determination of the regime of a specific system depends not only on the jet luminosity but also on its opening angle, ejecta properties and the delay time. Thus, even in case that we observe the jet and the ejecta, e.g., via the afterglow and kilonova emission, we may be still unable to determine whether the jet propagation was in the strong or weak regime unless we know what $t_d$ is.
In the following we provide an analytic solution to the jet evolution in the strong and the weak regimes.

\subsection{Strong jet - $\etacrit \gg 1$}
When $\etacrit \gg 1$ the jet starts propagating in the ejecta at $\tti-1 \ll 1$ with $\eta \gg 1$  (see Fig. \ref{fig:illustration}a). Thus, $\eta$ approaches its asymptotic value, $\eta_a$, from above, and as long as $\eta>1$, we can use the approximation of a jet that propagates in a static medium. However, at first the size of the cavity at $r<\rmin$ affects the collimation of the jet. Therefore, the propagation approaches the solution of \cite{Bromberg2011} only when the jet head reaches
\begin{equation}\label{eq:rh_col}
    \rhcol \approx \rmin \theta_{j,0}^{-2/3}~,
\end{equation}
and the volume of the bubble inflated by the cocoon, $\sim 2\pi r_h^3 \theta_{j,0}^2$ \citep{Bromberg2011}, is comparable to the volume of the cavity, $(4/3) \pi r_{\rm min}^3 $. Only at $r \gtrsim r_h$ the volume of the cavity has a minor effect, so the collimation process is complete and we can refer to the jet as being ``fully collimated". An example of a jet in that phase is shown in Fig. \ref{fig:illustration}b. Using equations A2 and A3 from \cite{Harrison2018} we estimate that at $\rhcol$,
\begin{equation}\label{eq:vh_col}
    \vhcol \approx v_{h,0} (\rhcol/\rmin)^{\frac{\al-2}{3}}~.
\end{equation}
Neglecting the ejecta motion between $t_d$ and the time that the jet head reaches $\rhcol$, we find that $\vej(\rhcol) \approx \vmin \rhcol/\rmin$ and therefore we can approximate,
\begin{equation}\label{eq:eta_col}
    \etac \equiv \eta(\rhcol) \approx \Ncol(\alpha)\frac{0.4}{(5-\alpha)^{1/3}} \eta_0 \theta_{j,0}^{\frac{2(5-\al)}{9}}~,
\end{equation}
where $ \Ncol (\alpha) $ is a numerical coefficient that we find from the simulations.
At $r_h>\rhcol$, as long as $\eta \gg 1$, the solution of a jet in a static media provides a good approximation and $v_h \propto (t-t_d)^\frac{\alpha-2}{5-\alpha}$. We can therefore relate the distance that the head traveled up to time $t$ and the distance traveled by the ejecta at the same location  
\begin{equation}\label{eq:vej_vh}
    \vej t \approx \int_{t_d}^t v_h(t') dt' \approx
    \frac{5-\alpha}{3} v_h(t) (t-t_d)~,
\end{equation}
where the last approximation is accurate for $r \gg \rhcol$. Rearranging Eq. \ref{eq:vej_vh} and adding a constant so $\eta(\tti \to \infty) \to \eta_a$ we obtain
\begin{equation}\label{eq:eta_delta}
    \eta \approx\frac{3}{5-\alpha}(\delta-1)+\eta_a=\frac{3\delta-2}{5-\alpha}~~~;~~~\eta<\etac,
\end{equation}
where
\begin{equation}\label{eq:delta_def}
    \delta \equiv \frac{t}{t-t_d}=\frac{\tti}{\tti-1} .
\end{equation}

Eq. \ref{eq:eta_delta} has two interesting implications. First, at  $r>\rhcol$ there is a universal solution for $\eta(\tti)$ which is applicable for all strong jets. The difference between different jets is the time and the value of $\eta$ at which they join this solution (given roughly by Eq. \ref{eq:eta_col}). Second, in all strong jets $\eta$ approaches its asymptotic value on a timescale that is not much larger than $t_d$ after the jet launching starts. 
We denote the time that the jet approaches $\eta_a$ as $t_a$. From Eq. \ref{eq:eta_delta} we see that $\eta_a<\eta(\tti>4) < 2\eta_a$, implying that in strong jets
\begin{equation}\label{eq:ta_strong}
    \ta \approx 4 ~~~;~~~\etacrit \gg 1.
\end{equation}

The evolution of $\eta$ before the jet becomes fully collimated is complex and among other things, it depends on the density in the cavity (whether it is negligible, as in an empty cavity that we consider in this section or not). We therefore take here the simplest possible approximation during that phase, a constant velocity. The entire evolution can be then approximated based on Eqs. \ref{eq:r_h}, \ref{eq:eta_a}, \ref{eq:rh_col} and \ref{eq:eta_delta}:
\begin{equation}\label{eq:rh_strong}
r_{h}(\tti) = \rmin
\begin{cases}
1+\frac{\theta_{j,0}^{-2/3}-1}{\tc-1}(\tti-1)& \tti < \tc \\
&\\
\theta_{j,0}^{-2/3}\bigg(\frac{\tti-1}{\tc-1}\bigg)^{\frac{3}{5-\alpha}} \Big(\frac{\tti}{\tc}\Big)^{\frac{3-\alpha}{5-\alpha}}& \tc < \tti
\end{cases} ~,
\end{equation}
where  
\begin{equation}\label{eq:t_col}
    \tc \approx \frac{3}{(5-\al)\etac}+1~,
\end{equation}
so Eq. \ref{eq:eta_delta} is approximately satisfied at $t>\tc$. As we show in \S\ref{sec:numerical}, using numerical simulations that scan a wide range of parameters, Eq. \ref{eq:rh_strong} provides a good approximation for $r_h(t)$ of jets with $\etacrit > \eta_a$. For very strong jets with $\etacrit \gg 1$ the approximation is very good, to within a factor of order unity in all our simulated range, while for jets with $\etacrit \sim 1$ it provides a fairly good approximation, to within a factor of 2.

\subsection{The asymptotic phase}\label{sec:asymptotic}
Regardless of whether the jet is strong or weak, after enough time the jet must approach the asymptotic phase. Moreover, since the evolution in this phase is independent of $t_d$ and $\vmin$, all jets, strong and weak, with the same value of $L_j/K$, $\qj$ and $\al$ must eventually converge to the same late time evolution. Thus, we can use the late time evolution of strong jets (at $t \gg t_d$) to find the 
location of the head during the asymptotic phase. Note, however, that in reality some jets break out before reaching the asymptotic phase. This is expected for example in BNS mergers where $\vmax/\vmin \approx 6$ and strong jets are expected to break out on a timescale $\lesssim t_d$. 

Taking the limit of $t \gg t_d$ in Eq. \ref{eq:rh_strong} we find that during the asymptotic phase
\begin{align}\label{eq:rh_asymptotic_full}
    r_h &=\left(\frac{2(5-\al)^\frac{2}{3}\Ncol}{15}\right)^\frac{3}{5-\al}
    \left(\frac{L_j}{K}\theta_{j,0}^{-4}\right)^\frac{1}{5-\al} t^\frac{6-\al}{5-\al} \notag\\
    & =  \left(\frac{2(2\pi(5-\al))^\frac{1}{3}\Ncol}{15}\right)^\frac{3}{5-\al}
    \left(\frac{L_j}{\Eejtot}\theta_{j,0}^{-4}\right)^\frac{1}{5-\al}\vmax t^\frac{6-\al}{5-\al}~,
\end{align}
where $\Eejtot$ is the total ejecta kinetic energy.
As expected this expression follows the scaling of Eq. \ref{eq:rh_asymptotic} and it is independent of $t_d$ and $\vmin$. It is a good approximation for strong jets at $t \gtrsim 4 t_d$.  An example of a strong jet that approaches the asymptotic phase can be seen in the left panels of Fig. \ref{fig:maps}. The time at which it becomes a good approximation for weak jets is found next.

\subsection{Weak jet - $\etacrit < \eta_a$}
As long as $\eta \ll 1$ the jet is too weak to breach the ejecta. Instead, the head propagates roughly at $\vmin$ and the volume of the cocoon is dominated by the cavity at $r<\vmin t$ so $\eta \approx \etacav$. Thus,  $\etacrit = \etacav(\tti=2) \ll 1$ and the jet starts to propagate into the ejecta only at $\tti \gg 1$ (i.e., $t \gg t_d$). An example of a weak jet that is stalled at the base of the ejecta on a timescale that is longer than $t_d$ is shown in Fig. \ref{fig:illustration}c. We can use Eq. \ref{eq:eta_cav} to approximate 
\begin{equation}\label{eq:eta_weak}
\eta (\tti) \approx
\etacrit (\tti-1)^{1/2} ~~~;~~~ \tti \ll \ta ~.
\end{equation}
As $\eta$ approaches $\eta_a$, the jet starts its propagation inside the ejecta and the collimation is not determined by the cavity alone so Eq. \ref{eq:eta_weak} is no longer accurate. An example of a weak jet that starts breaching the ejecta is shown in Fig. \ref{fig:illustration}d. We know that before the jet starts propagating in the ejecta the head velocity satisfies $v_h \approx \vmin$, while after it starts to propagate in the ejecta it approaches the asymptotic solution given in Eq. \ref{eq:rh_asymptotic_full}. Therefore we can approximate the head location of a weak jet as
\begin{equation}\label{eq:rh_weak}
r_h (\tti) \approx \rmin \tti 
\begin{cases}
1 & \tti <\ta \\
&\\
\Big(\frac{\tti}{\ta}\Big)^\frac{1}{5-\alpha} & \tti > \ta \\
\end{cases} ~,
\end{equation}
where 
\begin{equation}\label{eq:ta_weak}
    \tti_a = \left(\frac{15}{2(5-\al)^\frac{2}{3}\Ncol}\right)^3 \eta_0^{-3}~~~;~~~\etacrit \ll 1~
\end{equation}
is chosen so the propagation during the asymptotic phase satisfies Eq \ref{eq:rh_asymptotic_full}. An example of a weak jet that approaches the asymptotic phase is shown on the right panels of Fig. \ref{fig:maps}. 
Our numerical simulations show that Eq. \ref{eq:rh_weak} provides a very good approximation of $r_h$ for weak jets, to within a factor of order unity during the entire simulated range.

\subsubsection{Effective delay time of weak jets}\label{sec:td_eff}
Since a weak jet is unable to propagate a significant way within the  ejecta up to $t_a$, which is much longer than $t_d$, there is a little difference between a jet that starts being launched at $t_d \ll t_a$ and a jet that starts being launched at $t_d \approx t_a$. To see that consider two similar jets (in terms of luminosity and opening angle), one is a weak jet that starts being launched at $t_d \ll t_a$ and the other starts being launched at $t_d = t_a$. In the first jet the pressure in the cavity builds up and collimate the jet on a timescale comparable to $t_d$, but the ejecta is too dense so the jet is stalled at first. Only at $t \sim t_a$ the ejecta density drops enough so the jet can start its propagation within the ejecta. In the second jet, the cavity pressure starts building up only at $t_a$, but at $t \approx 2t_a$ it becomes comparable to that of the first jet at the same time, so the propagation of the two jets becomes similar. This implies that the propagation of weak jets is only weakly dependent on $t_d$ and it depends mostly on $t_a$ and all weak jets behave as if $t_d \sim t_a$. We therefore define $t_a$ as an effective delay time for weak jets. Following Eq. \ref{eq:ta_weak} 
\begin{equation}\label{eq:tdeff}
    \tdeff=t_a =\left(\frac{15}{2(5-\al)^\frac{2}{3}\Ncol}\right)^3
    \frac{\mmin v_{min}^2}{L_j}\theta_{j,0}^{4}~.
\end{equation}
This result has an interesting implication for BNS mergers. As $t_d$ depends on the post merger evolution, it is an important quantity that may teach us about the physics of mergers. Unfortunately, the discussion above shows that in case of weak jets the observations are almost independent of $t_d$.

\subsection{The structure of the jet-cocoon}\label{sec:cocoon}

	\begin{figure}
		\centering
		\includegraphics[scale=0.165]{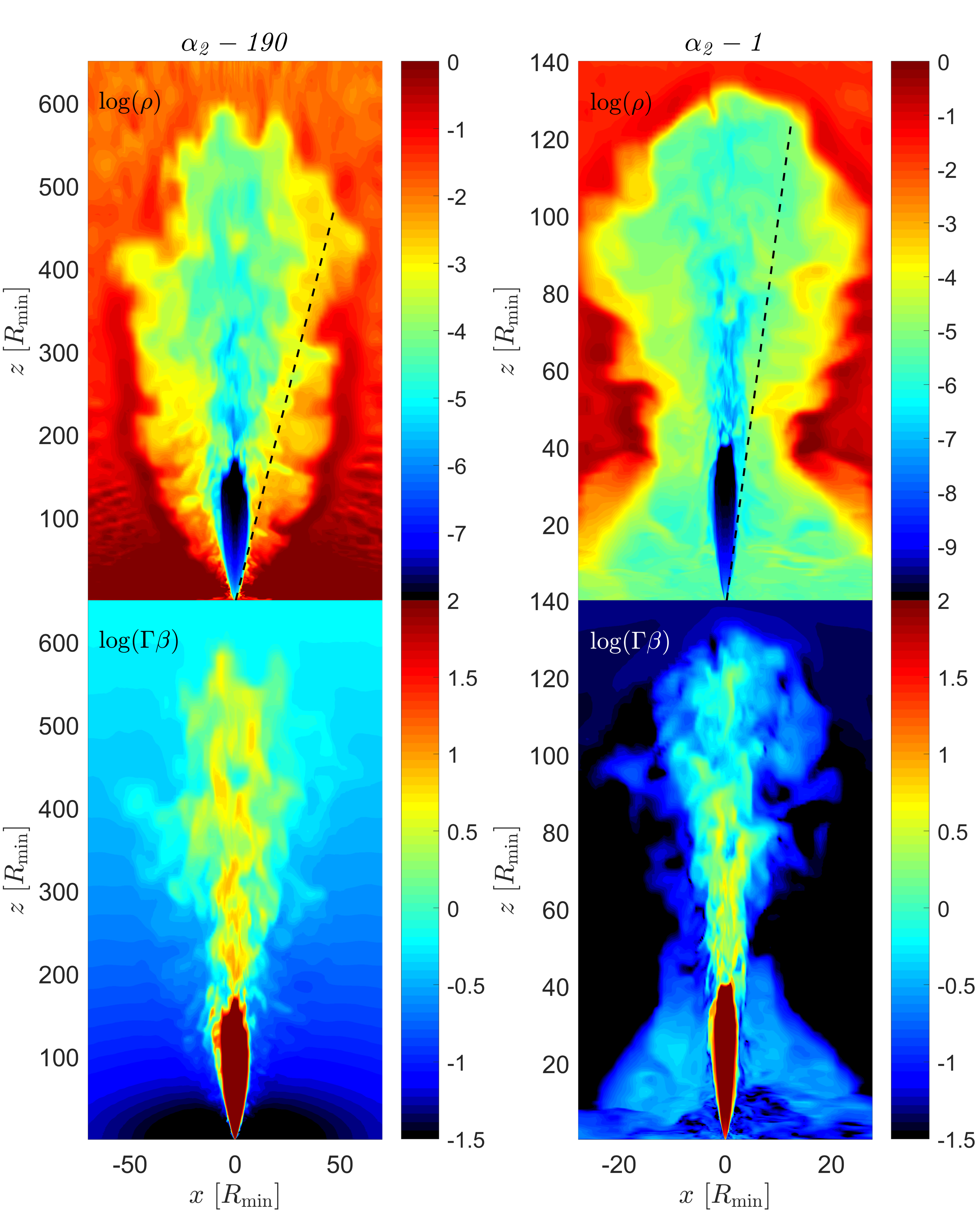}
		\caption[]{
		Meridional maps of our canonical strong $ \Ca $ (left) and marginal $ \Cc $ (right) jet models (taken from the low resolution simulations, see \S\ref{sec:numerical} for details) close to the asymptotic phase, at $ \tti = 5.7 $ and $ \tti = 41 $, respectively.
		Top panels: the logarithmic mass density (in arbitrary units) map, where a dashed black line is stretched from the origin at the jet opening. It shows that in both simulations the cocoon opening angle is $\theta_c=r_h/l_c \approx \qj$.
		Bottom panels: logarithmic proper-velocity.
		}
		\label{fig:maps}
	\end{figure}

The solution of \cite{Bromberg2011} to a jet in a static medium includes the full structure of the jet-cocoon system during the propagation. This includes the  cocoon opening angle, $\theta_c$, and the Lorentz factor of jet material after it is shocked by the collimation shock and before it reaches the head, $\Gamma_1$ (we use here the notation as in \citealt{Bromberg2011}). 
 $\theta_c$ is of interest since it defines the part of the ejecta that affects the jet propagation. $\Gamma_1$ is of interest since it can affect the emission from GRB jets (e.g., the efficiency of the photospheric emission, \citealt{Gottlieb2019}).
Here we extend the solution of these quantities to a jet in an expanding medium.

As in most other properties, the discussion of the cocoon opening angle  depends on the value of $\eta$. Here we focus on the asymptotic phase since in the regime of $\eta \ll 1$, the cocoon is confined to the cavity. If the cavity is empty, then the cocoon structure is trivial (approximately a sphere with radius $\vmin t$), otherwise the structure depends on the cavity details. On the other hand, for $\eta \gg 1$ the solution can be approximated by the static case and therefore the structure derived by \cite{Bromberg2011} is applicable. 

Similarly to the value of $\eta_a$, the evolution of the jet-cocoon structure during the asymptotic phase can be deduced from dimensional analysis. Since there is only a single combination of the system parameters that can give a length scale during the asymptotic phase (see \S\ref{sec:eta_a}), all length scales must have the same dependence on time. This includes the cocoon cylindrical radius, $l_c$, which evolves as   $l_c \propto r_h \propto t^{\frac{6-\al}{5-\al}}$. Thus, all length scale ratios, including $\theta_c= (l_c/r_h)$, remain constant during the asymptotic phase, at least until the head becomes relativistic. The way to evaluate $\theta_c$ is by noting that one of the pathways to the asymptotic phase is via a strong jet that starts with $\eta \gg 1$. According to the static solution, during this phase  the cocoon opening angle is always constant in time and it satisfies $\theta_c \approx \qj$. Thus, assuming that $\theta_c$ does not change much during the transition from $\eta \gg 1$ to $\eta =\eta_a$ we can approximate $\theta_c \approx \qj$ also during the asymptotic phase. We verify that this relation holds in the three simulations that we have that reach this phase (strong, marginal and weak), as can be seen in Fig. \ref{fig:maps} in two examples.
This result is of special interest in the context of BNS mergers where the ejecta is most likely not isotropic. It implies that during the entire evolution the jet-cocoon structure is sensitive only to the density within an opening angle $\sim \qj$ around the jet symmetry axis. Given that typical jet angles are considered to be of the order of $0.1$ rad (e.g., in GW170817 it is most likely $\approx 0.05$ rad), the solution presented here is applicable even if the ejecta is highly aspherical as long as its density and velocity distributions do not vary much over an angle of $\qj$ around the jet axis.

Another interesting property of the jet-cocoon structure is the Lorentz factor above the first collimation shock. \cite{Bromberg2011} showed that in static media $\Gamma_1 \approx 1/\qj$. This relation arises from the requirement that after the collimation the pressure in the shocked jet is similar to the cocoon pressure. The same equality holds also in expanding media in all of the various regimes, and therefore also here $\Gamma_1 \approx 1/\qj$. In Fig. \ref{fig:maps} we show maps of two jets that are approaching the asymptotic phase, one very strong and the other marginally strong. Those maps show that in both cases $\theta_c \approx \qj$ and $\Gamma_1 \approx 1/\qj$.

\section{A Breakout criterion and the dimensionless parameter $\E$}\label{sec:E_tilde_and_BO}
The ability of a jet to break out from the merger ejecta has profound observational implications. Therefore, it is most interesting to characterise the necessary conditions for breakout. One way to find analytically if a jet breaks out of the ejecta is to use the approximated formula for $r_h$ given above (Eqs. \ref{eq:rh_strong} and \ref{eq:rh_weak}) and check whether at the time that the engine stops the jet has broken out or not. However, as we show below if we are only interested in whether the jet breaks out or not, then there is a much simpler criterion that provides an answer. To find this criterion we define a new dimensionless parameter that we denote $\E$, which depends on the energy ratio between the jet and the ejecta and on the jet opening angle. Below we find first the relation between $\E$ and $\eta$.

\subsection{Definition of $\E$ and the relation with $\eta$}
\subsubsection{Fast jet head - $\eta \gg 1$}
When $\etacrit \gg 1$, the jet head is initially fast, namely $v_h \gg \vej$ and $\eta \gg 1$.
As long as this condition is satisfied, and after the jet is fully collimated ($\tti>\tc$) we can use the equations of propagation in a static medium as a good approximation. We start from equation A3 of \cite{Harrison2018} written in the instantaneous ejecta rest frame:

\begin{equation}\label{eq:velocity_difference}
    v_h-\vej = \bigg(\frac{\tilde{N}_s^5}{(5-\alpha)^{3-\alpha}}\frac{L_j}{A_\rho \theta_{j,0}^4}\bigg)^{\frac{1}{5-\alpha}}(t-t_d)^{\frac{\alpha-2}{5-\alpha}} ~,
\end{equation}

where $ \tilde{N}_s $ is a numerical coefficient that is related to, but different than, the coefficient $ N_s$ that was found for the case of a static medium \citep{Harrison2018} and later verified for an expanding ejecta \citep{Hamidani2020}. We discuss its role and calculate its value for various $\al$ in Appendix \ref{sec:static}.
Dividing both sides by $ \vej $ yields:
\small
\begin{equation}\label{eq:eta_tmp}
    \eta = \bigg(\frac{\tilde{N}_s^5}{(5-\alpha)^{3-\alpha}}\frac{L_j(t-t_d)^{\alpha-2}t^{5-\alpha}}{A_\rho \theta_{j,0}^4 r^{5-\alpha}}\bigg)^{\frac{1}{5-\alpha}}=\bigg(\frac{2\pi\tilde{N}_s^5}{(5-\alpha)^{4-\alpha}}\frac{\delta^{3-\alpha} E_j}{\Eej\theta_{j,0}^4}\bigg)^{\frac{1}{5-\alpha}} ~,
\end{equation}
\normalsize

where $ E_j = L_j(t-t_d) $ is the energy injected into the jet up to time $t$ and 
\begin{equation}\label{eq:Eej}
    \Eej(v < \vej) = \frac{2\pi}{5-\alpha}A_\rho r_h^{5-\alpha}t^{-2}
\end{equation}
is the kinetic energy carried by the ejecta\footnote{Note that we assume here spherically symmetric ejecta. For application to an ejecta that is not spherically symmetric see \S\ref{sec:nonisotropic}} within $ v < \vej = r_h/t $.

We then define a dimensionless parameter of the system that evolves as the jet propagates and depends mostly on the jet to ejecta instantaneous energy ratio and the jet opening angle:

\begin{equation}\label{eq:tilde_E}
    \E(\tti,v)\equiv  \frac{2N_E^5(\alpha)}{(5-\alpha)^{4-\alpha}}\frac{E_j(\tti)}{\Eej(\vej)}\theta_{j,0}^{-4} ~,
\end{equation}
 where $ N_E(\alpha) $ is a numerical coefficient (obtained from the simulations), which includes $ \tilde{N}_s $. Then Eq. \ref{eq:eta_tmp} can be written in a compact form:

\begin{equation}\label{eq:eta_E}
    \eta(\tti) \approx \big[\delta(\tti)^{3-\alpha}\E(\tti,v)\big]^{\frac{1}{5-\alpha}}  ~~~;~~~ \eta \gg 1~,~ \tc<\tti<\ta~.
\end{equation}
This approximation is valid up to the time where $\eta \sim 1$ and $\E \sim 1$. Note that Eqs.~\ref{eq:eta_tmp}-\ref{eq:eta_E} are all depending on the value of $\frac{E_j}{\Eej\theta_{j,0}^{4}}$, which shows a non-trivial dependence on $\qj$. The reason for this dependence is that the relevant parameter on which $\eta$ depends is the ratio of the jet and ejecta energies per unit of area along the path of the jet. Thus, if the collimation of the jet by the cocoon is (wrongfully) ignored, one obtains that Eqs.~\ref{eq:eta_tmp}-\ref{eq:eta_E}  depend on  $\frac{E_j}{\Eej\theta_{j,0}^{2}}\left(\approx \frac{E_{j,iso}}{\Eej}\right)$, in which case the dependence on $\qj$ corresponds to the energy per unit of area of an uncollimated jet. When the collimation, which by itself depend on $\qj$, is taken into account, one obtains the dependence on $\frac{E_j}{\Eej\theta_{j,0}^{4}}$.

\subsubsection{The asymptotic regime - $\eta = \eta_a$}
During the asymptotic phase the jet evolution is independent of whether it started strong or weak, and thus also the value of $\E$ during the asymptotic phase is similar for both types of jets. The evolution of $\E$ during this phase can be found analytically. When $\eta=\eta_a$ the head radius increases as $t^{1+\eta_a}$ implying that $\vej \propto t^{\eta_a}$ and therefore $\Eej(\vej) \propto \vej^{5-\alpha} \propto t^{(5-\alpha)\eta_a} \propto t$. Since $E_j \propto t$ as well, we obtain that $\E$ is constant during the asymptotic phase:
\begin{equation}\label{eq:E_eta_a}
    \E = \E_a ~~~;~~~ t \gg t_a .
\end{equation}
Below, we use numerical simulations to find the value of $\E_a$, which is indeed of order unity with some dependence on $\alpha$  (see Table \ref{tab:coefficients}).

\subsubsection{Slow jet head - $\eta < \eta_a$}
\label{sec:weak_jets}
When $\etacrit \ll \eta_a$, the jet head propagates very slowly up to $t \sim t_a$ without propagating a significant distance inside the ejecta. For our purpose, it is not very interesting to follow the value of $\E$ during this phase. At $t \gtrsim t_a$ the jet propagation merges with the asymptotic regime discussed above.

\subsection{Breakout criterion}
The conclusion of the previous section is that after the jet propagates a significant way in the ejecta (i.e., $\tti>\tc$ in strong jets and $\tti>\ta$ in weak jets), we can use the approximation
\begin{equation}\label{eq:tilde_E(t)}
    \E(\tti) \approx \delta^{\alpha-3}(\delta-1)^{5-\alpha}+\E_a \approx (\tti-1)^{-2}+\E_a~.
\end{equation}
In strong jets this approximation is obtained from Eqs. \ref{eq:eta_delta}, \ref{eq:eta_E} and \ref{eq:E_eta_a} when we neglect the coefficient that depends on $\alpha$ in Eq. \ref{eq:eta_delta} and use the fact that $\delta(\tti \gg 1) -1 \ll 1 $. In weak jets it is derived using the fact that the jet propagates in the ejecta only at $\tti \gtrsim \ta \gg 1$ when $\delta-1 \ll 1$ and $\E \approx \E_a$. 

Based on that approximation to $\E(\tti)$ we can find a breakout criterion. We define the following parameters:
\begin{equation}\label{eq:Etot}
    \Etot \equiv \frac{2N_E^5}{(5-\alpha)^{4-\alpha}}\frac{\Ejtot}{\Eejtot}\theta_{j,0}^{-{4}}= \frac{N_E^5}{(5-\alpha)^{4-\alpha}}\frac{\E_{j,{\rm iso,tot}}}{\Eejtot}\theta_{j,0}^{-{2}}~,
\end{equation}

where $\Ejtot=L_j t_e$ is the total jet energy, $\E_{j,{\rm iso,tot}}=2 \Ejtot \theta_{j,0}^{-2} $ is the total jet isotropic equivalent energy, and $t_e$ is the jet engine work-time. 
 The jet must propagate a significant way within the ejecta in order to break out and therefore Eq.~\ref{eq:tilde_E(t)} holds at the time of the breakout, when it can be written as $\E(\tti=\tti_{bo}+1) \approx \tti_{bo}^{-2}+\E_a~$
where $  \tti_{bo} \equiv t_{bo}/t_d $, and $t_{bo}$ is the duration that the engine works until the jet breakout (the jet breaks out at time $t_d+t_{bo}$ after the merger). A successful jet requires $t_e>t_{bo}$ and $\E_{\rm tot}>\E(\tti=\tti_{bo}+1)$. Therefore a criterion for a successful breakout is:
\begin{equation}\label{eq:Etot_crit}
\begin{aligned}
    \E_{\rm tot}> \NEtot(\alpha)\tti_e^{-2}+\E_a(\alpha)~,
\end{aligned}
\end{equation}
where $ \NEtot(\alpha)$ is another numerical coefficient obtained from the simulations. When all the numerical calibration factors are included we find a criterion that is accurate to within an order of magnitude to all the density profiles that we explored.
\begin{equation}\label{eq:Etot_crit_all}
 E_{j,{\rm iso,tot}} \gtrsim \left[\left(\frac{t_d}{t_e}\right)^2+2\right]
\begin{cases}
 1.5\Eejtot\left(\frac{\qj}{0.1{\rm~rad}}\right)^2 ~&unmagnetized \\
&\\
 0.2\Eejtot\left(\frac{\qj}{0.1{\rm~rad}}\right)^2  ~& magnetized
\end{cases} ~.
\end{equation}
The different factors of magnetized and unmagnetized jets are discussed in \S\ref{sec:magnetization}. This criterion is applicable only in strong jets with $t_e>t_{\rm col}-t_d$ and in weak jets with $t_e > t_a$. Any weak jet that breaks out must have $t_e > t_a$, since at $t<t_a$ the jet is unable to make a significant way within the ejecta. In strong jets the collimation takes place at $r_{h,col}$ (given in Eq. \ref{eq:rh_col}), and therefore the breakout takes place at $t>t_{\rm col}$ (i.e., after the jet is fully collimated) if $\vmax/\vmin > \theta_{j,0}^{-2/3}$, which is expected in BNS mergers (see discussion at \S\ref{sec:BNS}). If $\vmax/\vmin < \theta_{j,0}^{-2/3}$, then the breakout may take place before a full collimation is achieved, and then Eq.~\ref{eq:Etot_crit_all} provides a lower limit for the jet energy needed for breakout, owing to the faster propagation of collimated jets compared to uncollimated ones. Note that Eq. \ref{eq:Etot_crit_all} is independent of the exact 
ejecta density distribution and as we show in the next section it is also applicable to jets with luminosity that evolves with time, as long as the total jet energy can be approximated by $E_{j,tot} \sim L_j(t_e) t_e$. Therefore this criterion can be applied to a very wide range of setups.

In case that $t_e \gtrsim t_a$, as in strong jets with $t_e > t_d$, and weak jets with $t_e \gtrsim \tdeff$, the term $(t_d/t_e)^2$ in Eq. \ref{eq:Etot_crit_all} can be neglected and the criterion can be simplified:
\begin{equation}\label{eq:Etot_crit_simple}
 E_{j,{\rm iso,tot}} \gtrsim
\begin{cases}
 3 \Eejtot\left(\frac{\qj}{0.1{\rm~rad}}\right)^2~& t_e > t_a,~~unmagnetized \\
&\\
0.4 \Eejtot\left(\frac{\qj}{0.1{\rm~rad}}\right)^2~& t_e > t_a,~~magnetized
\end{cases} ~.
\end{equation}
If $t_e$ is unknown then this simplified criterion is necessary, but it may be insufficient.

We stress that Eq. \ref{eq:Etot_crit_all} and \ref{eq:Etot_crit_simple} apply to only subrelativistic heads, and given all the approximations we used, they are accurate to within an order of magnitude. Thus, if the two sides of Eq. \ref{eq:Etot_crit_all} are comparable to within an order of magnitude, a numerical simulation is probably required to find if the jet breaks out or not.

\subsection{Breakout time}\label{sec:breakout_time}
 In case of a successful breakout, the previous section can be used to estimate the breakout time. It is important to note that the breakout that we consider here is the emergence of the jet from the region where the ejecta velocity is $\vmax$, which is the maximal velocity up to which the ejecta distribution satisfies $\rho \propto v^{-\alpha}$ (we assume $\vmax \ll c$). If there is ejecta at higher velocities than $\vmax$, we assume that the density at these velocities drops faster than $v^{-5}$, so if the jet breaks out of $\vmax$ it is likely to  break out of the faster velocity material, even if the engine stops.  We stress this point since it is likely that in BNS mergers in addition to the bulk of the ejecta there is also a low-mass fast tail (see \citealt{Nakar2019} and references therein). Thus, while the observations of GW170817, as well as numerical simulations of BNS mergers, suggest $\vmax \approx 0.3$c, the numerical simulations also find a very steep density profile that continues up to mildly or even ultra-relativistic velocities. Thus, the breakout time that we find here (from $\vmax$) should not be confused with the breakout time from the fast tail which can be significantly longer. Yet, the breakout time found here is approximately the minimal time that the engine should work for the jet to break out of $\vmax$, and thus also most likely from the fast tail.
 
 We start from Eq. \ref{eq:tilde_E(t)} which at the time of the breakout can be written as 
 \begin{equation}\label{eq:t_bo1}
    \E_d \tti_{bo} \approx
    \NEtot\tti_{bo}^{-2}+\E_a~,
\end{equation}
where $  \tti_{bo} \equiv t_{bo}/t_d $, and $t_{bo}$ is the duration that the engine works until the jet breakout (the jet breaks out at time $t_d+t_{bo}$ after the merger). $\E_d$ is defined using Eq. \ref{eq:Etot} by replacing $\Ejtot$ with $L_j t_d$ (or  $\E_{j,{\rm iso,tot}}$ with $\Ljiso t_d$):
\begin{equation}\label{eq:Ed}
     \E_d \equiv \frac{N_E^5}{(5-\alpha)^{4-\alpha}}\frac{\Ljiso ~t_d}{\Eejtot}\theta_{j,0}^{-{2}}~,
\end{equation}
Eq. \ref{eq:t_bo1} can be solved directly to find $t_{bo}$, or without loss of much accuracy, the solution to $t_{bo}$ of a subrelativistic jet head can be approximated as:  
\begin{equation}\label{eq:t_bo2}
t_{bo} \approx
\begin{cases}
 \NEtot^{\frac{1}{3}} \E_d^{-\frac{1}{3}}t_d=
 \left(\frac{\NEtot(5-\alpha)^{4-\alpha}}{N_E^5}\frac{\Eejtot\theta_{j,0}^2}{\Ljiso} t_d^2\right)^{\frac{1}{3}}, & {\scriptstyle \E_d > \E_a~(t_{bo}<t_d)}\\
&\\
\frac{\E_a}{\E_d}t_d=\frac{\E_a(5-\alpha)^{4-\alpha}}{N_E^5}\frac{\Eejtot}{\Ljiso}\theta_{j,0}^2~~,& {\scriptstyle \E_d < \E_a~(t_{bo}>t_d)}
\end{cases} ~.
\end{equation}
Eq. \ref{eq:t_bo2} shows two interesting properties of $t_{bo}$. First, in strong jets with $\E_d>\E_a$ the dependence of $t_{bo}$ on the jet and ejecta parameters is relatively weak. Second, in weak jets where $\E_d<\E_a$, there is a relatively strong dependency on the jet parameters (linear or stronger), but no dependency on $t_d$. This implies that for a given ejecta and jet parameters where only $t_d$ can vary, the breakout time has a minimal value which is obtained for $\E_d \lesssim 1$ where $t_d<t_{bo}$. The physical reason for that is that when $t_d$, and thus $\rmin=\vmin t_d$, is small enough, the ejecta density is too high for the jet to propagate through, and the ejecta has to expand to radii that are much larger than $\rmin$ before the jet head can start propagating. This expansion time provides an effective delay, which sets the breakout time (see discussion in \S\ref{sec:td_eff}).

\subsection{Propagation in aspherical ejecta}\label{sec:nonisotropic}
In many systems, including binary mergers, the ejecta is expected to be nonspherical. Our model, which assumes spherically symmetric ejecta, is applicable to aspherical ejecta (with the minor adjustments listed below), as long as the properties of the ejecta around the jet axis do not considerably vary over an angle that is larger than $\qj$. The reason is that the entire jet-cocoon structure is confined to an angle $\qj$ during the jet propagation in the ejecta (\S\ref{sec:cocoon}). As a result the jet propagation is insensitive to the ejecta properties at $\theta \gg \qj$. If a nonspherical ejecta satisfy this criterion, then our analytic model can be used when the ejecta properties are taken as the isotropic equivalent properties of the ejecta near the poles. Specifically, one should use the following mapping:
\begin{equation}
    \begin{array}{ccc}
\rho_{ej}(v,t) & \to & \rho_{ej}(v,t,\theta=0) \\
v_{ej} & \to & v_{ej}(\theta=0) \\
\Eej(v) & \to & \Eej(v,\theta<\qj)\times 4\theta_{j,0}^{-2} \\
\Eejtot & \to &\Eejtot(\theta<\qj)\times 4\theta_{j,0}^{-2}
\end{array}
\end{equation}
where $\Eej(v,\theta<\qj)$ and $\Eejtot(\theta<\qj)$ are, respectively, the kinetic energy at velocities lower than $v$, and the total kinetic energy which is carried by material that is confined to $\theta<\qj$ (one sided).

\section{Time-dependent luminosity}\label{sec:varying_L}
In the previous sections, and in most previous studies, the jet luminosity is assumed to be constant. In GRBs the jet luminosity may vary on short time scales, but time averaged energy output seems to be relatively constant over the duration of the prompt emission phase. Yet, there are scenarios where the jet luminosity may vary continuously with time. 
It is straightforward to generalize the theory derived above for a luminosity that evolves as a power-law, as long as most of the jet energy at any given time is injected by recent (rather than early) engine activity, namely
\begin{equation}\label{eq:L_t_dependent}
    L \propto (t-t_d)^k~~~;~~~ k>-1~.
\end{equation}
We do not give here a complete solution for this case, nor do we test it numerically. Instead we provide the power-law dependence of $v_h$ and an order of magnitude breakout criterion. Note that if the luminosity drops faster than Eq. \ref{eq:L_t_dependent} (e.g., $k<-1$) then as long as $\al<5$ (and possibly for larger values of $\al$) the jet cannot support the propagation of the head and it is eventually stalled, even if it had been propagating in the ejecta at first.

For a strong jet, as long as $\eta \gg 1$ we use the static approximation. Following the derivation of \cite{Bromberg2011}, we find that the head propagation in the Newtonian regime (their equations B2-B11) can be generalized to accommodate a time evolving luminosity simply by replacing the constant $L_j$ with $L_j(t)$. This is accurate up to a constant correction of the normalization factors that depend on $k$. This corresponds to a head velocity that evolves as:
\begin{equation}\label{eq:v_h_Levolve}
    v_h \propto (t-t_d)^{\frac{2+k-\al}{5-\al}}~~~;~~~ \eta \gg 1~.
\end{equation}
The evolution of the velocity during the asymptotic phase can be derived using dimensional analysis. Following similar lines as in \S\ref{sec:asymptotic} we find (see also \citealt{Margalit2018}):
\begin{equation}
    v_h \propto t^{\frac{6+k-\al}{5-\al}}~~~;~~~t>t_a
\end{equation}
and
\begin{equation}\label{eq:eta_a_Levolve}
    \eta_a = \frac{1+k}{5-\al}
\end{equation}

We can now replace the time dependence in Eq. \ref{eq:velocity_difference} with the one derived in Eq. \ref{eq:v_h_Levolve} and following the same line of arguments as in \S\ref{sec:E_tilde_and_BO}, we find that Eq. \ref{eq:eta_E} is applicable (up to a constant factor that depends on $k$) to a time-dependent luminosity where we approximate $E_j(t) \approx L_j(t)(t-t_d)$. Similarly, using Eq. \ref{eq:eta_a_Levolve} we see that Eq. \ref{eq:E_eta_a} is also applicable to a time-dependent luminosity. Thus, both the breakout criterion (Eq. \ref{eq:Etot_crit_all}) and breakout time given in Eq. \ref{eq:t_bo2} are applicable to a time-dependent luminosity (as long as the $\int_{t_d}^t L dt \sim L(t) (t-t_d)$), where in the definition of $\E_d$ in Eq. \ref{eq:Ed} the luminosity $ \Ljiso $ is taken as $\Ljiso(t=2t_d)$.

\section{Relativistic head}\label{sec:relativistic}
In this paper we consider only Newtonian ejecta. Thus, if the head is relativistic then $\eta \gg 1$, and the static solution is an excellent approximation, so we can readily apply the results of \cite{Bromberg2011} and \cite{Harrison2018}. Note that since there is no study of the weakly magnetized jets in the relativistic regime, our results in this section are applicable to unmagnetized jets. First we find the condition for the jet head to be relativistic at the time of the breakout from $\vmax$. The transition from a Newtonian to a relativistic head takes place roughly when the solution of the Newtonian head (\S\ref{sec:analytic}) gives $v_h = c$. Since the jet head reaches the location where $\vej=\vmax$ after it is fully collimated, we can estimate the head velocity upon breakout from Eq. \ref{eq:vh0} where $\rmin$ is replaced with $\rmax=\vmax t_d$. Doing that and using the definition of $\eta_0$ (Eq. \ref{eq:eta0}) we obtain that the breakout is relativistic if
\begin{equation}\label{eq:rel_bo}
    \eta_0 \gtrsim 3 \frac{ w^{\frac{2-\alpha}{3}}}{\bmin}~~~;~~~ {\rm relativistic~breakout}~,
\end{equation}
where we ignore the weak dependence of the coefficient in Eq. \ref{eq:vh0} on $\alpha$ and use the definition
\begin{equation}
    w \equiv \frac{\vmax}{\vmin}~.
\end{equation}
The jet head is relativistic when it starts to propagate in the ejecta, even before it is collimated, if $\etauc > c/\vmin$ or:
\begin{equation}\label{eq:rel_start}
    \eta_0 \gtrsim \beta_{\rm min}^{-1/2} \theta_{j,0}^{-2/3} ~~~;~~~ {\rm relativistic~start}~.
\end{equation}
When both criteria are satisfied the head is relativistic during its entire propagation through the ejecta. If only one is satisfied, then the jet starts in one regime and crosses to the other during the propagation in the ejecta. Below we find the breakout criterion for a jet head that is relativistic at all times. Usually this criterion is applicable also when only the breakout is relativistic since most of the propagation time is spent at larger radii (after collimation). Yet, in this case it is safer to solve the entire evolution,  starting at the collimation phase, using the static approximation.

The location of a jet head when it is relativistic at all times is simply
\begin{equation}
    r_h \approx c(t-t_d)~.
\end{equation}
Unlike the case of a Newtonian head, a successful breakout is possible also in cases where the engine stops long before the breakout takes place. The criterion for a successful breakout of a relativistic head is that the jet engine time will be long enough to allow for the jet tail (the fluid elements ejected last) to break out from the ejecta before crossing the reverse shock at the jet head. Defining $t_{e,bo}$ as the engine work-time at the point that the fluid element that crosses the reverse shock upon breakout is launched, the breakout criterion is $t_e>t_{e,bo}$. Since the tail Lorentz factor is much larger than that of the head, the relative velocity between them at any time is $c/2\Gamma_h^2(t)$.  Assuming that the head Lorentz factor evolves as $\Gamma_h \propto t^m$, we obtain
\begin{equation}\label{eq:rel_bo_criterion}
    t_{e,bo} = \frac{R_{bo}}{2c(1-m) \Gamma_{h,bo}^2} = \frac{\beta_{\max}}{2(1-m) \Gamma_{h,bo}^2}t_d~,
\end{equation}
where $\Gamma_{h,bo}$ is the head Lorentz factor at the breakout radius, $R_{bo} \approx \vmax t_d$.

The dependence of $\Gh$ on the system parameters depends on whether the jet is collimated or not\footnote{Note that we consider breakout that takes place at $r>\rhcol$, so a jet with a relativistic head is uncollimated only if it is too strong for the cocoon pressure to collimate it (see \citealt{Bromberg2011}). This is not to be confused with a jet with a Newtonian head that may be uncollimated only at very early times, before the pressure in the cocoon builds up.}. Following equation 2 in \cite{Harrison2018} we find that the criterion for the jet to be collimated upon breakout is 
\begin{equation}\label{eq:rel_collimation}
    \eta_0 \lesssim \frac{w^{\frac{2-\alpha}{3}}}{\bmin}  \theta_{j,0}^{-10/9}~~~;~~~ {\rm collimated~jet}~.
\end{equation}
In the collimated regime we use equation A11 of \cite{Harrison2018} to obtain:
\begin{equation}
    \Gamma_{h,bo}^2 \approx \left(\frac{1}{10\pi}\frac{(3+\al)^2}{7-\al}\right)^\frac{1}{5} 
    \left(\frac{L_j \theta_{j,0}^{-4}}{A_\rho c^{5-\al}}\right)^\frac{1}{5}
    (\beta_{\max} t_d)^\frac{\al-2}{5}
\end{equation}
Using this relation together with Eqs. \ref{eq:rel_bo_criterion} and \ref{eq:Eej} we obtain an expression for relation between $t_{e,bo}$ and $t_d$. The requirement that $t_e>t_{e,bo}$ gives the following breakout criterion,
\begin{equation}\label{eq:rel_bo_criterion_col}
    t_e \gtrsim 0.7 \left(\frac{\Eejtot \theta_{j,0}^2}{\Ljiso}\right)^\frac{1}{5} \beta_{\max}^{\frac{2}{5}}~~ t_d^\frac{4}{5} ~~;~~ {\rm collimated~breakout}~,
\end{equation}
where we neglect a very weak dependence on $\al$. The same criterion can be written, using Eq. \ref{eq:rel_bo_criterion}, in terms of the total jet energy,
\small
\begin{equation}
      E_{j,{\rm iso,tot}} \gtrsim \frac{3(5-\al)}{(3+\al)^2} \Eejtot \left(\frac{\qj}{0.1{\rm~rad}}\right)^2 
      \left(\frac{\beta_{\max}}{0.3}\right)^{-2} \Gamma_{h,bo}^{8} \\
      ~~;~~ {\rm collimated~breakout}~.
\end{equation}
\normalsize

In the uncollimated regime equation A19  of \cite{Harrison2018} reads (after correcting a typo\footnote{Note that there is a typo in Eq. A19 of \cite{Harrison2018} (as well as in Eq. B26 of \citealt{Bromberg2011}), the dependence on the opening angle should be $\Gamma_h^2 \propto \theta_{j,0}^{-1}$.})
\begin{equation}
    \Gamma_{h,bo}^2 \approx \left(\frac{L_j \theta_{j,0}^{-2}}{4\pi A_\rho c^{5-\al}}\right)^\frac{1}{2}
    (\beta_{\max} t_d)^\frac{\al-2}{2} ~.
\end{equation}
Following the same steps as in the collimated regime we obtain the breakout criterion on $t_e$
\begin{equation}\label{eq:rel_bo_criterion_uncol}
    t_e \gtrsim \frac{5(5-\al)^\frac{1}{2}}{7-\al} \left(\frac{\Eejtot }{\Ljiso}\right)^\frac{1}{2} \beta_{\max}^{-\frac{1}{2}}~~ t_d^\frac{1}{2} ~~;~~ {\rm uncollimated~breakout}~,
\end{equation}
and the equivalent criterion on $ E_{j,{\rm iso,tot}} $:
\small
\begin{equation}
      E_{j,{\rm iso,tot}} \gtrsim 100 \frac{(5-\al)}{(7-\al)} \Eejtot 
      \left(\frac{\beta_{\max}}{0.3}\right)^{-2} \Gamma_{h,bo}^{2} \\
      ~~;~~ {\rm uncollimated~breakout}~.
\end{equation}
\normalsize

Note that in the uncollimated regime, the solution is applicable only for $\alpha<4$ (for $\alpha>4$ the head can propagate without support of the jet).

A comparison of the breakout criteria in the relativistic regimes to the breakout criterion in the Newtonian head regime (Eq.\ref{eq:Etot_crit_all}) shows that, for a given ejecta and initial jet opening angle, a breakout of a relativistic head  always requires more energy than the breakout of the Newtonian head. This implies that the necessary (but possibly insufficient) breakout criterion, Eq. \ref{eq:Etot_crit_simple}, is valid also when a relativistic head is considered.

\section{Numerical simulations}
\label{sec:numerical}

\begin{table}
	\setlength{\tabcolsep}{3.5pt}
	\centering
	\begin{tabular}{ l | c  c  c  c  c  c  c  c  c  }
		
		Model & $ \etacrit $ & $ \eta_0 $ & $ \etauc $ & $ \etac $ & $ \theta_{j,0} $ & $ \alpha $ & $ \rmin~[z_{\rm beg}] $ & $ w_f $ & $ \tilde{t}_f $ \\
		\hline
		$ \Ca $ & 190 & 240 & 26 & 10 & 0.1 & 2 & 3 & 42 & 2.2 \\
		$ \Nc $ & 240 & 450 & 26 & 12 & 0.05 & 2 & 4 & 61 & 2.2 \\
		$ \Wc $ & 390 & 290 & 40 & 16 & 0.18 & 2 & 4 & 52 & 2.2 \\
		$ \Cb $ & 1.7 & 11 & 0.4 & 0.4 & 0.1 & 2 & 3 & 2.3 & 16 \\
		$ \Cc $ & 1.0 & 7.4 & 0.2 & 0.3 & 0.1 & 2 & 3 & 3.6 & 53 \\
		$ \Cd $ & 0.5 & 4.6 & 0.1 & 0.2 & 0.1 & 2 & 3 & 1.8 & 11 \\
		$ \Ce $ & 0.15 & 2.1 & 0.04 & 0.08 & 0.1 & 2 & 3 & 1.7 & 72 \\
		$ \Cf $ & 0.05 & 1.0 & 0.01 & 0.04 & 0.1 & 2 & 3 & 1.1 & 29 \\
		$ \Oa $ & 370  & 380 & 60 & 11 & 0.1 & 1 & 8 & 18 & 1.7 \\
		$ \Ob $ & 1.0 & 7.4 & 0.2 & 0.2 & 0.1 & 1 & 3 & 2.2 & 17 \\
		$ \Ta $ & 39 & 84 & 4.5 & 4.8 & 0.1 & 3 & 3 & 16 & 2.3 \\
		$ \Tb $ & 1.0 & 7.4 & 0.2 & 0.4 & 0.1 & 3 & 3 & 2.3 & 9 \\
		$ \Tc $ & 0.05 & 1.0 & 0.01 & 0.05 & 0.1 & 3 & 3 & 1.1 & 27 \\
		$ \Fa $ & 50 & 100 & 5.8 & 11 & 0.1 & 4 & 3 & 27 & 1.6 \\
		$ \Fb $ & 1.0 & 7.3 & 0.2 & 0.8 & 0.1 & 4 & 3 & 3.3 & 12 \\

	\end{tabular}
	\hfill\break
	
	\caption{ Table of numerical models. Shown are $ \etacrit $ (Eq. \ref{eq:eta_crit}), $ \eta_0 $ (Eq. \ref{eq:eta0}), $ \etauc $ (Eq. \ref{eq:eta_uncol}) and $ \etac $ (Eq. \ref{eq:eta_col}), as well as the jet opening angle $ \theta_{j,0} $, the ejecta density power-law index $ \alpha $, and the minimal radius of the ejecta, $ \rmin $ in units of the jet injection height, $ z_{\rm beg} $, $ w_f $ is the ratio between the ejecta velocity at the jet head location at the end of the simulation and the minimal ejecta velocity, and $ \tilde{t}_f \equiv t_f/t_d $.
		}
\label{tab:models}
\end{table}
 
We calibrate the analytic model by performing a set of relativistic hydrodynamic simulations with \textsc{pluto} \citep{Mignone2007}. The simulations also test the model within the parameter phase space that they cover. In the simulations we use a relativistic ideal gas equation of state, both for the jet and for the ejecta, as appropriate for radiation dominated gas. We use piecewise parabolic reconstruction method and an HLL Riemann solver. As pointed out in the previous sections, the general jet evolution is dictated by three parameters:  $ \theta_{j,0} $, $ \alpha $ and $ \etacrit $ (one can use $\eta_0$ instead of $\etacrit$), and thus they compose our parameter space. The full list of simulations and their parameters is given in Table \ref{tab:models}. Once those three parameters are set, the specific choice for the rest of the parameters is degenerate (e.g., the degeneracy between $ L_j, A_\rho, \vmin, t_d $ in $ \etacrit $), and only affect the scaling of the problem.

In reality the engine has a finite working time, $t_e$, and the ejecta have a maximal velocity, $\vmax$, above which the density drops sharply (sharper than $\rho \propto v^{-5}$). In our simulations the jet engine operates throughout the entire time of the simulations and the jet never reaches the edge of the ejecta. Thus, the simulation ends at time $t_f$ where the head reached ejecta with velocity $\vej(t_f)$. This setup enables us to find out for each simulation whether the jet breaks out successfully for any 
value of $t_e \leq t_f-t_d$ and/or $\vmax \leq \vej(t_f)$. To see that, let us first consider $\vmax$. The evolution of the system at any given time is independent of ejecta at $ v > \vej$, i.e., at any time of the simulation $t$, one can define $ \vmax = \vej(t) $ and find the breakout time for this value of $ \vmax $. Next, consider an engine that stops working at time $t<t_f$ (so $t_e=t-t_d$). The jet is choked once the last fluid element launched by the engine crosses the reverse shock at the jet head. The relative velocity between the last fluid element and the jet head is $\approx c-\vejh(t)$. Since we simulate only Newtonian heads,  $r_h(t) \ll c (t-t_d)$ and the last fluid element that is launched at time $t$ reaches the head before it doubles the radius at which it was at time $t$, so the jet is choked roughly at $r_h(t)$. Therefore, for a given engine time $t_e$ the jet is choked approximately at $r_h(t_e+t_d)$. Thus, our simulations find the breakout time for all the systems with $ \theta_{j,0} $, $ \alpha $ and $ \etacrit $ that appear in Table \ref{tab:models} and with $\vmax/\vmin \leq \vej(\tti_f)/\vmin$ and  $t_e/t_d<\tti_f-1$.

\subsection{Numerical setup}

We recently demonstrated in \citet{Gottlieb2021a} that the propagation velocity of jets in dense media, such as the ejecta from the merger, is more accurate in 3D simulations than in 2D simulations. The two main differences between axisymmetric 2D and 3D models lie in the structure of the jet-cocoon interface and the jet head. Axisymmetric hydrodynamic 2D jets do not exhibit the development of local hydrodynamic instabilities along the jet-cocoon interface, which disrupt the jet's spine and subsequently slow its head down. At the same time, those 2D simulations are subject to a numerical artifact of heavy material that is accumulated on top of the jet head due to the axisymmetry imposition (known as the ``plug").
Before carrying all our simulations we first tested whether 2D simulations are accurate enough for our purposes. Unfortunately, we found that axisymmetric simulations yield significantly different jet propagation velocities than 3D ones, primarily in jets with low values of $ \etacrit $. The reason is that in weaker jets the plug plays a much more dominant role. Therefore, all our simulations are carried out in 3D. We note that all previous works that studied the jet propagation in BNS ejecta, verified and calibrated their results by comparing them with axisymmetric 2D simulations (\S\ref{sec:previous}).

In our 3D Cartesian grid, the jet is injected axisymmetrically as a top-hat jet along the $\hat{z}$ axis from the center of the lower boundary. The jet is injected through a nozzle with a cylindrical radius $ r_j = 10^7\cm $, at the lower boundary at $z=z_{beg}$, where  $z_{beg}$ is set by the jet opening angle such that $ z_{\rm beg} = r_j\theta_{j,0}^{-1} $.
The top-hat jet is injected hot (initial specific enthalpy, $ h = 100 \gg 1 $), with initial Lorentz factor $\Gamma_0$ such that it expands sideways to an angle of $ \theta_{j,0} = 1.4\Gamma_0^{-1} $ (there is no difference between this mode of injection and the injection of a conical cold jet with the same $\qj$; \citealt{Mizuta2013,Harrison2018}).

All simulation grids are constituted by three patches on the $ \hat{x}-\hat{y}-$ axes, a central patch of a uniform cell distribution with two outer logarithmic patches. The $ \hat{z} $-axis includes one uniform patch from $ z_{\rm beg} $ to $ 4\times 10^8 $ cm, after which the cell distribution becomes logarithmic.
The number of cells and the size of each patch vary between the cases of $ \etacrit < 1 $, in which the jet evolution takes place on small scales, and $ \etacrit > 1 $, where the jet reaches farther distances. For $ \etacrit < 1 $ ($ \etacrit > 1 $), the $ \hat{x}-\hat{y}-$ axes have one central patch inside the inner $ |6 \times 10^7| $ cm ($ |3 \times 10^7|~\theta_{j,0}/0.1 $ cm) with a uniform distribution of 100 ($ 50~\theta_{j,0}/0.1$) cells, and 160 (180) logarithmic cells up to $ |6\times 10^9| \cm $. The uniform and logarithmic patches on the $ \hat{z} $-axis have 200 and 400 (500) cells, respectively. The grid boundary on $\hat{z} $-axis is at $ 8\times 10^9 $ cm ($ 4\times 10^{10} $ cm).
In Appendix \ref{sec:convergence} we show that these grids are more than sufficient to reach convergence.

\subsection{Numerical results}
\label{sec:results}

\begin{table}
	\setlength{\tabcolsep}{4.2pt}
	\centering
	\begin{tabular}{  l | c  c  c  c  c  c  }
		
		Coefficient & Equation & $\alpha=1$ & $\alpha=2$ & $\alpha=3$ & $\alpha=4$ & Magnetized \\
		\hline
		$ \Ncrit $ & \ref{eq:eta_crit} & \multicolumn{4}{c}{0.5} & $\times 1$\\
		$ \Ncol $ & \ref{eq:eta_col} & 1.0 & 0.9 & \multicolumn{2}{c}{0.5} & $\times 2$ \\
		$ N_E $ & \ref{eq:tilde_E} & 0.7 & 0.57 & \multicolumn{2}{c}{0.31} & $\times 1.5$\\
		$ \E_a $ & \ref{eq:E_eta_a} & 1.0 & 2.0 & \multicolumn{2}{c}{1.0} & $\times 1$\\		$ \NEtot $ & \ref{eq:Etot_crit} & \multicolumn{2}{c}{0.8} & \multicolumn{2}{c}{0.5} & $\times 1$\\

	\end{tabular}
	\hfill\break
	
	\caption{The calibrated numerical coefficients from the simulations. The last column shows the factor by which one needs to multiply the coefficient to account for the effect of a non-negligible magnetic field $ \sigma \gtrsim 10^{-2} $ (see \S\ref{sec:magnetization}).
		}
\label{tab:coefficients}
\end{table}

\begin{figure*}
		\centering
		\includegraphics[scale=0.35]{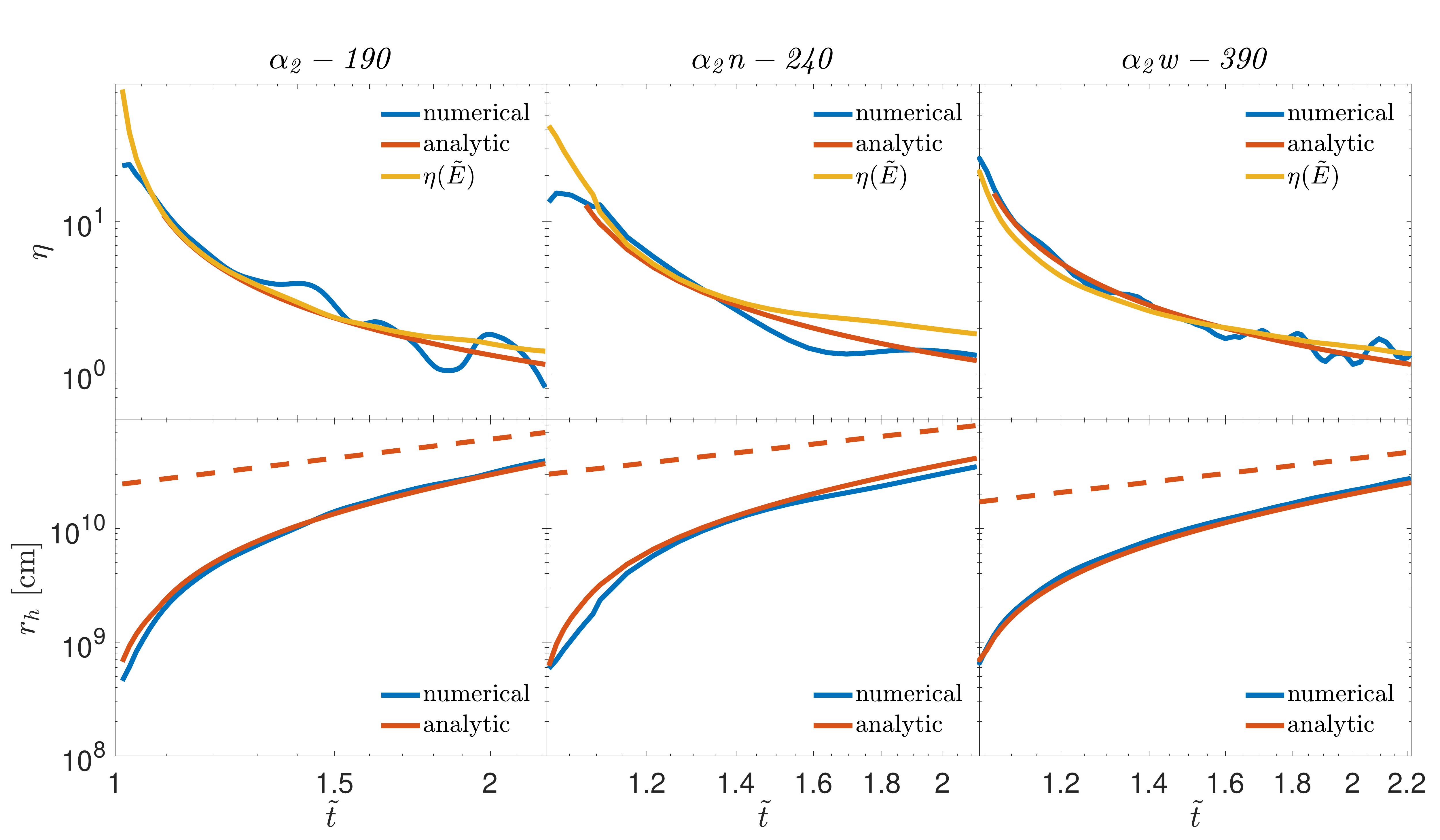}
		\includegraphics[scale=0.35]{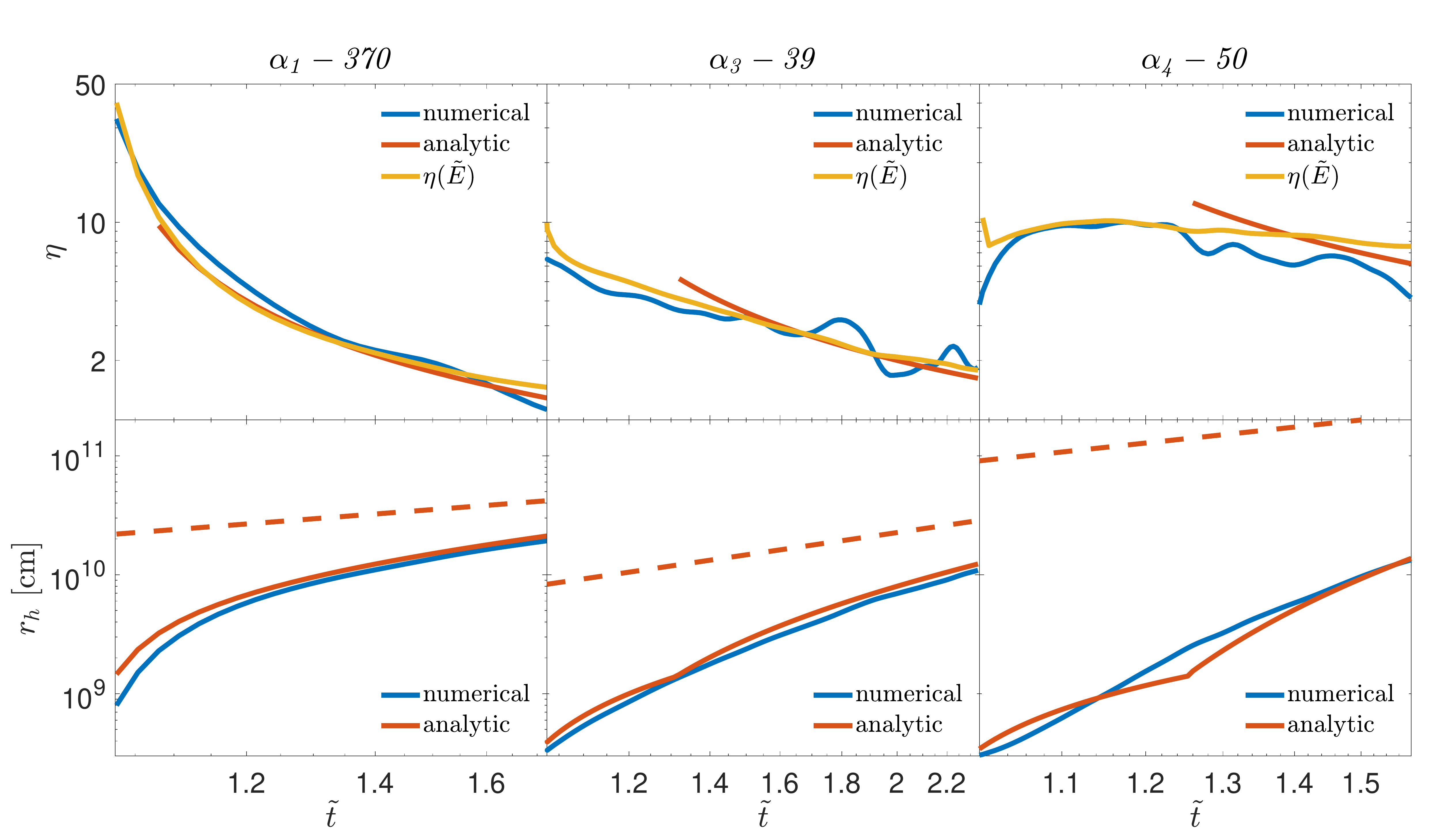}
		\caption[]{
		A comparison of the analytic expressions with the numerical results for $ \etacrit \gg 1 $ models, with our canonical power-law index, $ \alpha = 2 $ (top three models), and with different $ \alpha $ values (bottom three models).
		For each model we show in the upper panel how the numerical value of $ \eta $ (blue) compares with Eq. \ref{eq:eta_delta} (red), and with $ \eta(\E) $ (yellow, Eq. \ref{eq:eta_E}). The lower panel of each model shows the jet head radius, $ r_h $, from simulations (blue) and as calculated analytically (solid red, Eq. \ref{eq:rh_strong}). We also show $r_h$ in the asymptotic phase (Eq. \ref{eq:rh_asymptotic_full}, dashed red line) although none of the simulations shown in this figure reach this phase.
		}
		\label{fig:high_etacrit}
	\end{figure*}

\begin{figure*}
		\centering
		\includegraphics[scale=0.35]{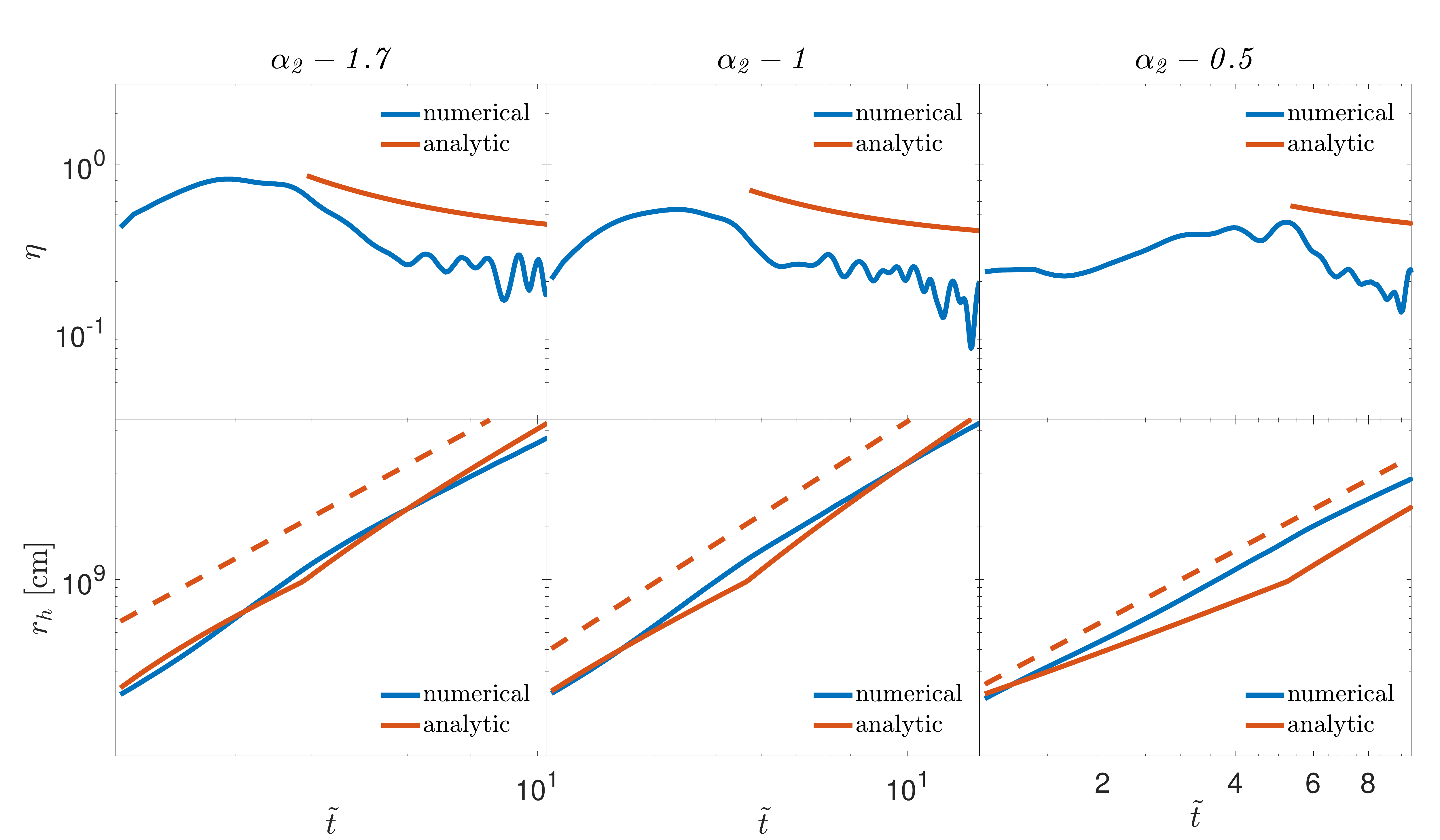}
		\includegraphics[scale=0.35]{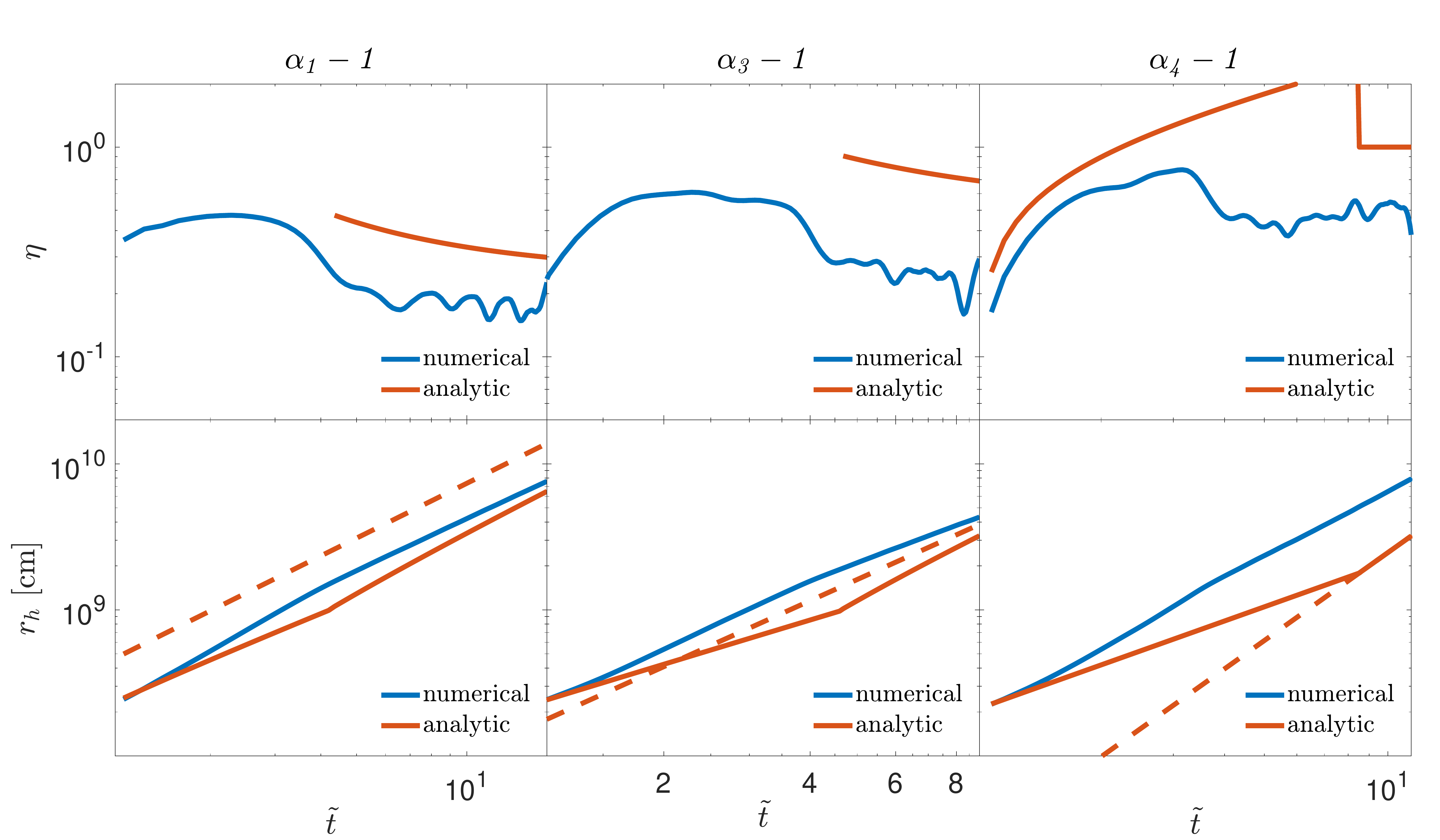}
		\caption[]{
		Same as Fig. \ref{fig:high_etacrit}, but for $ \etacrit \sim 1 $ models. Top: models with different $ \etacrit $ and $ \alpha =2$. Bottom: models with $ \etacrit = 1 $ and different $ \alpha $.
		}
		\label{fig:unity_etacrit}
	\end{figure*}

\begin{figure*}
		\centering
		\includegraphics[scale=0.35]{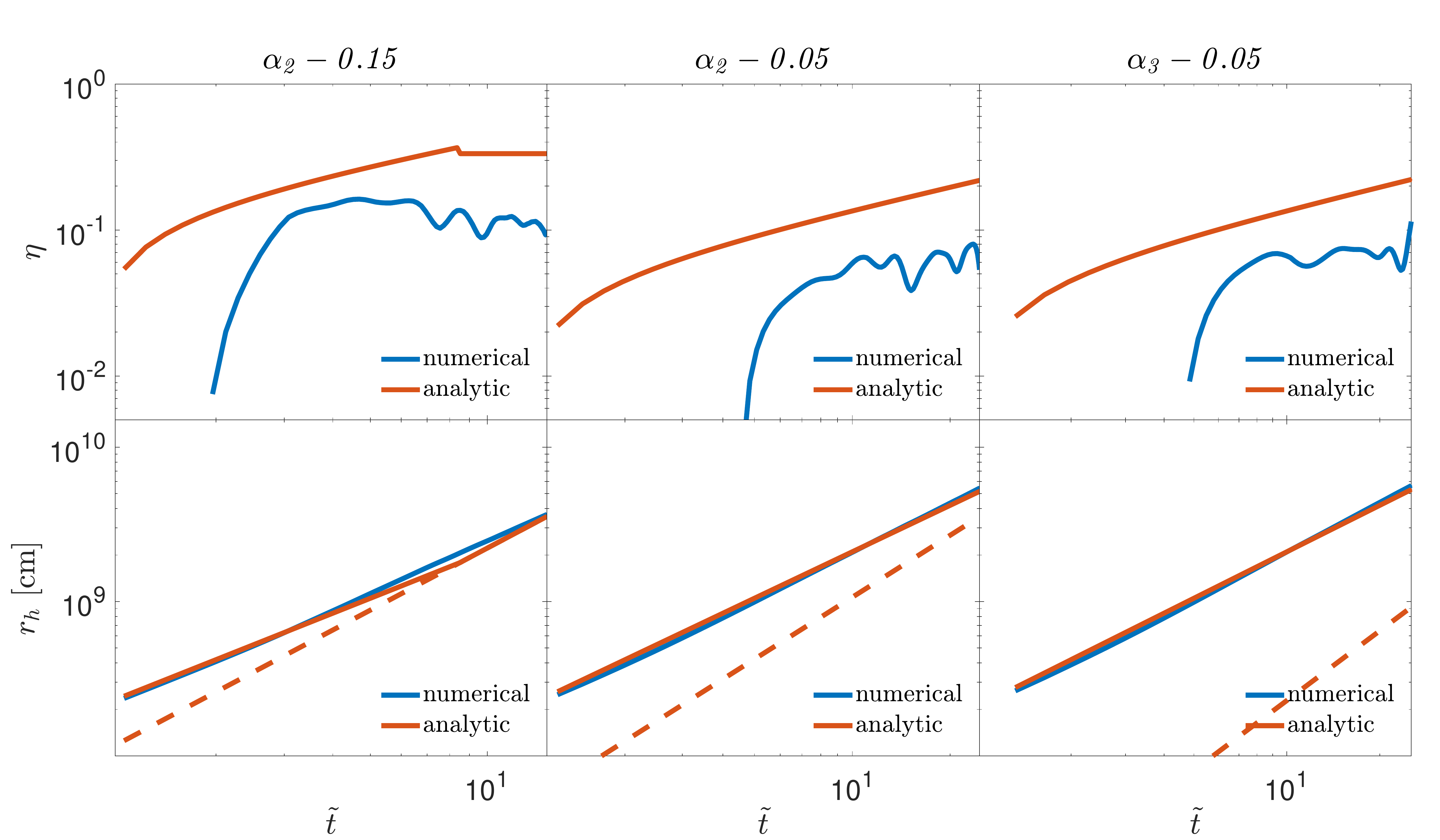}
		\caption[]{
		Same as Figs. \ref{fig:high_etacrit} and \ref{fig:unity_etacrit}, but for $ \etacrit \ll 1 $ models. The analytic curves of $ r_h $ are obtained from Eq. \ref{eq:rh_weak}.
		}
		\label{fig:low_etacrit}
	\end{figure*}

Figs. \ref{fig:high_etacrit}, \ref{fig:unity_etacrit}, and \ref{fig:low_etacrit} depict a comparison of the analytic expressions with the numerical results. The upper panel of each model shows how the numerical value of $ \eta $ (solid blue) compares with analytic approximations (solid red) as obtained by Eqs. \ref{eq:eta_delta} ($ \etacrit > \eta_a$, where the analytic fit starts at $\tc$) and \ref{eq:eta_weak} ($ \etacrit < \eta_a $). The lower panel compares the distance covered by the jet head, $ r_h $, as given by Eqs. \ref{eq:rh_strong} ($ \etacrit > \eta_a $) and \ref{eq:rh_weak} ($ \etacrit < \eta_a $) with its location in the numerical simulations. The asymptotic curve that all the jets eventually converge to (Eq. \ref{eq:rh_asymptotic_full}) is shown as well (dashed red). For very strong jet models (Fig.~\ref{fig:high_etacrit}) we also show $ \eta(\E) $ (yellow) from Eq. \ref{eq:eta_E}, which is the basis for the breakout criterion (Eq. \ref{eq:Etot_crit}) in those jets. The simulations presented in those figures were used to find the calibrating factors given in Table \ref{tab:coefficients}, and the analytic curves in the figures include these factors. The comparison of the simulations to the analytic approximation shows a good agreement. Not only does the general behavior of the jets agree with the analytic expectation, but also the numerical coefficients are in good agreement as all the calibrating factors that we find numerically are of order unity (Table \ref{tab:coefficients}). 

	\begin{figure*}
		\centering
		\includegraphics[scale=0.35]{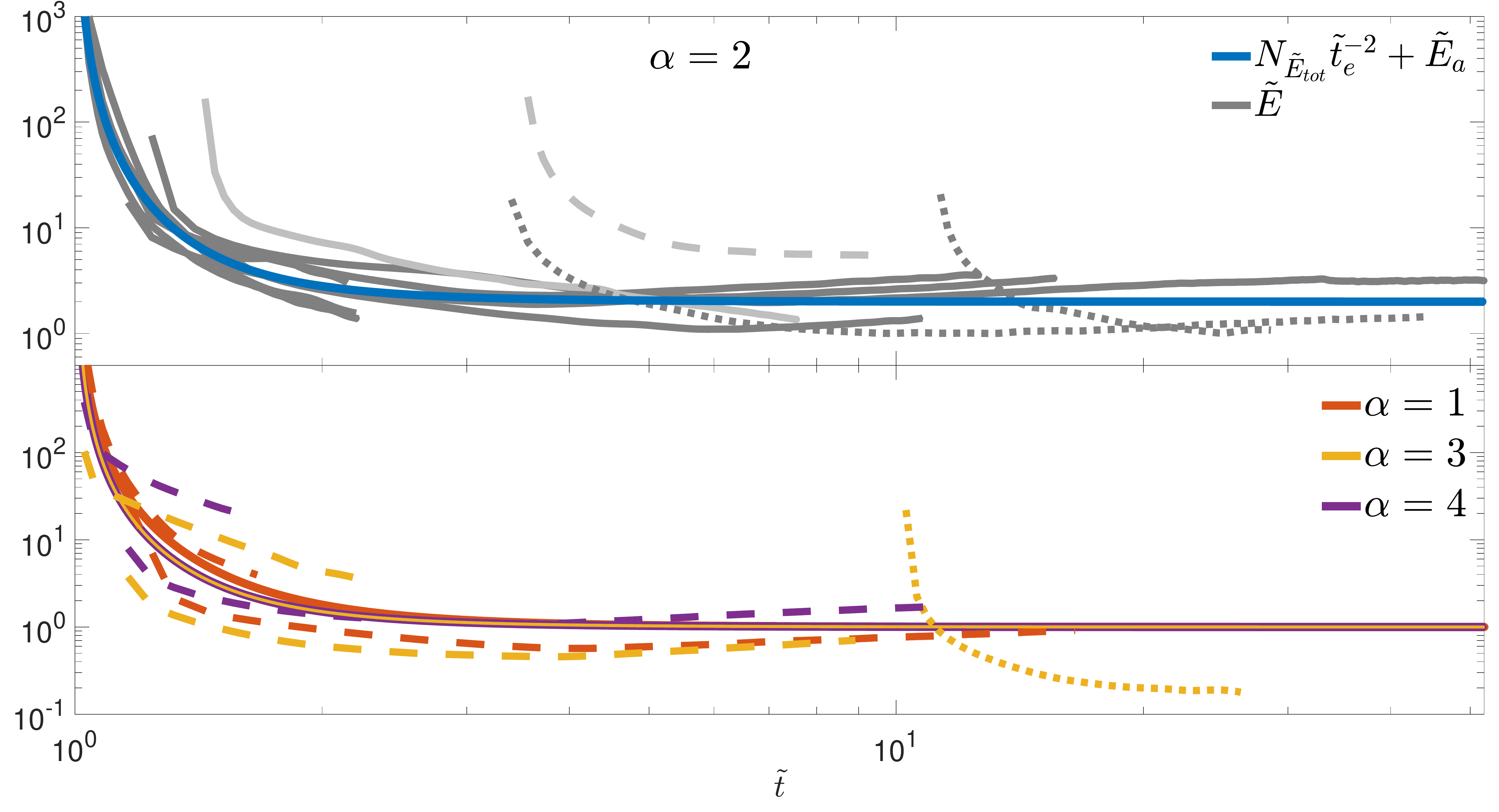}
		\caption[]{
		Top: A comparison between numerical curves of $ \E $ (Eq. \ref{eq:tilde_E}, grey) and the r.h.s. of the analytic expression in Eq. \ref{eq:Etot_crit} (blue). 
		The dotted grey lines are weak jet models $ \Ce $ and $ \Cf $, where $ \etacrit \ll 1 $.
		The light grey lines represent models with cavity of uniform density, with solid and dashed lines reflect magnetized and hydrodynamic jets, respectively (see discussion in \S\ref{sec:magnetization}).
		Bottom: The same comparison as in the top panel, but for $ \alpha = 1 $ (red), $ \alpha = 3 $ (yellow) and $ \alpha = 4 $ (purple).
		Dashed lines are the numerical curves (dotted is the weak jet model $ \Tc $), and solid lines are the analytic expressions.
		}
		\label{fig:Etot}
	\end{figure*}

\begin{figure}
		\centering
		\includegraphics[scale=0.25]{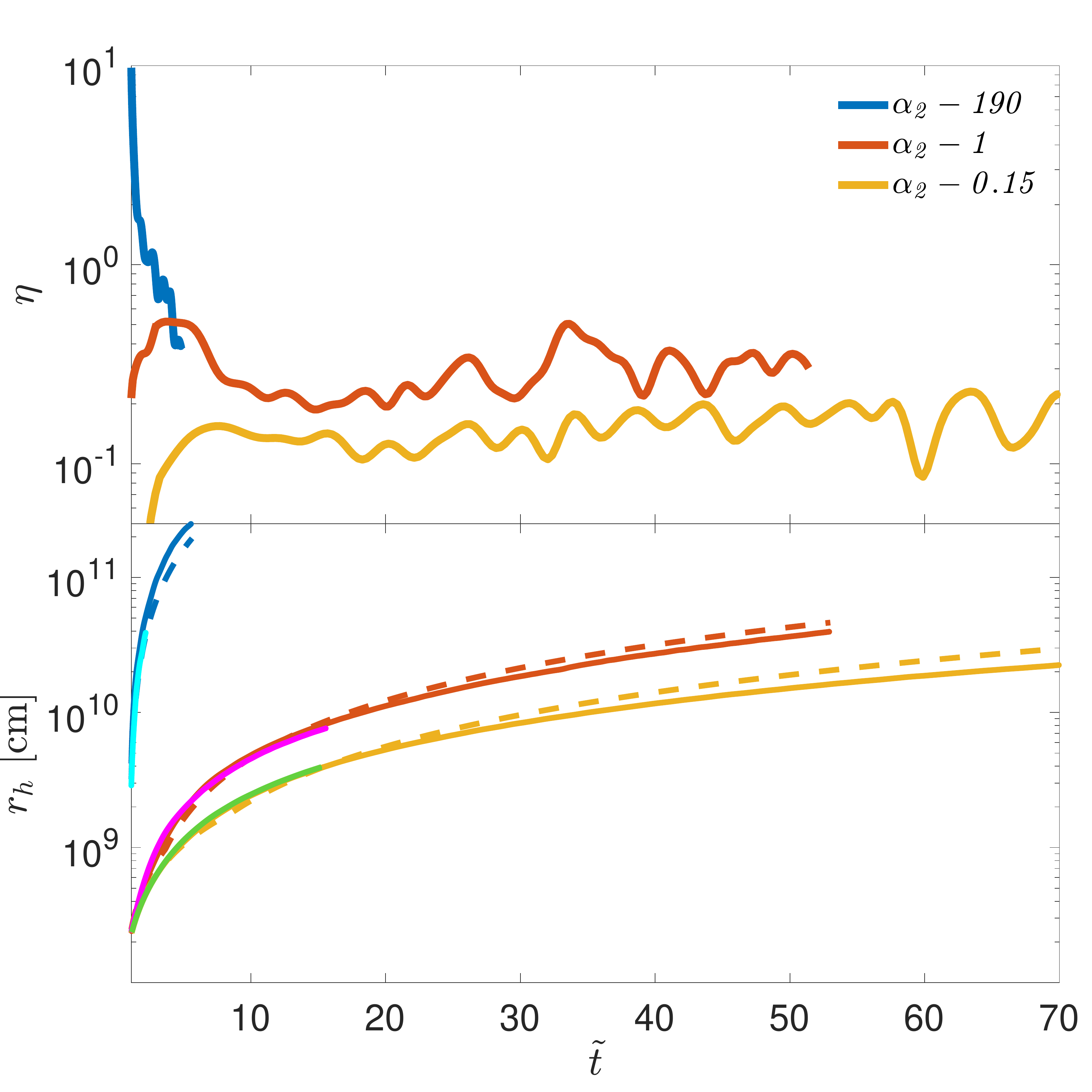}
		\caption[]{Long duration simulations of very strong jet ($ \Ca $; blue), marginally strong jet ($ \Cc $; red) and weak jet ($ \Ce $; yellow) that extend to the asymptotic phase. Solid lines mark the numerical results and dashed lines mark the analytic model (Eq. \ref{eq:rh_strong}).
		High resolution $ \Ca $ (cyan), $ \Cc $ (magenta) and $ \Ce $ (green) models, which are available only for a limited duration, are also shown in the bottom panel for comparison.
		}
		\label{fig:eta_a}
	\end{figure}

Fig. \ref{fig:high_etacrit} depicts very strong jets with $ \etacrit \gg 1 $. The top three models are with our canonical power-law index of $ \alpha = 2 $ and varying value of $\qj$. The three bottom models are with our canonical opening angle, $\qj=0.1$ rad and a variety of $ \alpha$ values. 
All models show a remarkable agreement between the numerical results and both estimates for $ \eta $, with the main deviation taking place with $ \eta(\tilde{E}) $ as $ \eta $ is approaching unity, where Eq. \ref{eq:eta_E} for $ \eta(\tilde{E})\gg1 $ can no longer be applied. Similarly, all the numerical and the analytic head locations are consistent with each other to within a factor of order unity. The simulations shown in Fig. \ref{fig:high_etacrit} do not reach the asymptotic phase. In all models $\eta$ shows a universal temporal evolution during which it drops to $\eta \sim 1$ by $\tti=2$ and we expect that after several ejecta dynamical times $\eta$ converges to its asymptotic value. However, our computational resources prevent us from continuing most of the strong jet simulations to later times. We carried out a single strong jet simulation for longer time, finding that  $\eta \approx \eta_a$ by $\tti=5$ (Fig. \ref{fig:eta_a}). At that time also the location of the jet head is roughly consistent with the analytic prediction for the asymptotic phase (Eq. \ref{eq:rh_asymptotic_full}). 

Fig. \ref{fig:unity_etacrit} shows marginally strong jets with $ \etacrit \approx 1$. The top three models are with $ \alpha = 2 $ and $ \etacrit=0.5, 1$ and $1.7$. The three bottom models are  all with  $ \etacrit = 1 $ and different $ \alpha $ values. The analytic approximation in these jets is good, but not as good as in the very strong jet case. This is expected since the strong jet model is based on the static approximation which is more accurate for larger values of $\etacrit$. Nevertheless, even for these marginal values of $\etacrit$ the approximation of $r_h$ is better than a factor of $2$ during the entire range of the simulations.
Fig. \ref{fig:low_etacrit} depicts weak jets with $ \etacrit < 1 $. The agreement between the analytic and numerical curves of $r_h$ is reasonably good (better than about a factor of 2 at all times).

Fig. \ref{fig:Etot} tests the accuracy of the calibrated analytic criterion for breakout (Eq. \ref{eq:Etot_crit}). It compares the analytic estimate of $\E(\tti)$ (Eq. \ref{eq:tilde_E(t)} with calibrated coefficients), which is the basis for the breakout criterion, with the instantaneous numerical value of $\E$ in each of our simulations. The numerical value of $\E(\tti)$ is extracted by finding $E_j$ and $\Eej$ at every time step of the simulations and plugging them into Eq. \ref{eq:tilde_E}.
In the top panel we present the cases of $ \alpha = 2 $, with $ \tilde{E} $ of all $ \alpha = 2 $ models (grey), compared with the analytic expression shown in blue. One can see that all $ \etacrit \gtrsim 1 $ models lie roughly on the curve of the analytic model, including the cases of full cavity (light grey lines) and magnetized jets (solid light grey, see \S\ref{sec:magnetization}), showing that the breakout criterion is accurate to all strong jet models. The dotted lines represent numerical models $ \Ce $, $ \Cf $ in which the jet is weak. In these models the curve of $ \tilde{E} $ begins as the jet first punches through the ejecta, which corresponds to the sharp drop as $ \Eej $ increases from zero. This brief stage is followed by a continuous gradual rise as the jet is still in the collimation phase, and thus is yet to reach its asymptotic $ \eta_a $.
Similar to strong jets, the two simulated weak jets also show a very good agreement with the breakout criterion.
The bottom panel shows the same comparison between analytic and numerical (dashed lines) results for $ \alpha \neq 2 $ models. Note that the scatter around the analytic estimate is larger for larger values of $ \alpha $, especially in the simulation of a weak jet that we carry for $\alpha=3$ (model $ \Tc $). Yet, for all our models, the analytic model of $\E$ and the simulations are in agreement to within an order of magnitude or better.
This agreement implies that our breakout criterion is applicable, at least in the whole phase-space that we studied numerically.

Finally, we run three additional simulations of strong, marginal and weak jets (models $ \Ca $, $ \Cc $ and $ \Ce $, respectively) for a longer time. Due to computational limitations these simulations were carried out in half of the resolution compared to the rest of the simulations. We verify that this resolution is high enough by a comparison to the full resolution simulations of the same models. The goal of these simulations is to see the transition to the asymptotic phase and the convergence of $\eta$ to its asymptotic value, which for $\al=2$ is predicted to be $\eta_a=1/3$ (Eq. \ref{eq:eta_a}).
In Fig. \ref{fig:eta_a} we present the behavior of $ \eta $ (top) and $ r_h $ (bottom) of the three long duration simulations. For the very strong jet (blue) the simulation is carried out up to $\tti=5$ at which time $\eta \approx \eta_a$. The marginal (red) and weak (yellow) jets are simulated up to $ \tti \approx 50 $ and $ \tti \approx 70 $, respectively. The high resolution simulations are depicted in cyan, magenta and green, showing a good agreement at all times with the low resolution simulations. In all marginal jets we see that within several dynamical times the value of $ \eta $ drops at first slightly below $\eta_a$ (see Fig. \ref{fig:unity_etacrit}). In Fig. \ref{fig:eta_a} we see that following this drop, $\eta$ of model $ \Cc $ rises slowly until it converges to $\eta_a$. The effect of these small variations of $\eta$  (where $\eta$ never drops far below $\eta_a$) on the analytic model of $r_h$ is not significant, as the jet head location is in a good agreement with the analytic estimate at all times. The weak jet ($\Ce $) features a slow monotonic increase, up to fluctuations in the jet head velocity, reaching $ \eta_a $ over a longer timescale of $ \tti \gtrsim 70 $.

\section{Non-empty cavity}\label{sec:cavity}
    
	\begin{figure}
		\centering
		\includegraphics[scale=0.33]{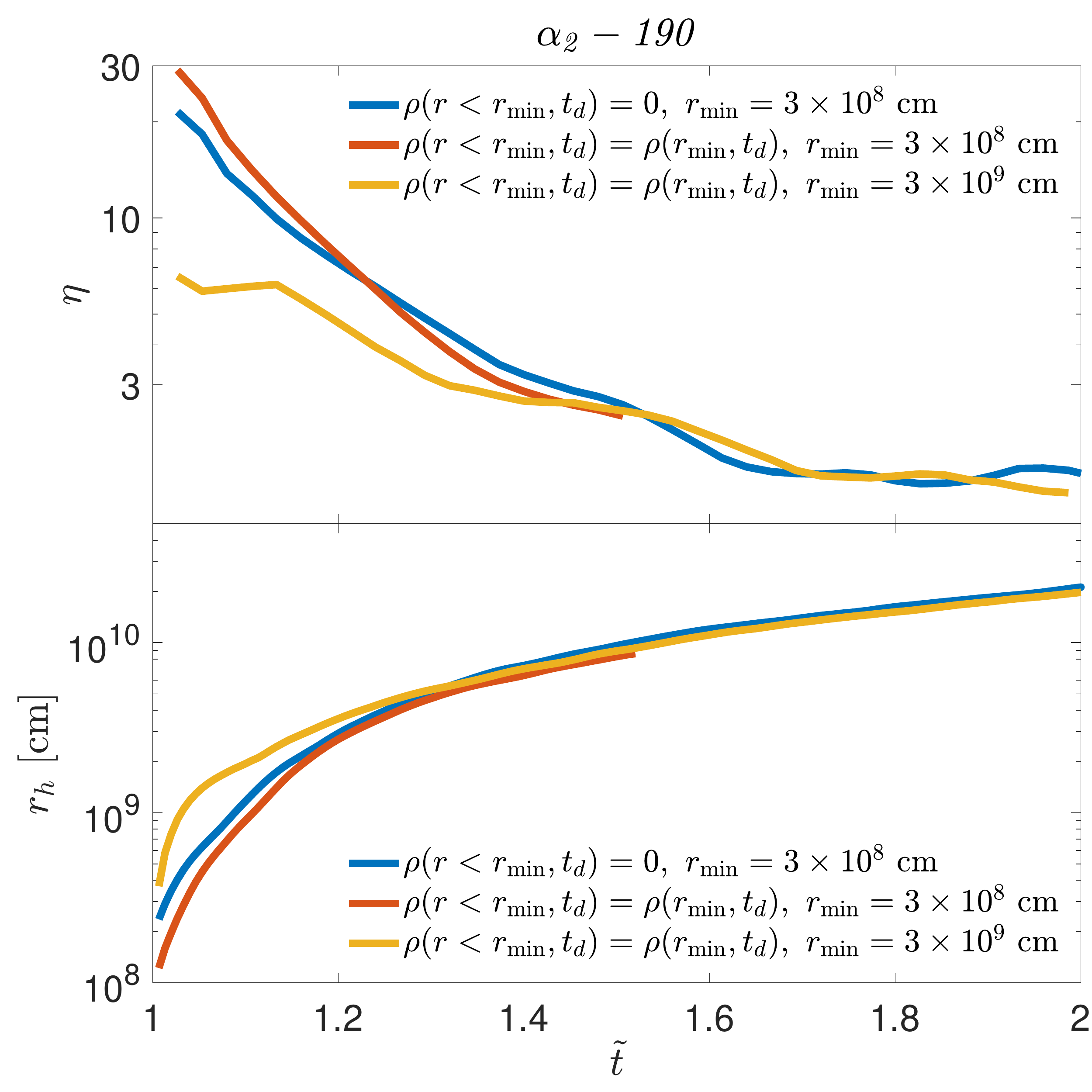}
		\caption[]{
		Comparison of model $ \Ca $ with two different types of cavity. Shown are the original model with an empty cavity (blue) and the same model with a cavity of uniform density that is equal to $ \rho(\rmin,t_d) $ (red). The figure also shows a model with a similar jet and a larger cavity ($\rmin=3 \times 10^9$ cm compared to $\rmin=3 \times 10^8$ cm in $ \Ca $, yellow). At $t_d$ the density at $r>3 \times 10^9$ cm is similar to the two other models, while at smaller radii $\rho(r<3 \times 10^9~\cm,t_d)=\rho(3 \times 10^9~\cm,t_d)$.}
		\label{fig:cavity}
	\end{figure}

In our analytic solution we assumed that at $t=t_d$ there is an empty cavity at $r<\rmin$. However, in a realistic scenario it is possible that the density in the cavity is not negligible. As discussed in \S\ref{sec:cavity_intro}, it is possible, for example, that there is ejecta with $v<\vmin$ with density that is low compared to extrapolation of $\rho \propto v^{-\alpha}$. Another possibility is that there is a continuous mass ejection so mass that was ejected at $t \sim t_d$ fills the cavity. In this section we consider the effect of a non-empty cavity on the propagation of the jet.

For a strong jet a non-negligible density at $r<\rmin$ is expected to have two opposite effects on the head velocity. The obvious one is that the density can be high enough to slow down the head velocity to be subrelativistic in the cavity, thereby delaying the time that the head reaches $\rmin$ compared to the empty cavity case. The second is that the higher density reduces the volume that the shocked jet can occupy, thereby increasing the pressure that the cocoon applies on the jet. The result is that the jet is at least partially collimated by the time that it reaches $\rmin$, and therefore its velocity at this point is faster than that of a jet that propagates in an empty cavity. Thus, when comparing the head propagation in empty and non-empty cavities we expect the former to propagate faster at first, but the latter should catch up at $r > \rmin$. The propagation in a non-empty cavity is bounded from above by the solution of empty cavity (Eq. \ref{eq:rh_strong}), and it can be bounded from below by a jet in a full cavity where at $t_d$ the density in the cavity is $\rho(r<\rmin)=\rho(\rmin) (r/\rmin)^{-\al} $. Our analytic solution already includes this scenario, since it is equivalent to taking Eq. \ref{eq:rh_strong} in the limit $\vmin \to 0$:
\begin{align}\label{eq:rh_full_cavity}
    r_h=& \left(\frac{2(2\pi(5-\al))^\frac{1}{3}\Ncol}{15}\right)^\frac{3}{5-\al}
    \left(\frac{L_j}{\Eejtot}\theta_{j,0}^{-4}\right)^\frac{1}{5-\al}\vmax 
    \notag\\
    &\times~(t-t_d)^\frac{3}{5-\al}t^\frac{3-\al}{5-\al}~~~;~~~{\rm strong~jet~in~ full~cavity},    
\end{align}
The two solutions (in empty and in full cavities) merge at $\tc$ of the empty cavity jet (since at $\tti>\tc$ both solutions follow the approximation of a fully collimated jet is a static medium), implying that at $\tti>\tc$ the jet location is independent of the density in the cavity. 

The above expectation is confirmed by numerical simulations\footnote{Here we are interested in comparing the effect of different cavities rather than the absolute behavior and calibration of the jet head motion. Thus, the simulations in this section are conducted in 2D, where $ r_h $ is smaller by $ \sim 1.5 $ compared to their 3D counterparts.}. Fig. \ref{fig:cavity} shows $\eta$ and $r_h$ of two simulations with similar setups, where the only difference is that in one the cavity is empty whereas in the other it is filled with a uniform density $\rho(r<\rmin,t_d)=\rho(\rmin,t_d)$. As expected, at first $r_h$ of the head in the empty cavity is larger, but quickly the velocity of the head in the filled cavity becomes larger, until at $\tti \approx 1.15$ it catches up, and from that time on the two heads propagate with similar $r_h(t)$. The jet in the empty cavity becomes fully collimated roughly at $\tc = 1.1$. 

Fig. \ref{fig:cavity} also shows a third model with a uniform density inside the cavity, $\rho(r<\rmin,t_d)=\rho(\rmin,t_d)$. It has the same jet parameters, $t_d$ and density normalization $A_\rho$ as the first two simulations, but its $\rmin=3 \times 10^9$ cm rather than $\rmin=3 \times 10^8$ cm of the first two (equivalent to $\vmin$ that is larger by a factor of 10). The head in the third model propagates at first at a faster  velocity than the other two, up to a radius larger than $10^9$ cm. But, as soon as it starts propagating inside the bulk of the ejecta, it significantly decelerates such that the three models converge at $\tti \sim 1.3$. At later times all the models have similar $r_h(t)$. Our conclusion is, that as expected, the density in the cavity affects the head location of strong jets up to a few times $\rmin$ and it has only a small effect on $r_h(t)$ once the head propagates a significant distance in the bulk of the ejecta (at $r>\rmin$). Naturally, the density in the cavity should have no effect on the propagation during the asymptotic phase.

\begin{figure}
		\centering
		\includegraphics[scale=0.33]{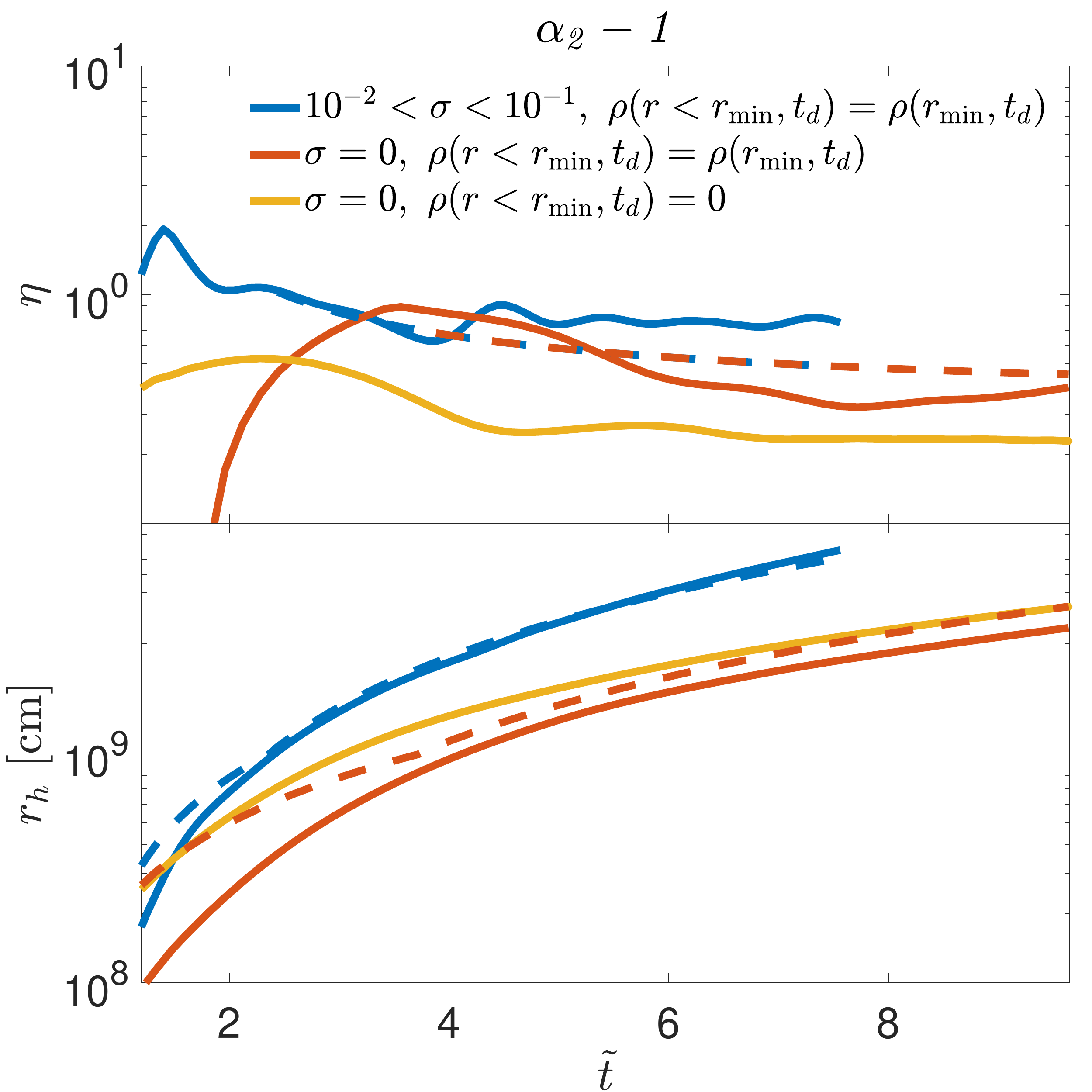}
		\caption[]{$ \eta $ (top) and $ r_h $ (bottom) plots of numerical (solid lines) and analytic (dashed lines) models of magnetized (blue) and hydrodynamic (red) jets in model $ \Cc $ with cavity of uniform density, compared with the canonical unmagnetized $ \Cc $ model (yellow) with an empty cavity. Magnetized jets with $ \sigma = 0.1 $ and $ \sigma = 0.01 $ yield the same results, and thus the blue curve represents the entire range in between.}
		\label{fig:magnetic}
	\end{figure}

For a weak jet a non-empty cavity may prevent the head from propagating in the cavity. Nevertheless, once the ejecta expands, the jet starts propagating in the cavity at $t \ll t_a$. Since at $t \gg t_a$ the jet propagates in the asymptotic phase where the initial conditions (e.g., $t_d$ and $\rmin$) are forgotten, it is expected that by that time the head of the jet with the non-empty cavity approaches the same $r_h(t)$ as that of the jet with the empty cavity. Fig. \ref{fig:magnetic} shows (in addition to the magnetic jet) two unmagnetized jets. One with an empty cavity and $\etacrit=1$ and the other with a constant density cavity. The latter propagates at first very slowly inside the dense cavity, but within several dynamical times it approaches the head location of the former.    

We conclude that in both strong and weak jets, the jet propagation is affected by the cavity density as long as the jet head is still at radii that are comparable to the cavity size. As the jet head enters the ejecta, the strong (weak) jet head is approaching Eq. \ref{eq:rh_strong} (\ref{eq:rh_weak}), regardless of the cavity density, and when $r_h \gg \vmin t$ it forgets the initial conditions in the cavity entirely.

\section{Effect of magnetization}\label{sec:magnetization}

An important jet property, which has never been addressed in the context of analytic description of jet propagation in expanding medium, is the jet's magnetization, $ \sigma \equiv \frac{B'^2}{4\pi h\rho c^2} $, where $ B' $ is the proper magnetic field. Even a subdominant\footnote{Studies of jets in stars found that even if the jets are launched as Poynting-flux dominated, magnetic dissipation at the collimation nozzle may reduce their magnetization to subdominant above the nozzle \citep{Bromberg2016,Gottlieb2022}.} magnetization can greatly alter the jet propagation in the medium, e.g., by a stabilizing the jet against the development of hydrodynamic instabilities on the jet-cocoon boundary \citep{Gottlieb2020b,Gottlieb2021b,Matsumoto2021}.
In a recent study, \citet{Gottlieb2020b} performed numerical simulations to compare hydrodynamic and magnetized jets with toroidal magnetic fields. They found that for a hydrodynamic jet which propagates at $ v \approx 0.2 $c in a star, introducing an initial magnetization of $ \sigma \sim 10^{-2}-10^{-1} $ to the jet increases its velocity by a factor of three. This result, albeit obtained in a static medium and for a specific set of jet and medium parameters, can be utilized for our results in the regime of strong jets with $ \etacrit \gg 1 $.

Here we carry out two simulations of a magnetized jet in the regime of $ \etacrit \sim 1 $, to complement the work of \citet{Gottlieb2020b} on propagation in static media, which is applicable to strong jets. We use a similar magnetic configuration of a toroidal field with initial $ \sigma = 10^{-1} $ and $ \sigma = 10^{-2} $ which we apply to model $ \Cc $\footnote{For the magnetic jet simulations we make use of the constrained transport method in \textsc{pluto}.}. To avoid numerical instability, we fill the cavity with uniform density such that $ \rho(r<\rmin,t_d)=\rho(\rmin,t_d) $. We also conduct a similar filled cavity simulation with a hydrodynamic jet for comparison (see \S\ref{sec:cavity} for a discussion about the effect of dense cavity on the jet propagation).

In Fig. \ref{fig:magnetic} we compare the numerical (solid lines) and analytic (dashed lines) results of magnetized (blue) and hydrodynamic (red) jets with filled cavity, and a hydrodynamic jet with an empty cavity (yellow). The analytic models are derived assuming an empty cavity, and therefore they are expected to converge with the filled cavity simulation only when $r_h$ reaches a few times $\vmin t$. We find that the magnetized jets with $ \sigma = 0.1$ and $ \sigma = 0.01 $ feature the same jet evolution, similar to what was found in \citet{Gottlieb2020b}, and thus the blue curve represents both values of $ \sigma $.
The effect of the magnetic stabilization of the instabilities along the jet boundary is that the jet head velocity is $ \sim 2-3 $ faster than that of hydrodynamic jets.  This implies, in turn, that the jet head location is also about 2-3 times farther than its hydrodynamic counterpart.
Therefore, the jet crosses the filled cavity and becomes collimated faster. We find that the required correction factor is doubling the value of $ \Ncol $. As can be seen in the bottom panel, taking $ \Ncol $ to be twice larger sets the analytic jet head location to fit  the simulations well.  The difference in the head velocity that we find here (a factor of 2-3) is similar to the one found in the case of a static medium \citep{Gottlieb2020b}, which is applicable for strong jets as long as $\eta \gg 1$. We conclude that both in the regime of $ \etacrit \sim 1 $ and $ \etacrit \gg 1 $ the jet head velocity propagates 2-3 times faster, and
taking  $\Ncol $  to be twice larger than in hydrodynamic jets is expected to provide a reasonable approximation to magnetized jets with $ \sigma \sim 10^{-2}-10^{-1} $. 

By the requirement that the jet head velocity is $ \sim $ 2 faster than its hydrodynamic counterpart, Eqs. \ref{eq:velocity_difference} and \ref{eq:tilde_E} dictate that $ N_E $ is 1.5 times larger than the value obtained for hydrodynamic jets. By applying the correction factor to $ N_E $, the magnetized jet model is found to coincide with the universal asymptotic behavior of $ \tilde{E} $ in Fig. \ref{fig:Etot} (solid light grey line in the top panel).
This follows a short delayed phase of $ \tilde{E} \gg 1 $ which comes as a result of the filled cavity which delays the onset of the jet propagation in the ejecta.
For comparison, we also present in Fig. \ref{fig:Etot}  a hydrodynamic jet with a filled cavity which exhibits the same delayed behavior (dashed light grey line in the top panel).
Finally, Eq. \ref{eq:Ed} shows that the increase in $ N_E $ for mildly-magnetized jets is equivalent to an increase of almost an order of magnitude in the jet luminosity. Similarly, Eq. \ref{eq:t_bo2} implies that the presence of a non-negligible magnetic field shortens the breakout time by a factor of two if the jet breaks out at $ t_{bo} < t_d $, and by almost an order of magnitude if $ t_{bo} > t_d $.

\section{Comparison with previous studies}
\label{sec:previous}
The propagation of relativistic jets in expanding media was studied by many authors over the past few years \citep[e.g.,][]{Murguia-Berthier2014,Murguia-Berthier2017,Murguia-Berthier2021,Duffell2018,Margalit2018,Matsumoto2018,Gill2019,Lazzati2019,Nakar2019,Beniamini2020,Lyutikov2020,Salafia2020,Hamidani2020,Hamidani2021}. Here we summarize the findings of previous studies and compare them to our results of the jet head propagation, the breakout time and the breakout criterion. We focus on studies that derived analytic expressions that can be compared to ours.

The most prominent difference between our work and previous studies is that our model was derived (and tested and calibrated numerically) for all types of jets  during the entire evolution (weak and strong, unmagnetized and magnetized, Newtonian and Relativistic). That is, we formulate the jet propagation for both strong and weak jets and we do that i) during the pre-collimation and the collimation phases; ii) during the collimated phase of a strong jet when $v_h \gg \vej$; and iii)  during the asymptotic phases as $v_h-\vej < \vej$. All previous analytic works considered only a single regime, either a strong jet with a Newtonian head and $ v_h \gg \vej $ or a Newtonian head in the asymptotic phase. 
As we discuss below, even in these phases there are some differences between our results and previous studies.

First attempts for deriving a breakout criterion for a jet in the Newtonian regime in expanding media were provided by \citet{Murguia-Berthier2014,Murguia-Berthier2017,Duffell2018}. These studies (and also \citealt{Lyutikov2020} later on) ignored the jet collimation. Namely, they assume (explicitly or implicitly) that the jet propagates in the ejecta with an opening angle that is similar to the one it was injected with, thus missing the factor of $ \qj^2 $ in the breakout criterion (e.g., in Eq. \ref{eq:Etot_crit_simple}). This is inconsistent with any of the jet regimes, since when the head is Newtonian the pressure, which builds up in the cocoon during the propagation, is always strong enough to collimate the jet. Simulations show that the only case where a jet with a Newtonian head remains uncollimated is when the initial jet opening angle is large, $\qj \gtrsim 0.5$ rad \citep[e.g.,][]{Gottlieb2018b}. In such jets the shocked ejecta is accumulated on top of the jet head since it does not have enough time to be pushed aside. The evolution of such jets is more similar to that of a segment from a spherical explosion,where Eq. \ref{eq:v_Lt}, which is used by all these studies, is not applicable. Therefore, the various results derived in studies that neglect the collimation by the cocoon of a jet with a Newtonian head, would not be applicable to a broad range of regimes of physical jets.

\begin{figure}
		\centering
		\includegraphics[scale=0.22]{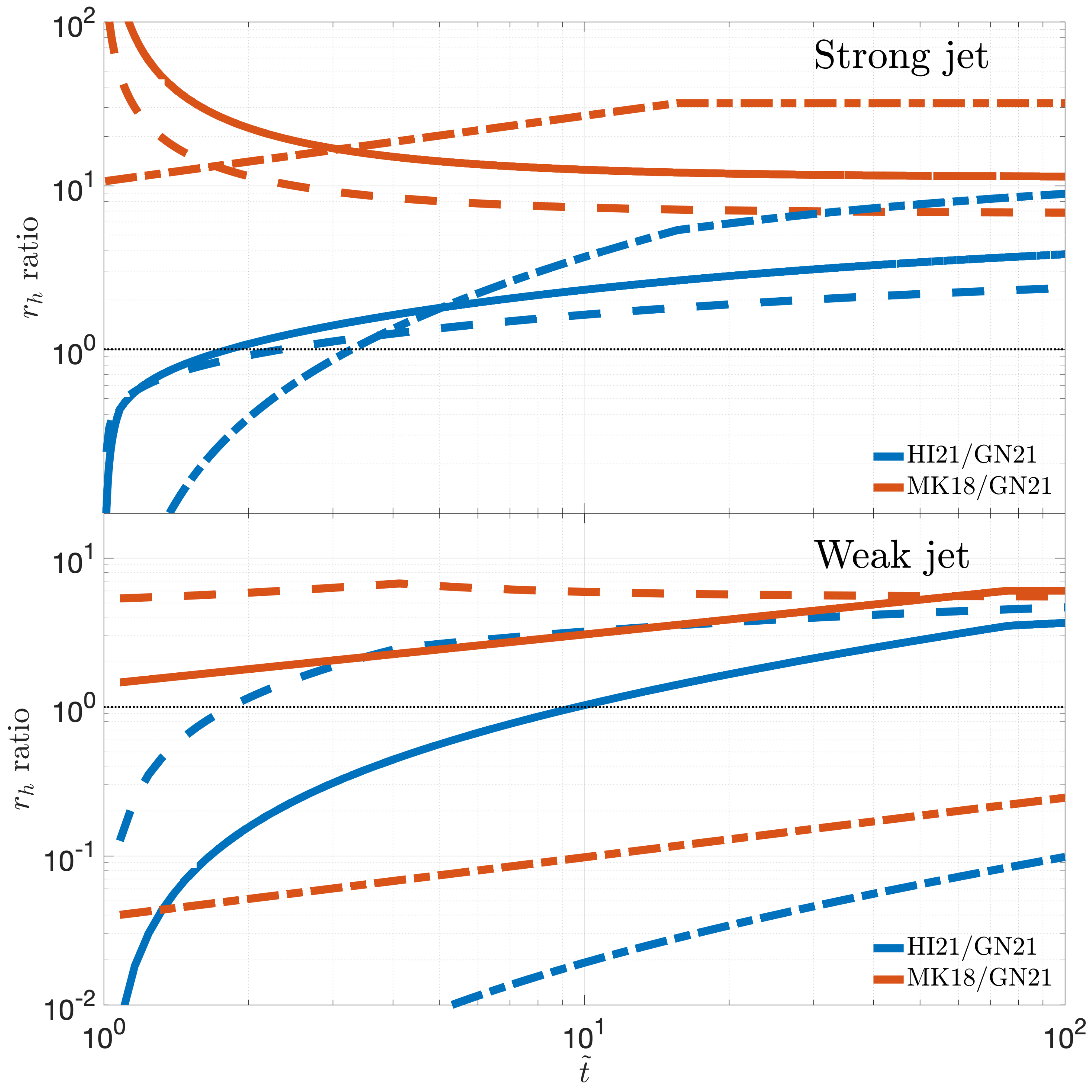}
		\caption[]{The ratio between the analytic solution to $ r_h $ of \citet[][]{Hamidani2021} (HI21; blue) and \citet[][]{Matsumoto2018} (MK18; red), for strong (top) and weak (bottom) jets, and our analytic solution for unmagnetized jets.
		Each set of models is shown for $ \alpha = 1 $ (dashed), $ \alpha = 2 $ (solid) and $ \alpha = 2.5 $ (dash-dotted). The model parameters that are used for the analytic solutions in the  strong (weak) jets with $ \alpha = 1,2, 2.5 $ are the same as those of $ \Oa $ ($ \Ob $), $ \Ca $ ($ \Cf $), and $ \Ta $ ($ \Tc $), respectively. Note that since the models of HI21 and MK18 are undefined at $ \alpha = 3 $, we use here the parameters of models $ \Ta $ and $ \Tc $ with the only modification of taking $ \alpha = 2.5 $.
		For the solution of HI21 we use their table 1 to set the following values for their parameters $ \langle\eta'\rangle=1/4 $ and $ \langle\chi\rangle=5/4$, as those are obtained in their model with an opening angle similar to ours, $ \qj = 0.1 $ rad.
		}
		\label{fig:comparison}
	\end{figure}

All other studies carried out some type of generalization of the model derived by \cite{Bromberg2011}. Some used the simplification of a single typical jet head velocity \cite[e.g.,][]{Gill2019,Lazzati2019,Nakar2019}, which is similar to taking $\alpha=2$. Others, considered a range of $\alpha$ values \citep[e.g.,][]{Margalit2018,Matsumoto2018,Hamidani2020,Hamidani2021}. \cite{Nakar2019} considered only the strong jet regime where $\eta>1$ obtaining analytic results that are consistent with ours in jets with $\etacrit \gg 1$ at $t \lesssim 2 t_d$. \citet{Margalit2018} and \cite{Matsumoto2018} provided an analytic expression for the evolution during the asymptotic phase, when $t \gg t_d$. The power-law of the temporal evolution that they found (equations 22 of \citealt{Margalit2018} and A10 of \citealt{Matsumoto2018}) is the same as the one that we derive (Eq. \ref{eq:rh_asymptotic}), but their normalizations of the asymptotic phase, which were not tested numerically, are different (e.g., in \citealt{Matsumoto2018} it is about an order of magnitude larger than in Eq. \ref{eq:rh_asymptotic_full} for $\al=2$). They also derived a breakout criterion for the asymptotic phase that has the same functional form as our Eq. \ref{eq:Etot_crit_simple}. However, also here their normalizations are significantly different. Their breakout criterion are lower than ours by about 2 orders of magnitude, which is a result of their overestimate of the jet head location.
\cite{Hamidani2021} carried out a careful analytic study with a comparison to 2D numerical simulations. Their analytic expression to the jet head in expanding media (their Eq. 42) includes some dependence on $t_d$ although it is derived assuming that $t \gg t_d$. At late times the temporal dependence of their expression is $r_h \propto t^\frac{6-\al}{5-\al} \ln(\tti)^{-\frac{2}{5-\al}}$, with a very weak dependence on $t_d$. This is compared to our expression $r_h \propto t^\frac{6-\al}{5-\al}$. Comparing the predicted value of $r_h$ for typical jet and ejecta parameters we find that the difference between the predictions depends mostly on $\al$ and on whether the jet is strong or weak, where in some of the regimes their equation 42 predicts $r_h$ that agree with ours to within an order of magnitude while in others their $r_h$ is larger by more than an order of magnitude.
Finally, for $\al \geq 3$ their expression is undefined and when $\al$ approaches 3 from below, their $r_h$ diverges and so does the ratio between their solution and ours.
In Fig. \ref{fig:comparison} we demonstrate the ratio between our solution to the jet head evolution in unmagnetized strong and weak jet models with different values of $ \al \leq 2.5 $, and those of \citet[][]{Hamidani2021} (HI21) and \citet[][]{Matsumoto2018} (MK18). These are the only works that derived an analytic solution to $ r_h(t) $. It shows that at early times ($\tti \lesssim 1$) the solutions of HI21 and MK18 often deviate from our solution by $\sim 2$ orders of magnitude. This is expected since their models were designed to fit only the asymptotic phase. Indeed at later times, ($\tti \gg 1$), the difference between the models is lower. The solution of HI21 deviates from ours by $ \sim 0.5-1 $ order of magnitude during the asymptotic phase of most of the presented scenarios, while MK18 model yields a similar ratio in weak jets, but features a larger ratio in strong jets, with a jet head location that is between 1 to 2 orders of magnitude farther than ours.

The results of \citet{Duffell2018} deserve some additional discussion. They report that for successful jets they find two regimes of jet breakout from the ejecta: early breakout on timescales shorter than the engine duration, and late breakout well after the engine shuts off. This seems to be in disagreement with the result that any jet fluid element that crosses the reverse shock at the head and spills into the cocoon is mixed with material from medium and cannot be part of any ultra-relativistic outflow. This result, which was found in a large number of simulations by different groups and different codes \citep[e.g.,][]{Mizuta2009,Mizuta2013,Harrison2018,Gottlieb2018a}, implies that when the jet head is Newtonian it is choked quickly after the engine stops (more precisely a time $r_h/c$ after the engine stops). A careful look at the setup used by \citet{Duffell2018} finds that they performed simulations of jets that are launched with a constant luminosity for a characteristic  time, $T$, where at $t \gg T$ the jet has a long lasting tail with a luminosity that drops as $t^{-2}$. Namely, their engine does not really shut off at any time. Thus, if their simulations are consistent with those of other groups, then in their simulations in which $t_{bo} \gg T$, it is the tail of the ejecta that is launched at $t \gtrsim t_{bo}$ that breaks out of the ejecta. Here it is important to note that the density distribution considered by \citet{Duffell2018} has two characteristic regimes. Their ejecta is homologous and it has a leading edge which at the beginning of the simulation is located at $R_0$. The initial distribution at $r \ll R_0$ is approximately a power-law in $r$ with $\alpha=2.5$. Near the leading edge, the distribution drops sharply as a power-law of the distance from the edge (instead of a power-law in $r$), $\rho_{ej} \propto (1-r/R_0)^{3/4}$. Our results show that a jet can propagate in a medium with $\alpha < 5$ only if the head is supported by a continuous flow of jet material through the reverse shock with a luminosity that drops slower than $t^{-1}$ (see also\footnote{As was recently shown by \cite{Govreen-Segal2021} a spherical Newtonian shock in expanding media always dies out when $\al<8$, unless there is a continuous energy injection into the shock. Since jets expand sideways, as the energy injection stops they cannot be sustained without a continuous energy injection even at $\al$ values that are larger than 8.} \citealt{Govreen-Segal2021}). The combination of their density profile and jet luminosity evolution leads to a head that can propagate at $t \lesssim T$ through the inner parts of the ejecta, and then be supported by a jet tail where $L_j(j) \propto t^{-2}$ as it propagates near the edge of the ejecta, where the density gradient is very sharp. This implies that the 'late breakout' is actually a breakout of the tail of the jet and it is not a generic property of jets in expanding ejecta in the sense that it is a result that depends of the specific setup chosen by \citet{Duffell2018}. Therefore, the various relations that they find in their simulations that refer to the `late breakout' are relevant only to their setup. For example, if the setup is changed so the jet launching is truly terminated at $T$ (i.e., $L_j(t>T)=0$), then there is no late jet breakout.

This result raises a more general interesting question. Is it possible that many of the observed GRBs are generated by jets that break out while their luminosity decays strongly with time (faster than $t^{-1}$) after propagating in sharp density media ($\frac{d\ln\rho}{d\ln v} \ll -5$), for which the breakout time that we derive here is incorrect? Current observations suggest that typical sGRBs are inconsistent with this picture. There are hundreds of sGRBs for which we have detailed observations of the prompt gamma-ray emission. The temporal properties of this emission were studied by many authors and it is found to be typically variable, with a complex temporal structure which is composed of many pulses, some of them overlapping and some separated \citep[e.g.,][]{McBreen2001,Nakar2002,Bhat2012,Li2020}. Interestingly, these studies did not find an evolution of the pulses properties with time. Instead, the prompt emission seems to last for a duration, $T$, during which the pulses' properties such as luminosity and duration do not show any specific evolution. This continues until at some point the gamma-ray emission stops over a time-scale that is comparable to the duration of a single pulse, which is often much shorter than the duration of the burst. Since the prompt emission reflects the activity of the central engine after the breakout, this result suggests that the central engine does not change its properties considerably for the duration of the burst. The gamma-ray emission stops either because the engine shuts off abruptly or the efficient generation of gamma-rays stops for some unknown reason. Otherwise, we would have seen the marks of the evolution of the jet properties on the prompt emission. For example, if the jet luminosity would have been decaying monotonically and significantly over the duration of the burst, we would expect that on average the pulses seen at the beginning of the burst would be brighter than those seen at the end of the burst. When combined with the fact that the engine is causally disconnected from the head and is unaware of the jet breakout, the observational properties of the prompt emission support a picture where the jet luminosity is roughly constant (up to possibly a short duration variability) during the entire propagation in the ejecta, through the breakout and the production of the gamma-rays. Thus, these observations are in tension with the late breakout scenario suggested by \cite{Duffell2018}.

\cite{Duffell2018} also found a breakout criterion that they calibrate numerically. The functional form of their criterion for breakout of jets with $t_{bo} < T$ (their equations 20 and 21) differs from ours (and from that of \citealt{Margalit2018} and \citealt{Matsumoto2018}). Their criterion is equivalent to our Eq. \ref{eq:Etot_crit_simple} without the dependence on the r.h.s. on $\theta_{j,0}^2$. The reason that they did not find this difference in their numerical tests is likely a result of the narrow range of $\qj$ (factor of 2) with which they tested the criterion and the effect of the late breakout on the test results. When we set the jet opening angle in our breakout criterion to the range in which they used to calibrate the numerical coefficient $\kappa$ of their breakout criterion, $\qj=0.07-0.14$ rad, their results are similar to ours to within an order of magnitude. In fact, we find that when our Eq. \ref{eq:Etot_crit_simple} is applied to their simulations it predicts that the transition from successful to failed jets should take place right at the phase-space region that \cite{Duffell2018} found a transition from 'early' to 'late' breakouts. This is consistent with our understanding that late breakouts corresponds to jets that almost broke free at $t \approx T$ and that are sustained longer by the support of the jet tail and/or the propagation in the very steep density gradient near the leading edge of the ejecta.

Finally, \citet{Beniamini2020} studied the impact of the propagation time on observations of sGRBs. They use it to obtain constraints on $t_d$. They study two regimes, static and dynamic, corresponding to $t_{bo}<t_d$ and $t_{bo}>t_d$, respectively. For the static regime they applied a similar analysis to that of \citet{Murguia-Berthier2017,Gill2019} and for the dynamic regime they adopted the solution by \citet{Duffell2018}. In \S\ref{sec:BNS} we discuss their results in view of our findings.

\section{Application to BNS mergers and short GRBs}\label{sec:BNS}
The results of the jet propagation have some important implications for BNS mergers and short GRBs. For this discussion we summarize first the main propagation parameters normalized to those inferred by observations of GW170817 (see \citealt{Nakar2019} and references therein), $\Ejtot \approx 10^{50}$ erg, $\qj \approx 0.05$ rad, $\Eejtot \approx 10^{51}$ erg, $\vmin \approx 0.05$c and\footnote{Note that $ \vmax $ is the maximal velocity of the bulk of the ejecta. Namely, it is the velocity where most of the kinetic energy of the ejecta is. It is believed that in GW170817 there was ejecta with velocities higher than $\sim 0.3$ c, but it carried a negligible fraction of the energy and did not have a significant effect on the jet propagation. See discussion in \S\ref{sec:breakout_time}.} $\vmax \approx 0.3$c, so $w \approx 6$. While the duration of the engine, $t_e$, and thus the jet isotropic equivalent luminosity, $\Ljiso$, are not constrained, we take a canonical value of $\Ljiso=10^{52}~{\rm erg~s^{-1}}$. We use $ \alpha = 3 $ since the amount of mass in low ($ v \approx 0.1 c $) and high ($v \approx 0.3 c $) velocities in GW170817 was found to be comparable, and under the assumption of isotropic mass distribution, one obtains $\alpha=3$. There are also numerical studies that find that $\alpha \approx3-4$ provide a reasonable description of the ejecta density profile \citep[e.g.,][]{Nagakura2014}. When the parameters of unmagnetized and weakly magnetized jets are different we give both where the weakly magnetized values are in `[]'. 

The dimensionless parameters for BNS mergers are:
\small
\begin{equation}\label{eq:eta0_BNS}
    \eta_0 = 80 \left(\frac{\Eejtot}{10^{51}{\rm~erg}}\right)^{-\frac{1}{3}}
    \left(\frac{w}{6}\right)^{\frac{2}{3}}
    \left(\frac{\Ljiso}{10^{52}{\rm~erg~s^{-1}}}\right)^{\frac{1}{3}}
    \left(\frac{\qj}{0.05{\rm~rad}}\right)^{-\frac{2}{3}}
    \left(\frac{t_d}{1{\rm~s}}\right)^{\frac{1}{3}}~,
\end{equation}
\normalsize
\begin{equation}\label{eq:etacrit_BNS}
    \etacrit = 17 \left(\frac{\Eejtot}{10^{51}{\rm~erg}}\right)^{-\frac{1}{2}}
      \left(\frac{w}{6}\right)
    \left(\frac{\Ljiso}{10^{52}{\rm~erg~s^{-1}}}\right)^{\frac{1}{2}}
    \left(\frac{t_d}{1{\rm~s}}\right)^{\frac{1}{2}}~,
\end{equation}
\small
\begin{equation}\label{eq:etacol_BNS}
    \etac = 4[8] \left(\frac{\Eejtot}{10^{51}{\rm~erg}}\right)^{-\frac{1}{3}}
    \left(\frac{w}{6}\right)^{\frac{2}{3}}
    \left(\frac{\Ljiso}{10^{52}{\rm~erg~s^{-1}}}\right)^{\frac{1}{3}}
    \left(\frac{\qj}{0.05{\rm~rad}}\right)^{-\frac{2}{9}}
    \left(\frac{t_d}{1{\rm~s}}\right)^{\frac{1}{3}}~,
\end{equation}
\normalsize
\begin{equation}\label{eq:Etilde_BNS}
    \E = 8[60] \left(\frac{\Eejtot}{10^{51}{\rm~erg}}\right)^{-1}
    \left(\frac{\Ejiso}{10^{52}{\rm~erg}}\right)  
    \left(\frac{\qj}{0.05{\rm~rad}}\right)^{-2}~,
\end{equation}
and
\begin{equation}\label{eq:Ed_BNS}
    \E_d = 8[60] \left(\frac{\Eejtot}{10^{51}{\rm~erg}}\right)^{-1}
    \left(\frac{\Ljiso}{10^{52}{\rm~erg~s^{-1}}}\right)  
    \left(\frac{\qj}{0.05{\rm~rad}}\right)^{-2}
     \left(\frac{t_d}{1{\rm~s}}\right)~.
\end{equation}
The corresponding breakout time in the Newtonian regime is:
\small
\begin{equation}\label{eq:tbo_BNS}
t_{bo} \approx
\begin{cases}
 0.4[0.2]~{\rm s}~\left(\frac{\Eejtot}{10^{51}{\rm erg}}\right)^{\frac{1}{3}}
    \left(\frac{\Ljiso}{10^{52}{\rm erg~s^{-1}}}\right)^{-\frac{1}{3}}  
    \left(\frac{\qj}{0.05{\rm rad}}\right)^{\frac{2}{3}}
     \left(\frac{t_d}{1{\rm s}}\right)^{\frac{2}{3}} & {\scriptstyle \E_d > 1~(t_{bo}<t_d)}\\
&\\
0.1[0.02]~{\rm s}~\left(\frac{\Eejtot}{10^{51}{\rm~erg}}\right)
    \left(\frac{\Ljiso}{10^{52}{\rm~erg~s^{-1}}}\right)^{-1} 
    \left(\frac{\qj}{0.05{\rm~rad}}\right)^{2}~& {\scriptstyle \E_d < 1 ~(t_{bo}>t_d)}
\end{cases} ~.
\end{equation}
\normalsize
The minimal energy work-time needed for breakout in the Newtonian regime is $t_{e,bo} \approx t_{bo}$. 

The transition to the relativistic regime can be estimated from the condition for a relativistic breakout (Eq. \ref{eq:rel_bo}):
\begin{equation}
    t_{d,{\rm rel}} = 0.1~{\rm s}~ \left(\frac{\Eejtot}{10^{51}{\rm~erg}}\right)
    \left(\frac{\Ljiso}{10^{52}{\rm~erg~s^{-1}}}\right)^{-1}
    \left(\frac{\qj}{0.05{\rm~rad}}\right)^{2}
    \left(\frac{\beta_{\max}}{0.3}\right)^3~.
\end{equation}
For $t_d \gtrsim t_{d,{\rm rel}}$ the head becomes relativistic before it breaks out of the ejecta. In that case, as long as the jet is collimated, the minimal engine work-time required for a successful breakout is (Eq. \ref{eq:rel_bo_criterion_col}):
\small
\begin{equation}\label{eq:rel_tbo_BNS}
    t_{e,bo} \approx 0.1~{\rm s}~  
    \left(\frac{\Eejtot}{10^{51}{\rm~erg}}\right)^{\frac{1}{5}}
    \left(\frac{\Ljiso}{10^{52}{\rm~erg~s^{-1}}}\right)^{-\frac{1}{5}}
    \left(\frac{\qj}{0.05{\rm~rad}}\right)^{\frac{2}{5}}
    \left(\frac{\beta_{\max}}{0.3}\right)^{\frac{2}{5}}
    \left(\frac{t_d}{1{\rm~s}}\right)^{\frac{4}{5}}~.
\end{equation}
\normalsize
The transition to an uncollimated jet takes place at (see Eq. \ref{eq:rel_collimation})
\small
\begin{equation}
    t_{d,{\rm uncol}} = 50~{\rm s}~ \left(\frac{\Eejtot}{10^{51}{\rm~erg}}\right)
    \left(\frac{\Ljiso}{10^{52}{\rm~erg~s^{-1}}}\right)^{-1}
    \left(\frac{\qj}{0.05{\rm~rad}}\right)^{-\frac{4}{3}}
    \left(\frac{\beta_{\max}}{0.3}\right)^{-3}~,
\end{equation}
\normalsize
where for $t_d>t_{d,{\rm uncol}}$ the jet is uncollimated (and surely the head is ultra-relativistic). The minimal engine work-time required for a successful breakout in the uncollimated regime is (Eq. \ref{eq:rel_bo_criterion_uncol}):
\begin{equation}\label{eq:rel_tbo_BNS_uncol}
    t_{e,bo} \approx 2~{\rm s}~  
    \left(\frac{\Eejtot}{10^{51}{\rm~erg}}\right)^{\frac{1}{2}}
    \left(\frac{\Ljiso}{10^{52}{\rm~erg~s^{-1}}}\right)^{-\frac{1}{2}}
    \left(\frac{\beta_{\max}}{0.3}\right)^{-\frac{1}{2}}
    \left(\frac{t_d}{50{\rm~s}}\right)^{\frac{1}{2}}~.
\end{equation}
Note that there  is a slight disagreement  by a factor of a few in the predicted value of $t_{e,bo}$ by Eqs. \ref{eq:tbo_BNS} and \ref{eq:rel_tbo_BNS} in the mildly relativistic region where $t_d \sim t_{d,{\rm rel}}$. This is not surprising since both equations are inaccurate in this regime. The reason for the difference is that the numerical correction factor $N_E$ is different between the Newtonian and relativistic regimes. For the relativistic regime we use the equations from \citet{Harrison2018} with no numerical correction for both unmagnetized and magnetized jets (see \citealt{Harrison2018} for a discussion of numerical corrections in the relativistic regime).

	\begin{figure}
		\centering
		\includegraphics[scale=0.28]{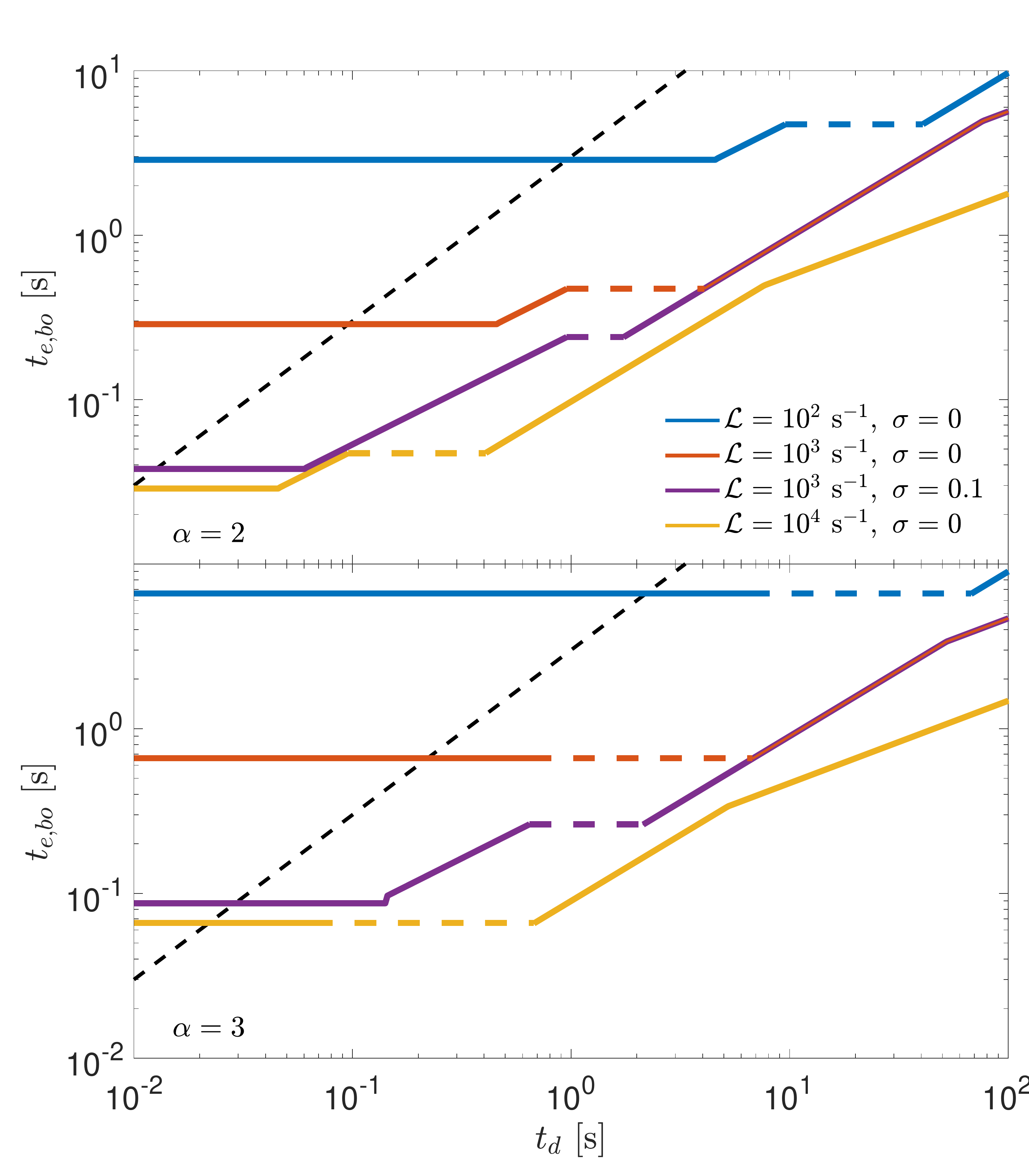}
		\caption[]{Required jet engine times for breakout as a function of $ t_d $, for $ \alpha =2 $ (top panel) and $ \alpha = 3 $ (bottom panel). Shown are different values of $ {\cal L} \equiv \Ljiso\qj^{-2}/\Eejtot$ and magnetizations, while keeping $ \bmin = 0.05 $ and $ \bmax = 0.3 $ fixed.
		The dashed black lines reflect the minimal engine time for breakout in the asymptotic phase, assuming it takes place at $ \ta = 4 $.
		Five regimes are present (from small to large $ t_d $):
		i) at small $ t_d $, $ \tilde{E}_d < \tilde{E_a} $; $ t_e $ is independent of $ t_d $.
		ii) $ \tilde{E}_d > \tilde{E_a} $; $ t_e \propto t_d^{2/3} $.
		iii) transition to relativistic velocities (dashed lines).
		iv) the ejecta expanded sufficiently to allow the jet reaching a relativistic breakout; $ t_e \propto t_d^{4/5} $.
		v) ejecta is too dilute to collimate the jet, and a relativistic uncollimated jet breaks out; $ t_e \propto t_d^{1/2} $.
		}
		\label{fig:t_bo}
	\end{figure}

\subsection{GW170817}
In GW170817 there was a delay of 1.7 s between the merger and the $ \gamma $-ray flash. The afterglow observations show that a jet with the properties given above broke out of the ejecta successfully and the $\gamma$-rays where presumably generated by the breakout of the cocoon driven by this jet from the ejecta or by the jet itself. In either case the observed delay of the gamma-rays sets an upper limit on the delay in the jet launching and the time it takes the jet to break out of the ejecta. If the jet is Newtonian, then the observed delay implies simply $t_d+t_{bo}<1.7$ s. If the head is relativistic, then the limit is $t_d+t_{bo}/2\Gamma_{bo}^2<1.7$ s. Since the minimal required engine work-time for breakout is $t_{e,bo} \approx t_{bo}$ in the Newtonian regime and  $t_{e,bo} \approx t_{bo}/2\Gamma_{bo}^2$ in the relativistic regime, we conclude that the observations of GW170817 imply that $t_d+t_{e,bo} <1.7$ s.
Eqs. \ref{eq:eta0_BNS}-\ref{eq:rel_tbo_BNS_uncol} show that the jet properties, which are inferred from the afterglow emission ($\Ejtot$, $\qj$), and the ejecta properties which are measured from the kilonova ($ \Eejtot $, $\vmin$, $\vmax$), are consistent with this limit. 

However, an interesting question is whether we can use the observations to obtain a better constraint on $t_d$ than the obvious one ($t_d<1.7$ s). Fig. \ref{fig:t_bo} depicts the minimal engine work-time needed for a successful breakout of the jet from the ejecta, $t_{e,bo}$, as a function of the parameter $ {\cal L} \equiv \Ljiso\qj^{-2}/\Eejtot$ and the delay time $t_d$. 
For GW170817 ${\cal L} \approx 10^3-10^4~{\rm s^{-1}}$, and Fig. \ref{fig:t_bo} shows that for every value of $t_d \lesssim 1$ s the minimal engine work-time is in the range of $t_{e,bo} \sim 0.1-1$ s. This implies that the observed delay is consistent with being dominated by any of the various processes involved. Namely, it can be dominated either by $t_d$ (e.g., $t_d \approx 1.5$ s and $t_{e,bo} \ll 1$ s), or by $t_{e,bo}$ (e.g., $t_d \ll 1$ s and $t_{e,bo} \approx 1.5$ s), or by other factors such as geometrical effects, so $t_d,t_{e,bo} \ll 1$ s, or by any combination of these processes. Therefore, every value $t_d<1.7$ is consistent with the observed delay between the GWs and $ \gamma $-rays and we cannot constrain $t_d$ from the observation of GW170817, apart for the obvious upper limit.

In the shock breakout model for the $ \gamma $-ray emission from GW170817, the breakout emission is released when the cocoon driven by the jet breaks out of a low mass fast tail that runs ahead of the core of the ejecta \citep{Kasliwal2017,Gottlieb2018b,Beloborodov2020}. The core of the ejecta is relatively massive ($\sim 0.05{\rm~M_\odot}$)  and it has $\alpha<5$.  In contrast, the fast tail density is expected to be much lower and most likely it has $\alpha$ much larger than 5, so if the jet breaks out of the core of the ejecta then it almost certain to break out from the fast tail as well. However, the breakout from the tail takes place at a much larger radius and at a much later time than the breakout from the core. 
In the breakout model the observed $ \gamma $-rays determine the radius and Lorentz factors of the ejecta and of the shock at the time of the breakout from the fast tail. This, in turn, determines the time at which the jet should break out of the core of the ejecta so the shock driven by the cocoon can overtake the fast tail at the correct radius \citep{Nakar2019}. In GW170817 the breakout from the core of the ejecta should be $0.1-1$ s after the merger for the $ \gamma $-rays to be consistent with the shock breakout emission. This range is consistent with the predicted breakout time derived above based on the jet and ejecta properties.

\subsection{Limits on the jet and ejecta parameters from sGRB observations}
Fig. \ref{fig:t_bo} has also interesting implications to sGRBs. The engine work-time of sGRBs cannot be much longer than the observed duration of the prompt emission \citep{Bromberg2013}, implying that $t_{e,bo} \lesssim 1$ s in sGRBs. Additionally, there is also an observational evidence that typically $t_{e,bo} \approx 0.1-1$ s \citep{Moharana2017}. Therefore, Fig. \ref{fig:t_bo} shows that a successful sGRB requires ${\cal L} \gtrsim 10^3~[10^2]~{\rm s^{-1}}$ for unmagnetized [magnetized] jets. The jet of GW170817 is strong and narrow enough to easily satisfy this criterion with ejecta mass of $\sim 0.05~{\rm M_\odot}$. However, typical sGRB jets that point towards Earth are much weaker and most likely also wider. The energy and opening angle of typical sGRB jets can be estimated based on the properties of their prompt $ \gamma $-rays. The prompt emission is efficient and the total energy carried by the jet is typically thought to be comparable to the one seen in $\gamma$-rays, or at most larger by a factor of a few  \citep[][ and references therein]{Nakar2007}.
Studies of the sGRB luminosity function show that the most common sGRBs, by volumetric rate, are much fainter than the jet seen in GW170817. In fact, the volumetric rate\footnote{Note that the luminosity function constrains the rate of sGRBs whose jets are pointing towards Earth. Thus, if the typical opening angle of more luminous jets is narrower than the opening angle of less luminous jets then the total rate of luminous and faint jets, including the beaming correction, may be comparable. The fact that the first BNS merger that was detected by GWs had a jet with unusually high luminosity compared to typical sGRBs suggests that this is the case. See \citep{Nakar2019} for discussion.} of sGRBs with a peak isotropic equivalent $ \gamma $-ray luminosity $L_{\gamma,{\rm iso}}^{\rm peak} \sim 10^{50}~{\rm erg~s^{-1}}$ is roughly $\sim 5 {\rm~Gpc^{-3}yr^{-1}}$ \citep[][ and references therein]{Wanderman2015}. This rate is at least 100 times more frequent than that of sGRBs with $L_{\gamma,{\rm iso}}^{\rm peak} \sim 10^{52}~{\rm erg~s^{-1}}$. Given that the jet luminosity is expected to be larger by a factor of a few than the $ \gamma $-ray luminosity, and that the average luminosity is a factor of a few lower than the peak luminosity we can approximate $ \Ljiso \sim L_{\gamma,{\rm iso}}^{\rm peak}$.
Since the sGRB luminosity function counts only jets that point towards us, rate considerations show that the opening angle of these jets cannot be very small. Given that the total rate of BNS mergers, as found by their gravitational waves, is $\sim 1000 {\rm~Gpc^{-3}yr^{-1}}$ \citep{Abbott2019}, jets with $ \Ljiso \sim 10^{50}~{\rm erg~s^{-1}}$ should have typically $\qj \gtrsim 0.1$ rad. Combining these results with the lower limit on ${\cal L}$ of sGRBs implies that the energy in the ejecta of sGRBs with $L_{\gamma,{\rm iso}} \sim 10^{50}~{\rm erg~s^{-1}}$ cannot exceed $10^{49}~[10^{50}]$ erg for unmagnetized [magnetized] jets. Since the jet is sensitive only to the ejecta in its opening angle, it sets limits on the isotropic equivalent energy of the ejecta along the poles. Since the velocities of the ejecta in various mergers are expected to be rather similar, this result indicates that the isotropically equivalent mass ejected along the poles in those events is $\lesssim 10^{-3} M_\odot$, significantly lower than the ejecta mass of GW170817, unless the ejecta along the pole in GW170817 was significantly lower than the ejecta closer to the equator \citep[see also discussion in][]{Gottlieb2021a}.

Fig. \ref{fig:t_bo} also enables us to learn during which phase the breakout in sGRBs typically takes place. We have seen that strong jets approach the asymptotic phase at $t \gtrsim 4t_d$. Weak jets break out at $t \gg t_d$ and always after reaching the asymptotic phase. Therefore every jet with $t_e>t_{e,bo} \gtrsim 3t_d$ breaks out successfully after approaching the asymptotic phase, if its head is Newtonian at the breakout (note that a Newtonian breakout takes place at $t_d+t_{e,bo}$). The dashed black line in Fig. \ref{fig:t_bo} marks $t_{e,bo} \gtrsim 3t_d$. Thus, only events with $t_d$ and $t_{e,bo}$ that puts it to the left and above this line break out during the asymptotic phase. In GW070817 this certainly requires $t_d \lesssim 0.1$ s and possibly even $t_d < 20$ ms if the jet is weakly magnetized or if its ${\cal L} \approx 10^4 {\rm~s^{-1}}$. For typical sGRBs with lower luminosities, the breakout upper limit of $ \sim 1 $ s dictates that the breakout can take place in the asymptotic regime only if $t_d \lesssim 0.2$ s. We therefore conclude that it is possible that many BNS mergers launch jets that break out or choke before reaching the asymptotic phase. 

Another interesting question is whether the breakout takes place before or after the jet is fully collimated. In \S\ref{sec:analytic} we showed that if the cavity is empty, then a full collimation is obtained roughly when $\frac{\vej}{\vmin} \approx 5 \left(\frac{\qj}{0.1{\rm~rad}}\right)^{-2/3}$. If the cavity is not empty, the collimation is obtained earlier. The expectation, which is also supported by the observation of GW170817, is that in BNS mergers $\vmax/\vmin \approx 6$. This implies that the jet is likely to spend a significant part of its propagation before being fully collimated, but it approaches full collimation near the time of the breakout. As we showed in \S\ref{sec:cavity}, this result implies that the initial conditions inside the cavity should have at most a minor effect on the breakout time (and thus on the breakout criterion) in BNS mergers. 

Finally, Fig. \ref{fig:t_bo} can also be used to set an upper limit on $t_d$ in sGRBs. It shows that a successful breakout with $t_{e,bo}<1$ s requires at least ${\cal L} \approx 10^3~[10^2]~{\rm s^{-1}}$ for unmagnetized [magnetized] jet, and that for such values of ${\cal L}$ the delay time has to satisfy $t_d \lesssim 5$ s (regardless of the jet magnetization). If ${\cal L}$ is larger by an order of magnitude, then $t_{e,bo} < 1$ s as long as $t_d \lesssim 50$ s. Assuming that the ejected mass from typical BNS mergers that produce sGRBs does not fall below $\sim 10^{-3}~\msun$, we conclude that the observations of sGRBs imply $t_d \lesssim 10$ s.
Note that recently \citet{Beniamini2020} used similar considerations to conclude that the observations require $t_d \lesssim 0.1$s. The two main reasons for the difference between our conclusions and those of \citet{Beniamini2020}, are: i) their study depends in part on \citet{Duffell2018} (which as we show in \S\ref{sec:previous}, cannot be used for an accurate estimate of $t_{e,bo}$); ii) they ignored the relativistic regime, which as we show here is the relevant one at large delay times.

\section{Summary}
\label{sec:summary}

Jet-medium interaction plays a key role in the evolution of jets. There are astrophysical systems, such as BNS mergers and possibly some types of superluminous SNe, where a relativistic jet encounters an expanding medium. The jet propagation is then different than in the simple case of a static medium. Recently, primarily motivated by GW detection of BNS mergers, several authors have addressed the jet evolution in expanding medium. However, these studies explored a limited set of configurations and a large area of the parameter phase space, some of which is directly relevant to BNS mergers, remained unexplored.
In this work we present an analytic model of the jet propagation in homologous expanding media, which covers a wide range of jet and ejecta parameters, including the first calculations of propagation in expanding medium of low-luminosity jets, mildly magnetized jets, jets with a time-dependent luminosity and jets with a relativistic head.

Our solution makes several assumptions. First, we assume a Newtonian expanding homologous ejecta with a power-law density profile with index $\al<5$, so most of the energy is carried by the fast material. The assumption of homologous flow is certainly valid if all the ejecta is launched before the jet injection starts. We show that it is valid also in cases where ejecta is continued to be launched simultaneously with the jet, as long as most of the ejecta energy is carried by material that was launched before the jet injection starts. The ejecta does not have to be spherically symmetric, but the solution assumes that it can be approximated as such over the region where the jet is passing. Namely, that the ejecta density and velocity distributions do not vary significantly inside an angle significantly larger than $ \qj$ around the jet symmetry axis. We further assume that the jet is launched with an opening angle that does not vary with time and that the energy launched into the jet grows with time (i.e., if $L_j$ drops with time, it is slower than $t^{-1}$). In order to derive this solution we also expand the solution of jet propagation in a static medium (see Appendix \ref{sec:static}), which was until now limited to a density profile with $\al<3$, to include steeper density profiles of $3 \leq \al<5$. We calibrate our model using a series of 3D RMHD simulations.

We find that the jet evolution is mainly dictated by a single dimensionless parameter (that may evolve with time), $ \eta =(v_h-\vej)/\vej$. Namely, the ratio between the jet head velocity, as measured in the local ejecta frame, and the ejecta velocity at the location of the jet head. Different values of $\eta$ define three main regimes of the evolution:\\
1. $ \eta \ll 1 $: The head of the jet is roughly at rest in the local reference frame of the ejecta. The velocity of the head in this regime can be approximated as a constant. The jet cannot remain in this regime indefinitely and the value of $\eta$ must eventually grow with time.\\
2. $ \eta \gg 1 $: The head velocity is much larger than the ejecta velocity at the head location. The ejecta velocity can be neglected in the calculation of the head velocity, which can be approximated very well by the theory of jet propagation in static media \citep{Bromberg2011}. The value of $\eta$ in this regime is dropping continuously.\\
3. $ \eta = \eta_a $ (the asymptotic regime): Regardless of the initial conditions, the value of $\eta$ always converges after enough time to a constant value\footnote{If the luminosity varies with time as $L \propto t^k$ where $k>-1$ then $\eta_a=(1+k)/(5-\al)$.} $\eta_a=1/(5-\al)$.  

This three regimes define two different types of jets, strong and weak. The 
type of the jet is determined by the system initial conditions, and it depends on another parameter that we define, $\etacrit$:\\
(i) Strong jet ($\etacrit>\eta_a$): The jet is strong enough to start its propagation in the ejecta, either before or after being collimated by the pressure that builds up in the cavity, within a timescale that is shorter than $t_d$. Note that the delay between the merger and the jet launching, $t_d$, is also the dynamical time of the ejecta when the jet launching starts. Such jets propagate at first with $\eta \gtrsim 1$ and within several dynamical times of the ejecta ($t \sim 4 t_d$) $\eta$ approaches its asymptotic value, regardless of how large $\eta$ is initially.\\
(ii) Weak jet ($\etacrit<\eta_a$): The jet is too weak to breach the ejecta and it is stalled at first at the base of the ejecta. Only after the ejecta expands significantly and its density becomes low enough, the jet can break through and propagate inside it. In these jets, initially $\eta \ll 1$, and it grows over a timescale that is much longer than $t_d$ until approaching $\eta_a$. As a result, the breakout time is independent of $t_d$, as the effective delay time is the time at which the jet starts propagating in the ejecta. 

We provide simple analytic approximations to the time-dependent location of a subrelativistic jet head in strong (Eq. \ref{eq:rh_strong}) and weak (Eq. \ref{eq:rh_weak}) jets. We derive a simple breakout criterion that can be applied to any jet with known $t_d$ and $t_e$, as well as the jet and ejecta total energies and the jet opening angle (Eq. \ref{eq:Etot_crit_all}).  The same criterion can be applied to a wide range of density profiles and to jets with time evolving luminosity (as long as  $E_{j,tot}\approx L_j(t=t_d+t_e)\cdot t_e$), implying that this criterion is largely independent of the exact density profile and the luminosity temporal evolution.
We also derive a criterion, Eq. \ref{eq:Etot_crit_simple}, that is necessary but may be insufficient for jet breakout, where $E_j$ is known, but neither $t_e$, $L_j$ and/or $t_d$ are unknown (e.g., such as in GW170817 where the afterglow reveals only $E_j$). If we know that $t_e \gtrsim t_d$ in a strong jet or that $t_e \gtrsim t_a$ in a weak jet , then satisfying Eq. \ref{eq:Etot_crit_simple} is sufficient for a successful breakout. All the analytic formulae that we provide include calibration coefficients that were obtained via numerical simulations. These coefficients, which depend on $\al$, are given in Table \ref{tab:coefficients}.


Our approximation for the head location includes the early collimation phase of the jet. This phase depends on the density between the jet launching site and the slow end of the bulk of the ejecta (moving at $\vmin$). In our analytic expressions we assume that there is an empty cavity at $r<\rmin$. If this is not the case, the evolution at early times can deviate from our solution for some time, but we show that after the jet is collimated, the head location becomes independent of the initial density in the cavity.

We explore the effect of subdominant magnetic field and find that if the jet maintains a toroidal magnetic field with a magnetization of $ \sigma \sim 0.01-0.1 $, it propagates 2-3 times faster than an unmagnetized jet. That is equivalent to an increase of about an order of magnitude in $ {\cal L}  = \Ljiso\theta_{j,0}^{-2}/\Eejtot $. Furthermore, such a magnetic field considerably shortens the breakout time, by an order of magnitude if the breakout time is $ t_{bo} > t_d $, and by a factor of two if $ t_{bo} < t_d $.
Lower magnetizations of $ 10^{-3} \lesssim \sigma \lesssim 10^{-2} $ are anticipated to be equivalent to a smaller correction in $ {\cal L} $.
As the jet is likely to have some degree of magnetization, at least before the collimation shock, this result implies that magnetic fields are likely to play a central role in the breakout of jets from dense media.

Applying our model to GW170817 we find that the breakout time of the jet from the ejecta was most likely $0.1-1$ s, with a very weak dependence on $t_d$. Thus, any $ t_d < 1.7 $ s is consistent with the observed delayed $ \gamma $-ray emission. When considering the sample of observed sGRBs, we find that in order for an sGRB jet to break out within an engine work-time of $ t_e \lesssim 1 $ s (as required by the observations), it must satisfy two conditions. First, the system must maintain $ {\cal L}\gtrsim 10^3~[10^2]~{\rm s^{-1}} $ for unmagnetized [magnetized] jets. We show that this implies that for the most common sGRBs with $\gamma$-ray luminosity of $10^{50}~{\rm erg~s^{-1}}$, the ejecta mass must be lower than $\sim 10^{-3} {\rm~M_\odot}$. If the ejecta is highly anisotropic then this limit is applicable for the isotropic equivalent ejecta mass along the jet axis. Second, unless $ {\cal L} \gg 10^3~{\rm s^{-1}}$, the jet must be launched after the merger within $t_d \lesssim 10$ s.
 
	\section*{Acknowledgements}
	We thank Tsvi Piran, Hamid Hamidani, Kunihito Ioka, Paz Beniamini and Ariadna Murguia-Berthier for useful comments.
	This research was partially supported  by a consolidator ERC grant (JetNS) and an ISF grant 1995/21.
	
	\section*{Data Availability}
	
	The data underlying this article will be shared on reasonable request to the corresponding author.

\bibliographystyle{mnras}
\bibliography{prop}

\appendix

\section{Jet propagation in static media with $3 \leq \al<5$}\label{sec:static}

\cite{Bromberg2011} limited their solution of jet propagation in static medium to $\al <3$. As we show here the solution is actually valid (with minor revisions) in the range  $3 \leq \al<5$ as well. We then complete the numerical calibration of the jet propagation in a static medium  \citep{Harrison2018} for $ \alpha = 3 $ and $ \alpha = 4 $.

The limit $\al <3$ that \cite{Bromberg2011} set on the solution originates in their calculation of the average cocoon density under the approximation of the cocoon shape as a cylinder. Then, if $\al \geq 3$ the mass in the cocoon diverges and there is no solution. This approximation, however, is irrelevant for large values of $\al$. The reason is that the shape of the cocoon depends on $\al$, where for $\al < 2$ the cocoon is wide at the base and narrow at the front (the shape of an arrowhead) and for $\al>2$ it is wide at the front and narrow at the base (an inverted arrowhead, see figure 3 of \citealt{Harrison2018}). Only for $\al=2$ the cocoon looks like a barrel and can be approximated by a cylinder. Physically, it is clear that the pressure in the cocoon, which determines the propagation of the head, should not be affected by the density distribution at radii that are much smaller than the smallest structure of the system, which is the collimation shock. Therefore, there is no reason why the solution should not be applicable for density profiles where the mass diverges at small radii. In what follows we derive the propagation equations in such density profiles.

	\begin{figure}
		\centering
		\includegraphics[scale=0.22]{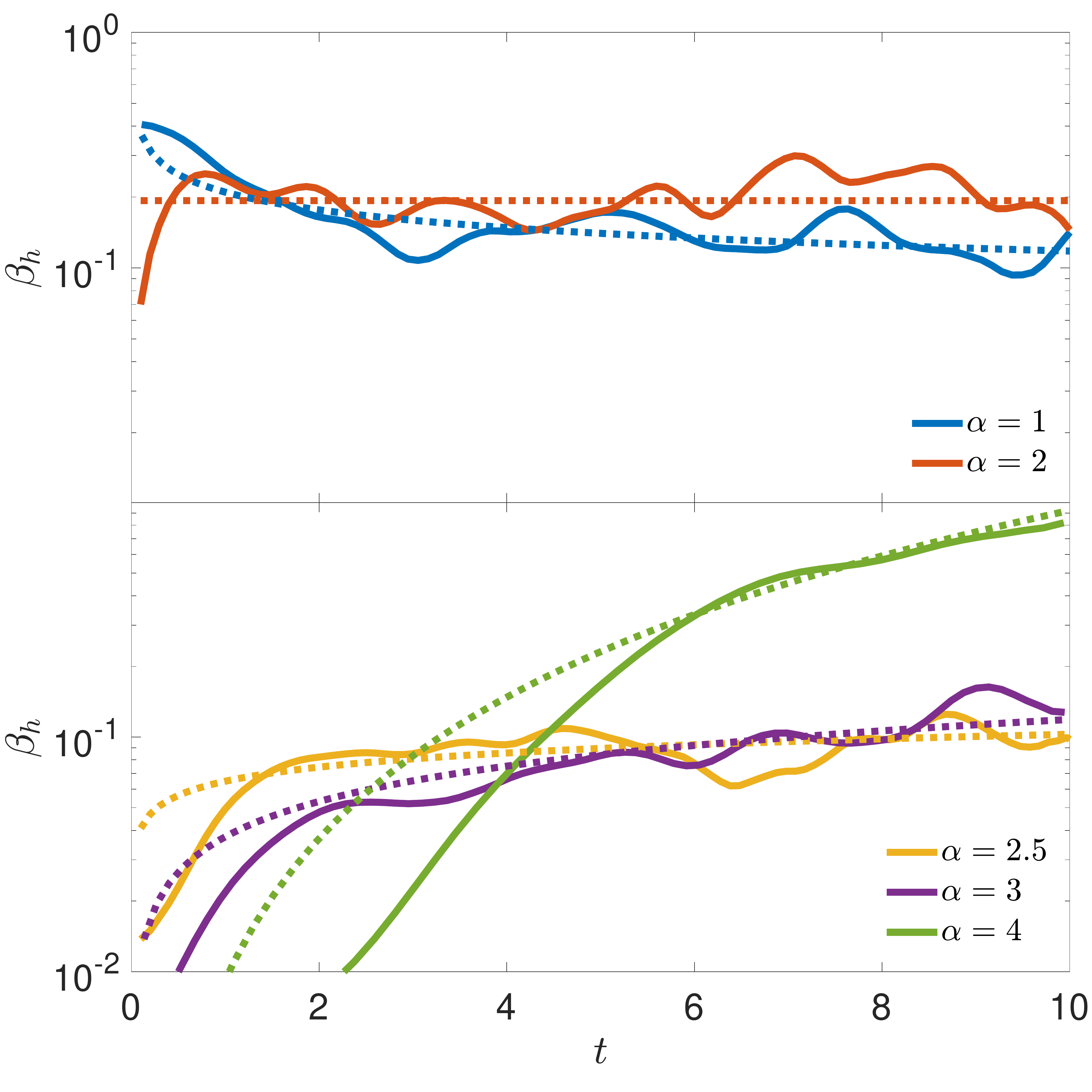}
		\caption[]{
		Calibration of the analytic expression of the jet head dimensionless velocity in a static medium with $ \alpha = 1 $ and $ \alpha = 2 $ (top), and $ \alpha = 2.5 $, $ \alpha = 3 $ and $ \alpha = 4 $ (bottom). Solid lines mark numerical results and dotted lines are the analytic curves with best fit $ \tilde{N}_s $ from Table \ref{tab:Ns}. Time is given in arbitrary units.}
		\label{fig:static}
	\end{figure}

We start from \citet{Harrison2018}, who derived the jet head dimensionless velocity in the Newtonian regime (their equation A3):
\begin{equation}\label{beta_h_static}
    \beta_h = \frac{N_s^{\frac{5}{5-\alpha}}}{c}\left(\frac{\Ljiso}{A_\rho\theta_{j,0}^2}\right)^\frac{1}{5-\alpha}\left(\frac{3^{3-\alpha}}{3-\alpha}\right)^\frac{1}{5-\alpha}\left(\frac{4(5-\alpha)^{\alpha-3}}{\pi}\right)^\frac{1}{5-\alpha}t^\frac{\alpha-2}{5-\alpha}~,
\end{equation}
where $ N_s $ is a numerical coefficient that was calibrated by their numerical simulations (Note that in \citealt{Harrison2018} the notation $L_j$ refers to a one-sided luminosity rather than the two-sided luminosity used here).
The factor $ \left(\frac{3^{3-\alpha}}{3-\alpha}\right)^\frac{1}{5-\alpha} $ appears in this equation due to the assumption of \cite{Bromberg2011} on the cocoon shape described above and it reflects the divergence of the mass at small radii for $\al >3$. Thus, during our derivation of Eq. \ref{eq:velocity_difference}, we omitted this factor and introduced the notation $ \tilde{N}_s $ instead of $ N_s = 0.35 $, to account for the missing factor:
\begin{equation}\label{beta_h_static}
    \beta_h = \frac{\tilde{N}_s^{\frac{5}{5-\alpha}}(5-\alpha)^{\frac{\alpha-3}{5-\alpha}}}{c}\left(\frac{\Ljiso}{A_\rho\theta_{j,0}^2}\right)^\frac{1}{5-\alpha}t^\frac{\alpha-2}{5-\alpha}~.
\end{equation}
This notation enables us to extend the solution to $\al<5$.

\begin{table}
	\setlength{\tabcolsep}{10.5pt}
	\centering
	\begin{tabular}{c c c c c c}
		
		 & $\alpha=1$ & $\alpha=2$ & $\alpha = 2.5 $ & $\alpha=3$ & $\alpha=4$ \\
		\hline
        $ \tilde{N}_s $ & 0.47 & 0.38 & 0.25 & 0.19 & 0.16\\
        
	\end{tabular}
	\hfill\break
	
	\caption{The numerical coefficient $ \tilde{N}_s $ for jet propagation in a static medium with power-law index $ \alpha $.
		}
\label{tab:Ns}
\end{table}

We carry out simulations of relativistic jets in static media with $ \al=1,2,2.5,3,4 $ to calibrate $ \tilde{N}_s $ (Table \ref{tab:Ns}). We use $ \alpha = 1,2, 2.5 $ to verify consistency with the results of \citet{Harrison2018}, and $ \alpha = 3,4 $ to extend the calibration to $ \tilde{N}_s $ of $ \al\geq 3 $.
The four simulations with integer values of $ \alpha $ are identical to $ \Oa $, $ \Ca $, $ \Ta $ and $ \Fa $ (and the one with $ \alpha = 2.5 $ has $ \etacrit = 140, \eta_0 = 200 $), but for the static case, i.e., $ v_{ej} = 0 $, $ t_d = 0 $ and $ \rmin = z_{\rm beg} $.
Fig. \ref{fig:static} depicts the comparison of the analytic (dotted lines) and the numerical (solid lines) velocities.
When the power-law index is $ \alpha = 1~(2) $, the best fit is obtained for $ \tilde{N}_s = 0.47~(0.38) $, indicating that $ N_s =0.33~(0.29) $ which is rather consistent with $ N_s = 0.35 $ of \citet{Harrison2018}. Similarly, for $ \alpha = 2.5 $ we find $ \tilde{N}_s = 0.25 $ which corresponds to $ N_s = 0.19 $. This implies that $ N_s $ (and $ \tilde{N}_s $) is not a constant when $\al$ is approaching 3, but decreasing monotonically with $ \alpha $.
When $ \alpha = 3 $, Eq. \ref{beta_h_static} shows that the jet head velocity accelerates as $ t^{0.5} $, in agreement with the numerical simulation. The best fit coefficient is found to be $ \tilde{N}_s = 0.19 $.
When $ \alpha = 4 $, the head accelerates fast as $ t^2 $ to relativistic velocities, and thus we can fit it over a limited range. Nevertheless, for $ \tilde{N}_s = 0.16 $ the analytic and the numerical curves coincide between $ t = 6 $ and $ t = 10 $ ($ t $ is shown in arbitrary units), including when the jet head becomes relativistic.

\section{Convergence tests}\label{sec:convergence}
    
    \begin{figure*}
		\centering
		\includegraphics[scale=0.35]{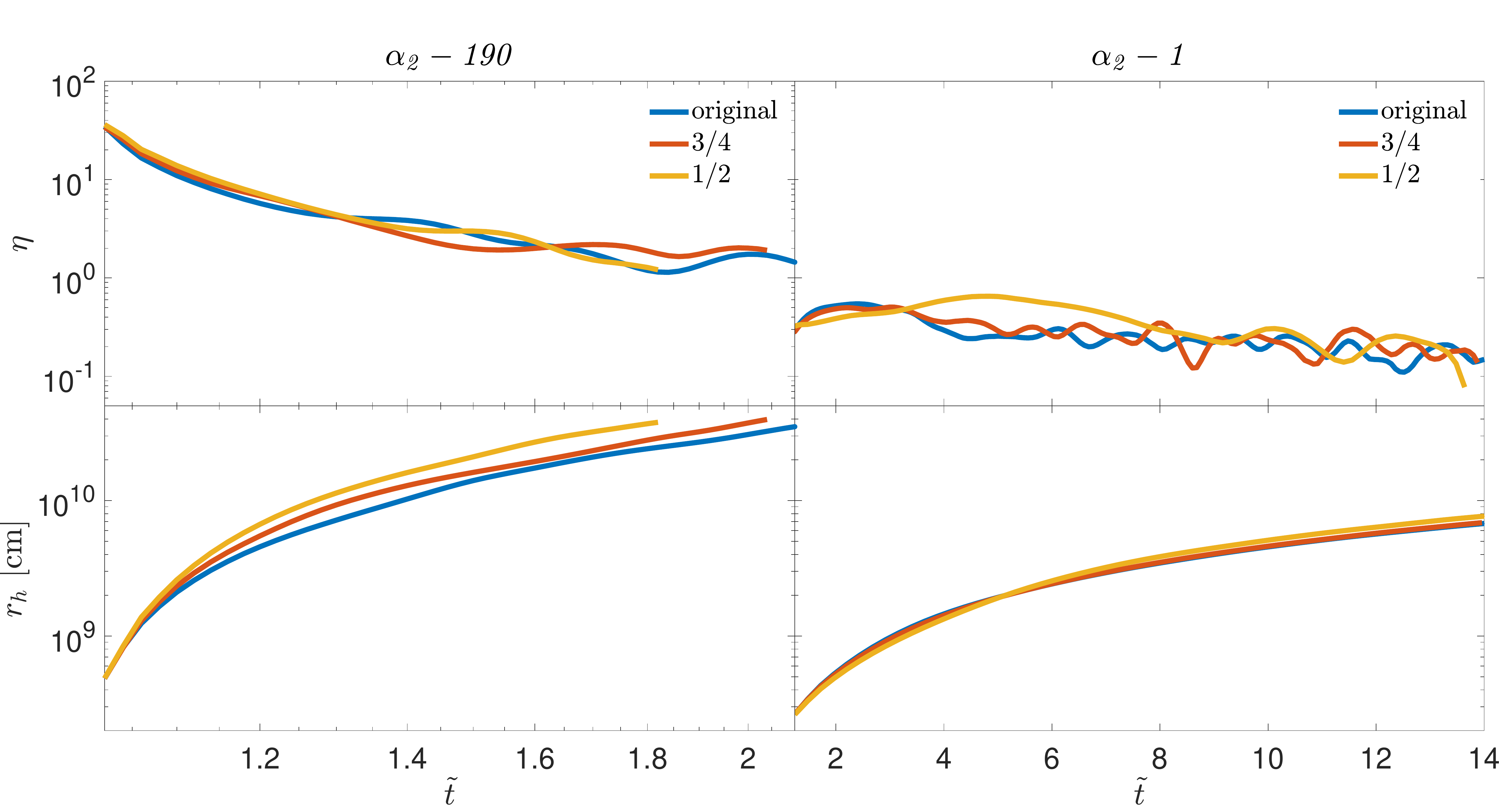}
		\caption[]{
		Convergence tests for high $ \etacrit $ model $ \Ca $ (left) and low $ \etacrit $ model $ \Cc $ (right). Shown are comparisons of $ \eta $ (top panels) and $ r_h $ (bottom panels) between original grid resolution (blue), 3/4 the number of cells (red), and 1/2 the number of cells in the grid (yellow).
		}
		\label{fig:convergence}
	\end{figure*}
    
    Highly demanding 3D grids are naturally run in lower grid resolutions compared to 2D simulations. Thus, it is particularly important to verify that the results are independent of the grid resolution chosen for the simulations. We present two convergence tests for high and low $ \etacrit $ values, for which different grids were employed. For the numerical convergence tests we choose our canonical $ \alpha = 2 $ models: $ \Ca $ and $ \Cc $ for high and low $ \etacrit $ models, respectively. For each of those models we run additional two simulations whose grids are distributed in the same way of our original grids, but their number of cells is 3/4 and 1/2 of that in the original grid.
    Fig. \ref{fig:convergence} depicts the comparison of $ \eta $ (top) and $ r_h $ (bottom) for models $ \Ca $ (left) and $ \Cc $ (right) between the different simulations.
    Both quantities show a remarkable agreement between the different grid resolutions.
    The $ \eta $ plots show full convergence between all resolutions. Model $ \Ca $ shows no clear trend between one resolution to another, and model $ \Cc $ indicates that it takes more time for lower resolutions to merge with the higher resolution curve. The $ r_h $ plots show very small deviations, with somewhat convergence trends towards our original resolution which moves somewhat slower, as expected from higher resolution simulations \citep{Harrison2018}.

\label{lastpage}
\end{document}